\documentclass[useAMS,usenatbib]{mn2e}
\usepackage[latin1]{inputenc}
\usepackage{verbatim}
\usepackage{aas_macros}
\usepackage{graphicx}
\usepackage{amsmath}
\usepackage{color}

\title[Biases and systematics in observationally-derived galaxy properties]
{Biases and systematics in the observational derivation of galaxy properties: comparing
different techniques on synthetic observations of simulated galaxies.}
\author[Guidi, Scannapieco \& Walcher]{Giovanni Guidi$^{1}$, Cecilia Scannapieco$^{1}$ and C. Jakob Walcher$^{1}$\\
$^1$ Leibniz-Institut f\"ur Astrophysik Potsdam (AIP), An der Sternwarte 16, D-14482, Potsdam, Germany}

\begin{document}

\date{Accepted \today \ Received ...; in original form ...}

\pagerange{\pageref{firstpage}--\pageref{lastpage}} \pubyear{2014}

\maketitle

\begin{abstract}
We study the sources of biases and systematics in the derivation
of galaxy properties of observational studies, focusing on
stellar masses, star formation rates, gas/stellar metallicities,
stellar ages and magnitudes/colors. We use
hydrodynamical cosmological simulations of galaxy formation,
for which the real quantities are known, and apply observational
techniques to derive the observables. We also make an
analysis of biases that are relevant for a proper
comparison between simulations and observations. 
For our study, we post-process the simulation outputs to calculate
the galaxies' spectral energy
distributions (SEDs) using Stellar Population
Synthesis models and also generating the
fully-consistent far UV-submillimeter 
wavelength SEDs with the radiative transfer code {\sc sunrise}.
We compared the direct results of simulations with the
observationally-derived quantities obtained in various ways, 
and found that systematic differences in all studied galaxy properties
appear, which are caused by: (1) purely observational biases, 
(2) the use of mass-weighted/luminosity-weighted quantities, 
with preferential sampling of more massive/luminous regions, 
(3) the different ways to construct the template of models when a fit to
the spectra is performed, and 
(4) variations due to the use of different calibrations, 
most notably in the
cases of the gas metallicities and star 
formation rates.
Our results show that large differences can appear 
depending on the technique used to derive galaxy properties.
Understanding these differences is of primary importance both for simulators,
to allow a better judgement on similarities/differences
with observations, and for observers, to allow a proper
interpretation of the data.
\end{abstract}


\section{Introduction}

In recent years, large galaxy surveys such as the 2dFGRS (Two-degree-field Galaxy
Redshift Survey, \citealt{Colles99}), SDSS (Sloan Digital Sky Survey,
\citealt{Abazajian03}) and 2MASS  (Two Micron All-Sky Survey, \citealt{Skrutskie06}),
have opened up the possibility to statistically 
study the properties of galaxies in the Local Universe,
revealing their great diversity:  even for a narrow
range in stellar mass, galaxies appear
in a large variety of morphologies, gas fractions,
star formation rates (SFRs) and chemical abundances.
These observations have also allowed to identify
important relations such as the mass-metallicity
\citep{Garnett87, Tremonti04}, and to  measure  the
corresponding scatter which encodes relevant
information on the galaxies' evolution. 
This wealth of data gives important insight on the process of  galaxy
formation and evolution, revealing the action of physical
mechanisms occurring in galaxies, 
both internal -- e.g. feedback \citep{Fabian12}, cooling \citep{Thoul95} -- and
in relation to larger-scale mechanisms -- mergers \citep{Naab07, Naab09}, 
interactions \citep{Scudder12, Stierwalt15}, accretion \citep{Putman12}. 
All these leave imprints on the shape of the spectral energy distributions (SEDs)
which constitute the primary source of information from large galaxy 
surveys.
In fact, in recent years it became possible to obtain
the  full SEDs of  galaxies at 
wavelengths from the X-Ray
to the radio. In particular for galaxy studies, 
wavelengths from the ultraviolet to the far infrared
are the most relevant as the light coming from the stars
and interstellar gas/dust dominates the spectra in this
wavelength range.

In addition to observations, numerical simulations are also a useful tool 
to study the formation
and evolution of galaxies and to investigate the links between
a galaxy's formation, merger and accretion history and its final
properties. Understanding these relations is relevant for
a correct interpretation of observations of 
galaxies at different cosmic epochs, and for a reconstruction
of the galaxy's histories.
Progress on the simulation of realistic galaxies
has been presented in many recent works (e.g. \citealt{Governato07, Scannapieco08, Aumer13, Vogelsberger14, Nelson15, Schaye15})
that, together with advances in computational resources,
are starting to allow the simulation of
relatively large cosmological volumes, recreating
a virtual universe where galaxies are naturally diverse as the result
of their individual evolution.

Despite the individual progress in observations and simulations,
much uncertainty remains in the comparison between them
even when these are used to decide on the successes and
failures of the models and are recognized as an important
aspect of the interpretation of observational results. 
In fact, previous works using simulations
have shown that, as the methods usually applied to derive the properties of 
simulated galaxies are very different from observational techniques, 
several biases might be introduced making the comparisons 
unreliable 
\citep[e.g.][]{Abadi03, Governato09, Scannapieco10, Snyder11, Munshi13, 
Christensen14}. 
For example, \cite{Scannapieco10} showed that disc-to-total ratios
can vary up to a factor of 250$\%$ when analysed using 
kinematic or photometric disc-bulge(-bar) decompositions.
In addition, some authors pointed out the 
effects of the stochastic 
sampling of the Initial Mass Function (IMF) in observations, 
in particular when properties are
derived fitting Stellar Populations Synthesis (SPS) models 
(e.g. \citealt{Krumholz15}).
In order to properly judge the agreement
between simulations and observations and make it possible
to better understand observational results, it is of primary
importance that these biases are well understood.

In this work, we do a systematic analysis of the sources
of biases in the comparison between simulated and observed galaxies,
creating synthetic SEDs for our simulated galaxies and
then deriving the galaxy properties using different 
techniques that mimic those used in galaxy surveys.
As the {\it true} values for properties such as SFRs, metallicities,
ages and stellar masses are known in the simulations, 
we can quantify how close the values obtained observationally
are to the real ones.
To create the synthetic SEDs, we use three methods with increasing
complexity: first, we use dust-free
SPS models; second, we include
dust using a 
simple parametrization; and third, we
calculate the transfer 
of the galaxy light through 
the Interstellar Medium (ISM). In this last case, we 
post-process the simulation outputs with the 3D polychromatic Monte Carlo radiative transfer 
code {\sc sunrise} \citep{Jonsson09}, which simulates the propagation of photons 
in a dusty ISM.
Using the synthetic SEDs, we derive stellar masses, stellar ages, stellar
and gas metallicities and SFRs. By comparing the direct
results of the simulations with the observationally-obtained quantities,
we study ($i$) how reliable and meaningful simple comparisons between
simulated and observed galaxies are, ($ii$) how  the different
assumptions involved in the creation of the SED affect the derived
galaxy properties, and  ($iii$) how we can
reliably  test the agreement between observations and models.

In this paper, we describe the techniques used
to create the synthetic spectra of our simulations and discuss
differences between the different methods of estimating
galaxy properties. In 
a companion paper (Guidi et al., in prep., PaperII hereafter), we
focus on the comparison of our simulated galaxies with data from the SDSS dataset.
This paper is structured as follows. In Section 2 we describe the 
simulations and the simulated galaxy sample,
Section 3 explains the techniques to create their SEDs
and discusses uncertainties and biases affecting them,
and in Section 4 we compare the magnitudes,
stellar masses, gas/stellar metallicities, stellar ages
and SFRs obtained using various methods.
Finally, in Section 5, we give our conclusions.

\section{The simulations}
\label{sec:simulations}

In this work, we use three sets of  galaxy simulations consisting
in total of fifteen galaxies formed in a $\Lambda$CDM universe. 
Each set comprises the same five galaxies but adopts 
a different modelling of chemical enrichment and feedback,
which is known to introduce differences in the final
properties of galaxies and in their evolution.
The focus of this work is to identify whether the analysis
techniques used to extract galaxy properties from simulations
and observations introduces important biases that make
the comparisons unreliable. Our set of fifteen galaxies is
ideal for this purpose as they all have similar total mass
but span a wide range in gas/stellar metallicities, stellar
ages and star formation rates.

As the goal of this work is not to decide how realistic the galaxies
are (which we discuss in PaperII), 
covering a variety
of galaxy properties allows one to test the biases properly,
unaffected by particular details of a given implementation. 
We use our sample to test minimum and maximum biases introduced
in the conversion of simulations into observables, as this
conversion is expected to primarily depend on the age of the stellar
populations, the stellar masses, the metallicities and the galaxy morphologies.
The latter is important as, on one side,  dust will influence differently
observations of face-on and edge-on galaxies and, on the other hand,
the presence of gradients in galaxy properties might affect
the derivation of their global properties from the observationally-obtained 
fiber quantities, since fiber spectrographs such as those used
for SDSS sample only the inner region
of galaxies.

The initial conditions correspond to five galaxies
which are the hydrodynamical counterparts of (a subset of) the
Aquarius halos \citep{Springel08}. These are similar in
mass to the Milky Way and formed in isolated environments (no neighbour exceeding
half their mass within 1.4 Mpc at redshift $z=0$), but
have different merger and accretion histories, as discussed
in \cite{Scannapieco09}. The galaxies have virial masses between $0.7$ and $1.7\times 10^{12}$M$_\odot$
(calculated within the radius where the density contrast is 200 times
the critical density),
stellar masses of $1-10\times 10^{10}$M$_\odot$,
and gas masses of $3-10\times 10^{10}$M$_\odot$.

For our first
set of simulations, 
we have used the  extended version of the Tree-PM SPH code
Gadget-3 \citep{Springel05}, which includes star formation,
chemical enrichment, supernova Type Ia and TypeII feedback,
metal-dependent cooling and a multiphase
model for the gas component which allows the coexistence of  dense and 
diffuse phases \citep{Scannapieco05,Scannapieco06}.
These simulations have been first presented in \citet{Scannapieco09} and further
analysed in \citet{Scannapieco10} and  \citet{Scannapieco11}, and will 
be called throughout this
paper  {\it A(-E)-CS} or {\it CS} galaxies.
The Scannapieco et al. code has been extensively used for
simulating galaxies of a wide range of total/stellar masses,
 and shown to be successful in reproducing
the formation of galaxy discs from cosmological initial conditions of Milky
Way-mass galaxies
(\citealt{Scannapieco08}, see also
\citealt{Saw10, Scannapieco12, Nuza14, Creasey15, Scannapieco15}).

For our second simulation set, we used an updated version of the
model of Scannapieco et al., in relation to the
treatment of chemical enrichment (Poulhazan et al., in prep).
These simulations will be referred to as {\it A(-E)-CS$^+$} or {\it CS$^+$}.
The updated code implements a different Initial Mass Function
(Chabrier instead of Salpeter), chemical yields from \citealt{Portinari98} 
(while the CS model uses the \citealt{Woosley95} yields), 
and the treatment of feedback
from stars in the AGB phase, which
contribute significant amounts of given chemical elements, such
as carbon and nitrogen. For these reasons, the CS$^+$ galaxies
have systematically higher chemical abundances compared
to those in the CS sample.
The modelling of energy feedback is the same as in the standard
Scannapieco et al. code.

The final set of simulations, which will be referred to as
{\it A(-E)-MA} or the {MA} galaxies,  have
used the \cite{Aumer13} code, which is an independent update
to the \citet{Scannapieco06}  
model. This code has a different set  of chemical
choices in relation to the initial mass function and chemical
yields, includes also stars in the AGB phase, and assumes a
different cooling function.
More importantly, it has a different treatment of supernova
energy feedback: unlike in the Scannapieco et al. code, where
feedback is purely thermal, in Aumer et al. 
the supernova energy is divided into a thermal and a kinetic part,
and also includes feedback from radiation pressure.
This results in stronger feedback effects compared to the 
Scannapieco et al. model 
and, as a consequence, produces galaxies that are more disk dominated,
younger and more metal rich compared to the rest of our simulations.
We refer the interested reader to \cite{Aumer13} for full
details on this implementation.

The assumed cosmological parameters of the simulations are as follows: 
$\Omega_{\rm m} = 0.25 $, $\Omega_{\Lambda} = 0.75$,  $\Omega_{\rm b} = 0.04$, 
$\sigma_8 = 0.9$ and $\text{H}_0 = 100 \, h \, \text{km} \text{ s}^{-1} \text{Mpc}^{-1}$,
with $h=0.73$.
All simulations have similar mass resolution ($2-5\times 10^{5}$M$_\odot$ for stellar/gas particles and $1-2\times 10^{6}$M$_\odot$ for dark matter particles)
and adopt similar gravitational softenings ($300-700$ pc).

\begin{figure*}
  \centering
  {\textbf{A-CS\hspace{5.cm}A-CS$^+$\hspace{5.cm}A-MA}\par\medskip\vspace{-0.2cm}\includegraphics[width=2.7cm]{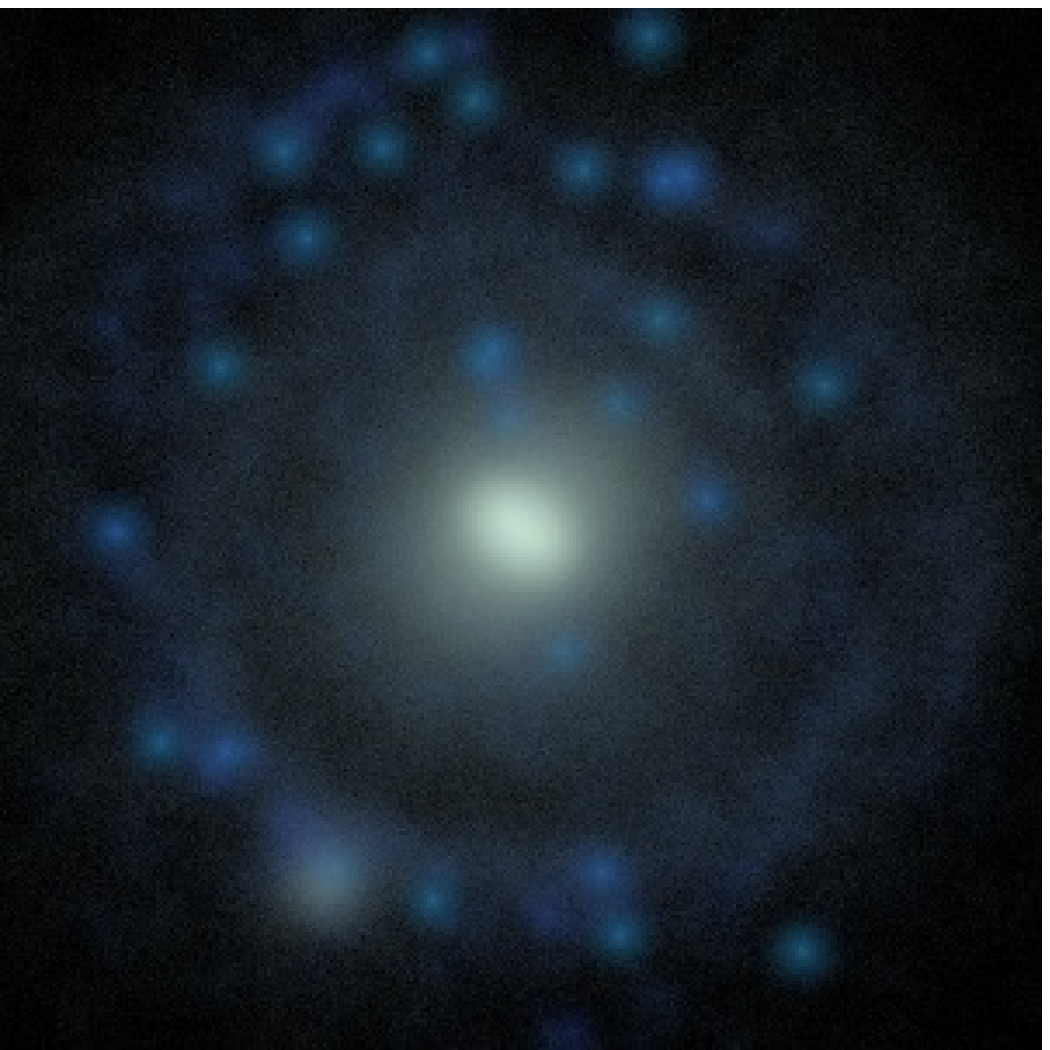}\includegraphics[width=2.7cm]{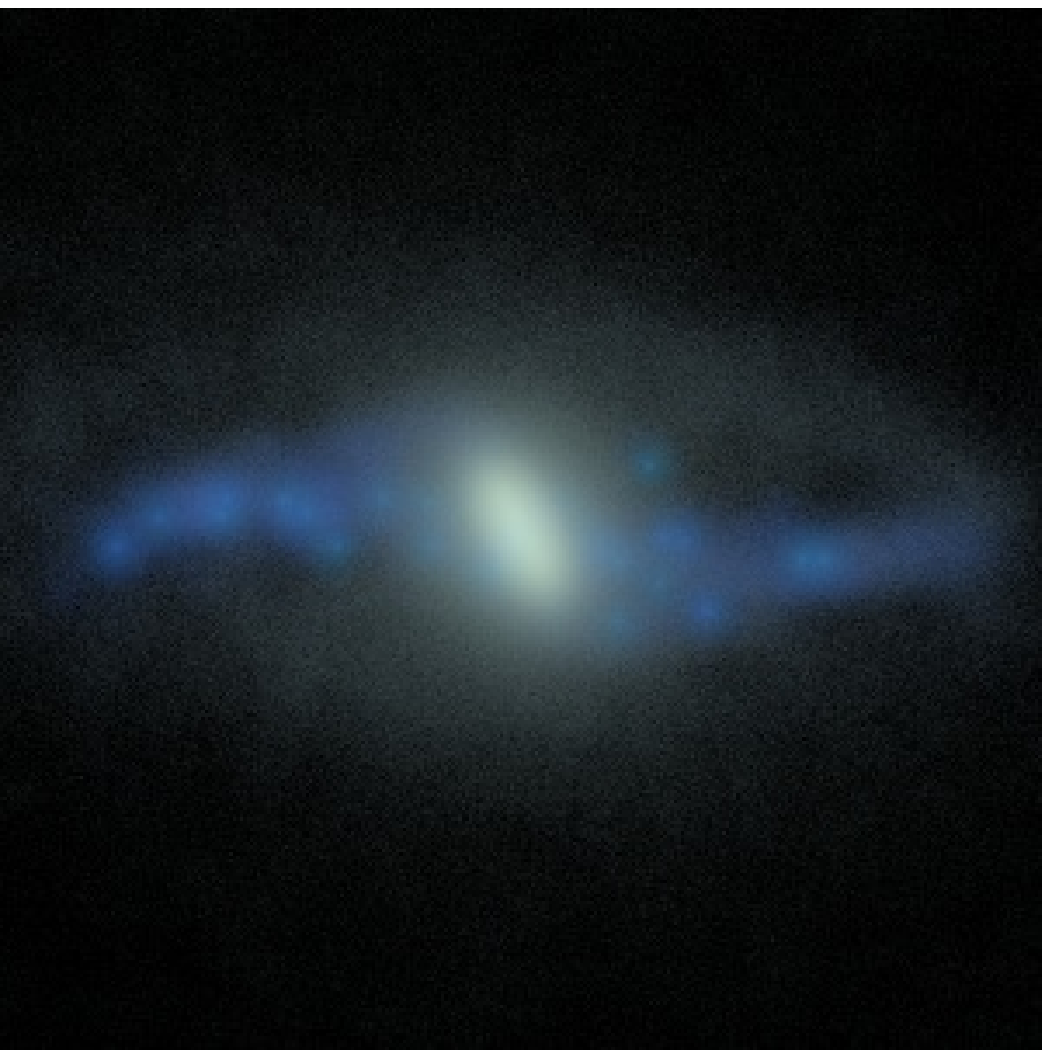}}\hspace{0.1cm}{\includegraphics[width=2.7cm]{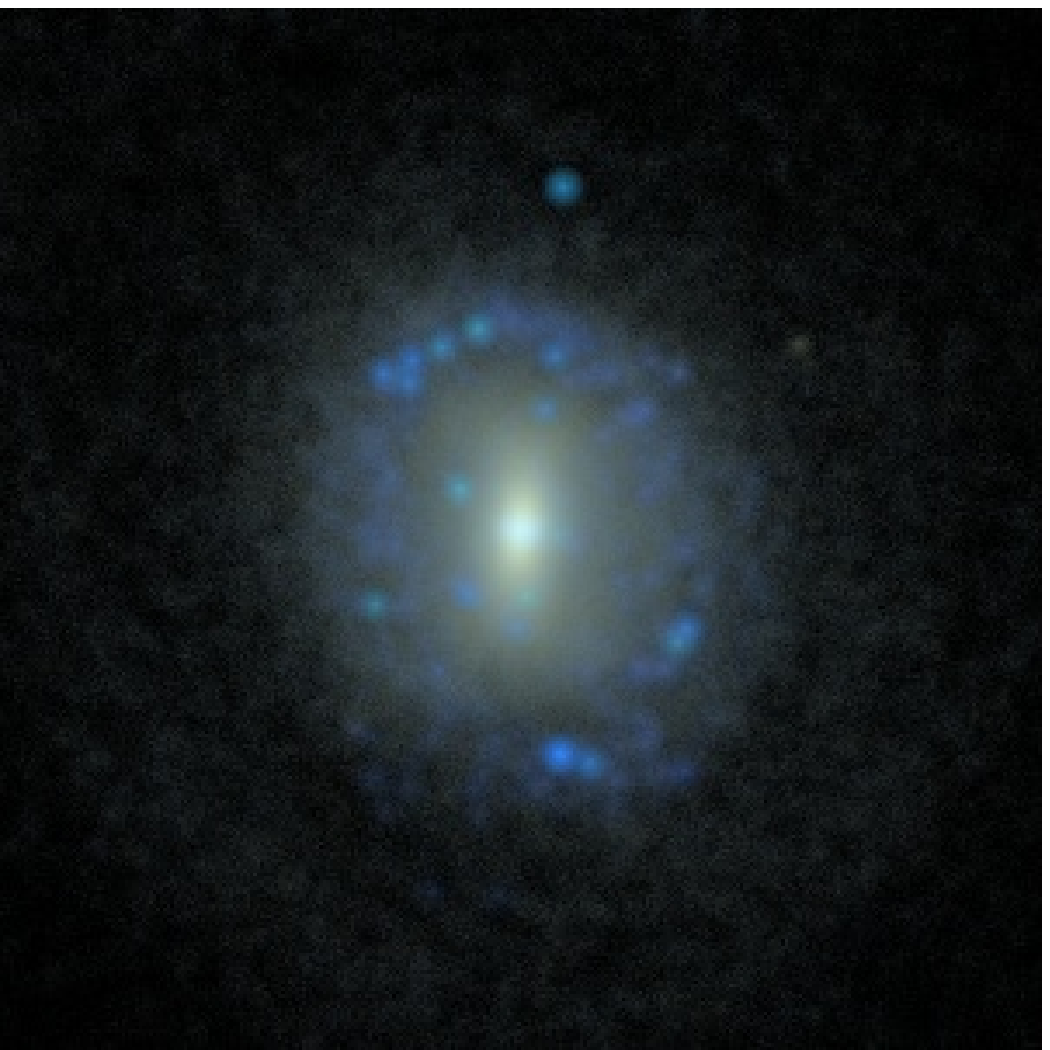}\includegraphics[width=2.7cm]{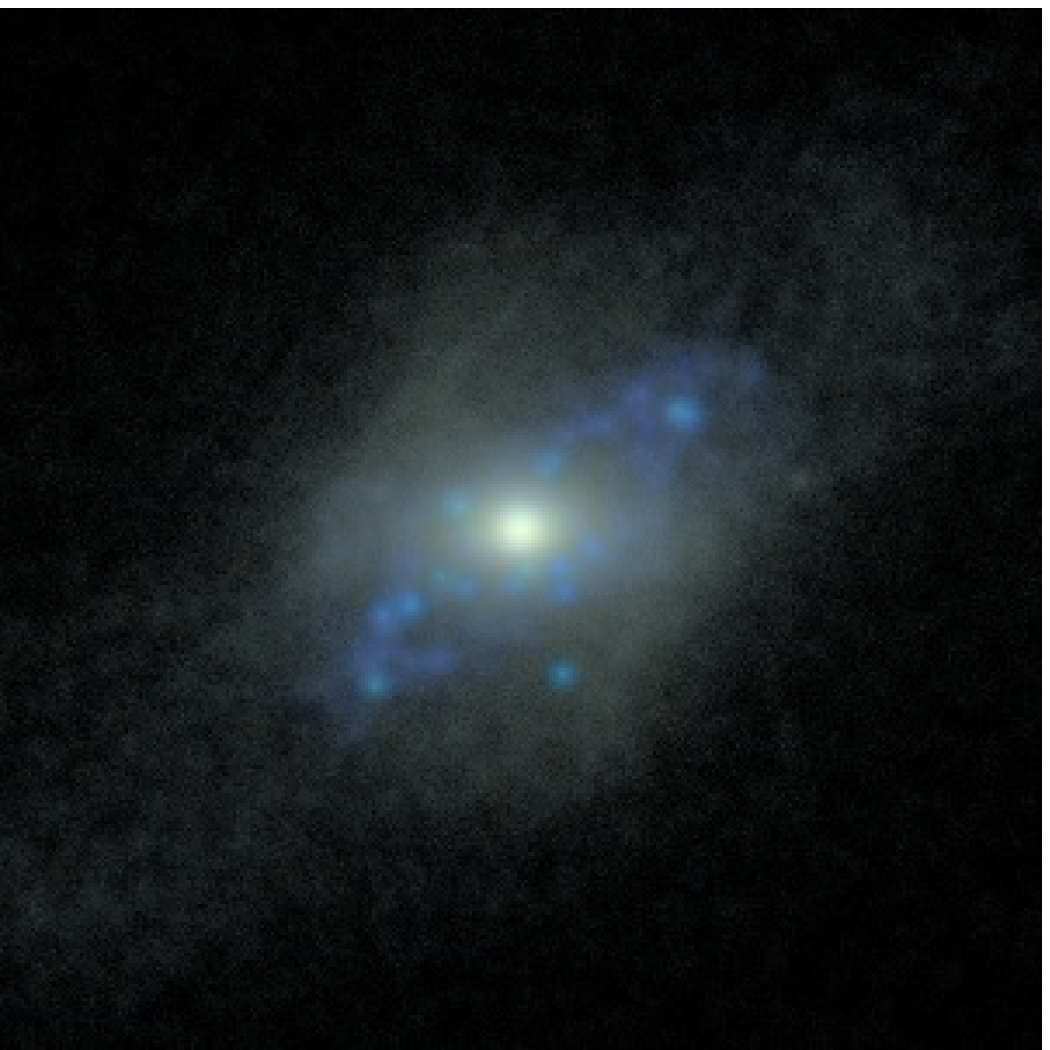}\hspace{0.1cm}\includegraphics[width=2.7cm]{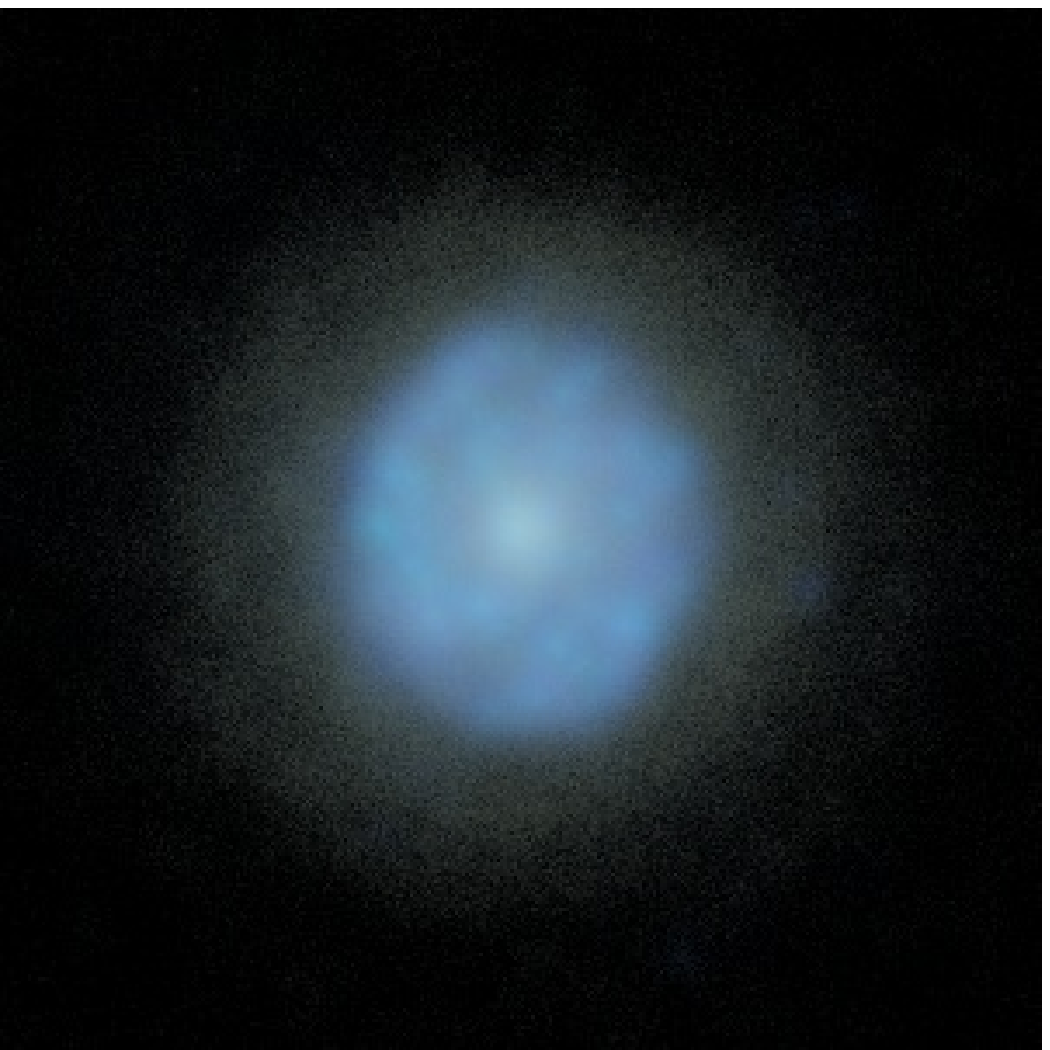}\includegraphics[width=2.7cm]{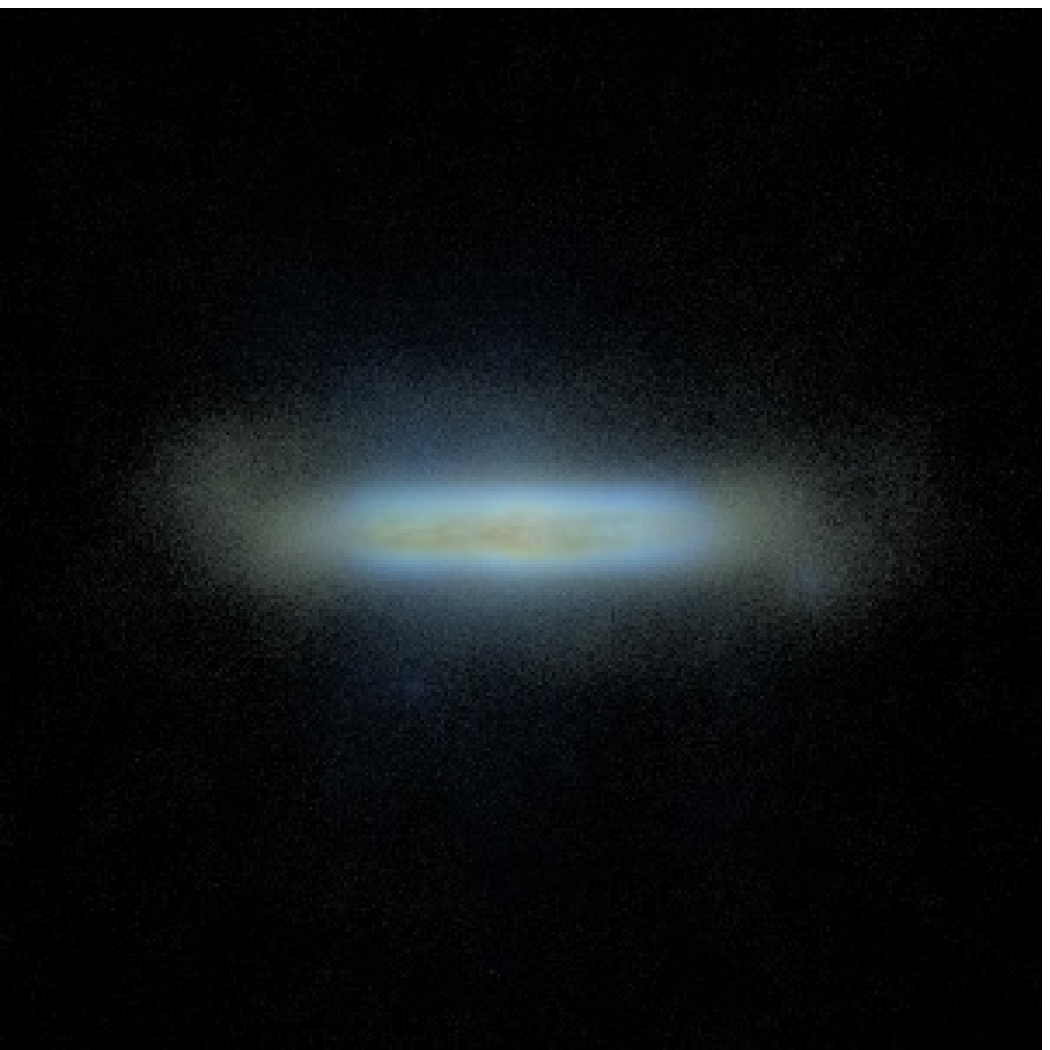}}

\vspace{0.2cm}

  {\textbf{B-CS\hspace{5.cm}B-CS$^+$\hspace{5.cm}B-MA}\par\medskip\vspace{-0.2cm}\includegraphics[width=2.7cm]{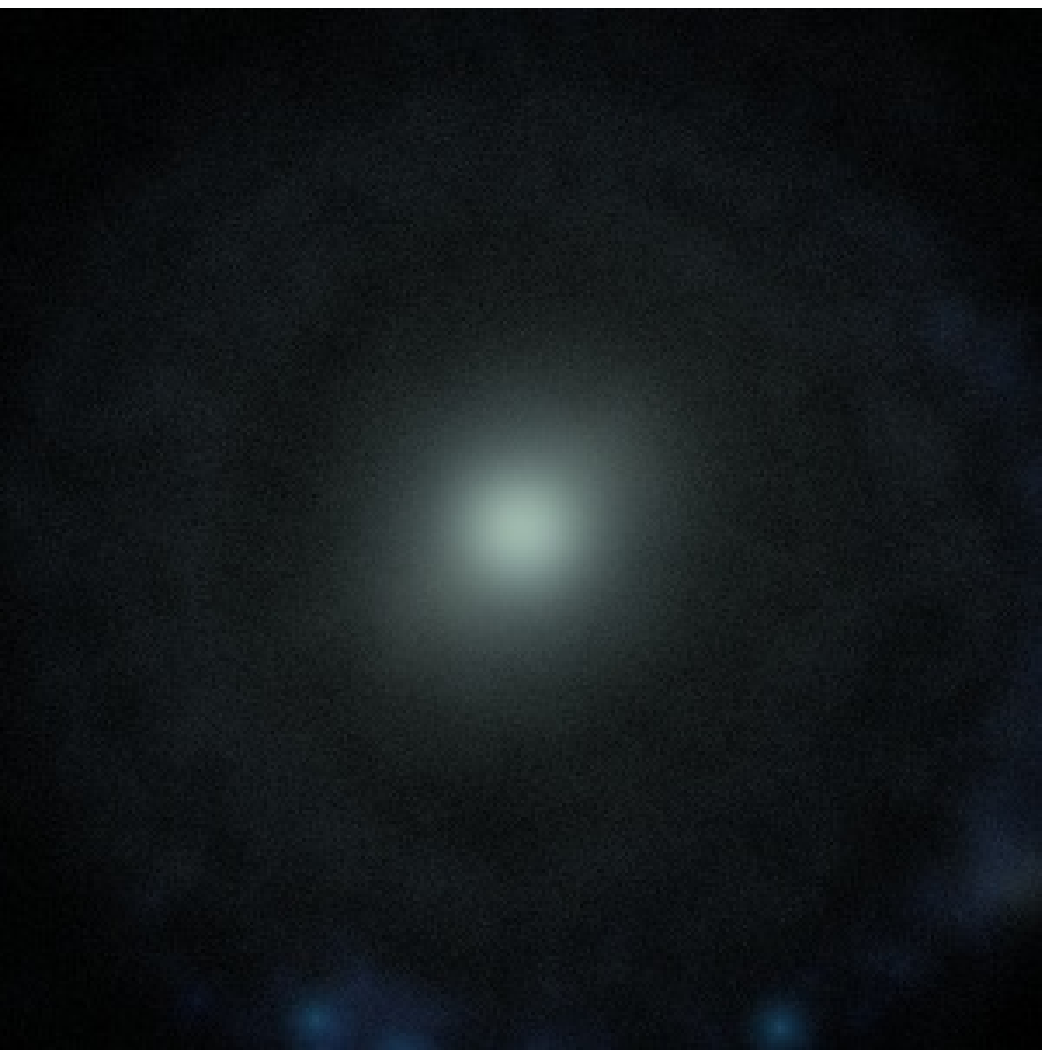}\includegraphics[width=2.7cm]{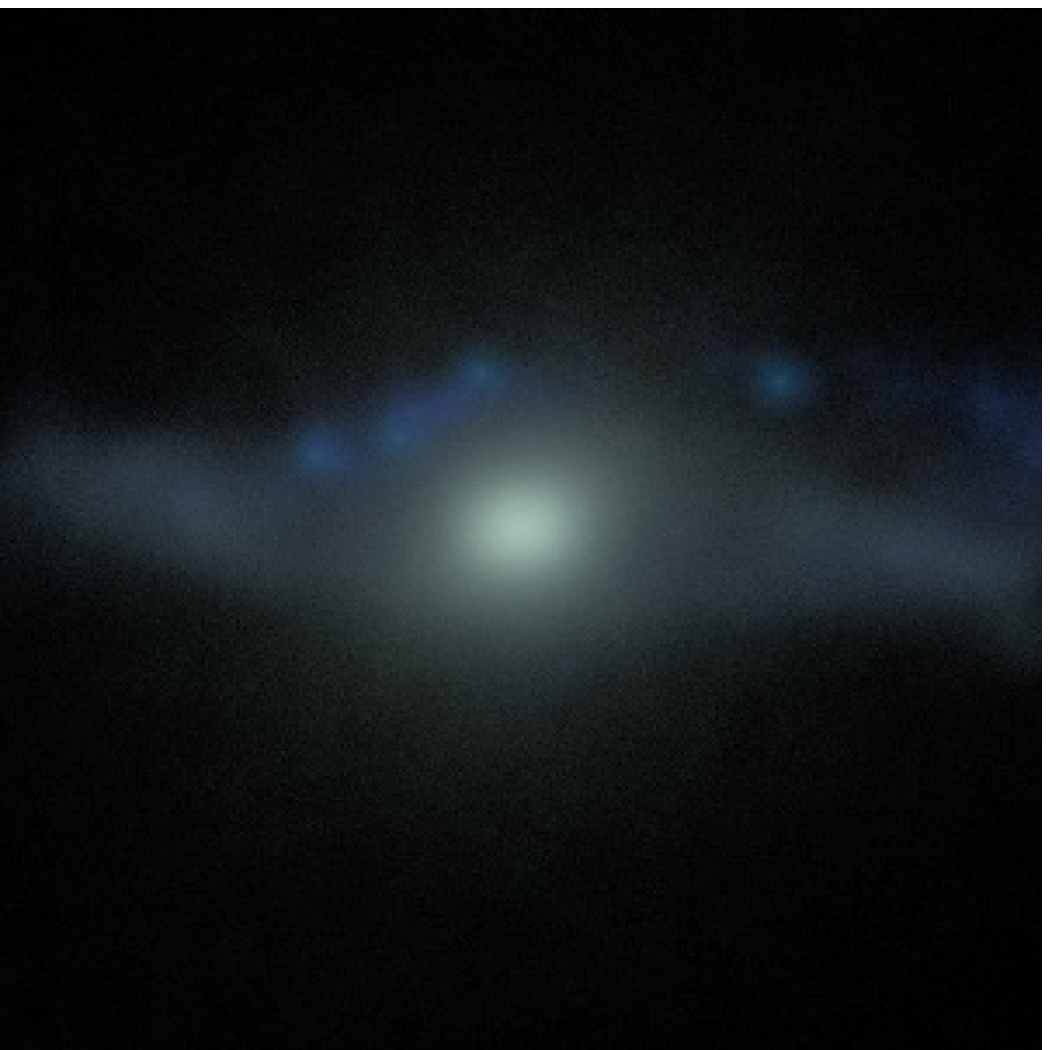}}\hspace{0.1cm}{\includegraphics[width=2.7cm]{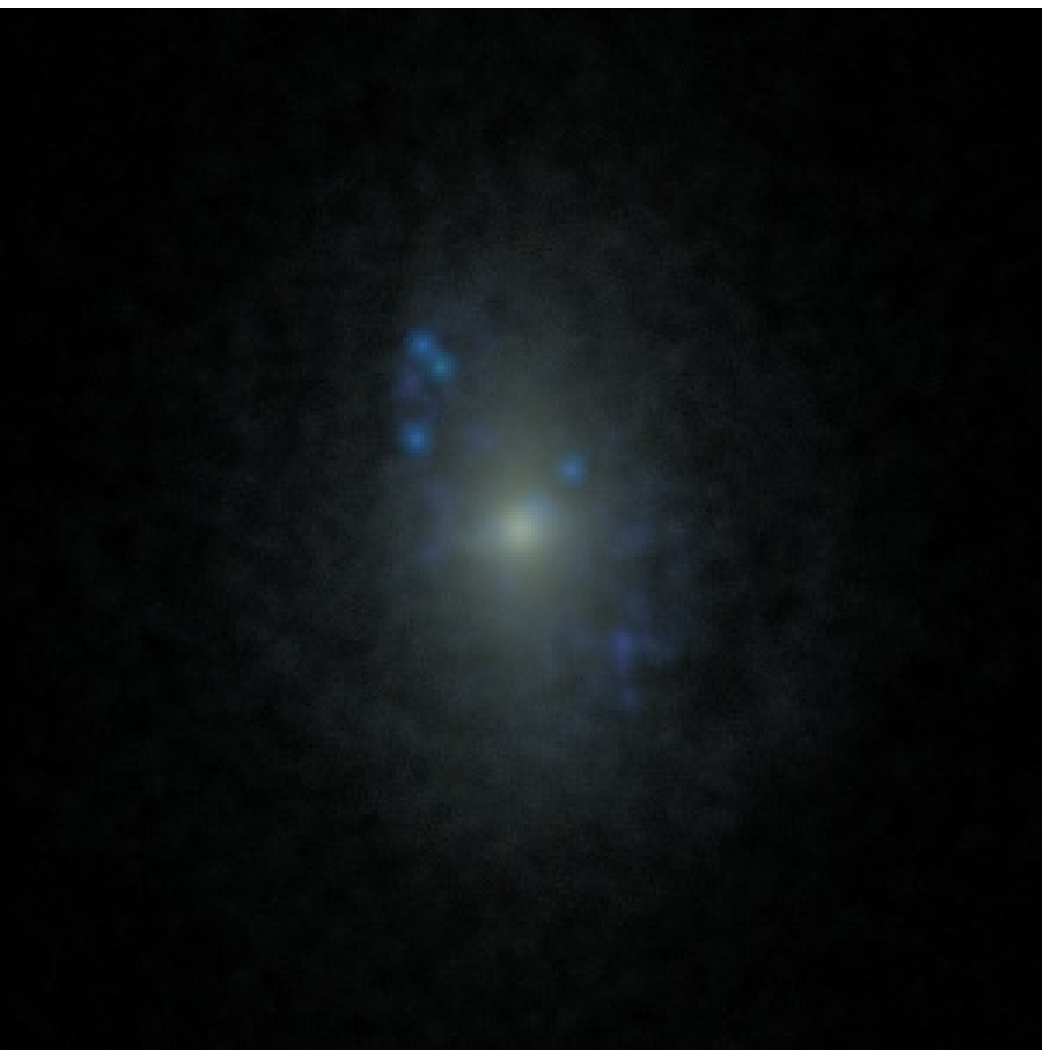}\includegraphics[width=2.7cm]{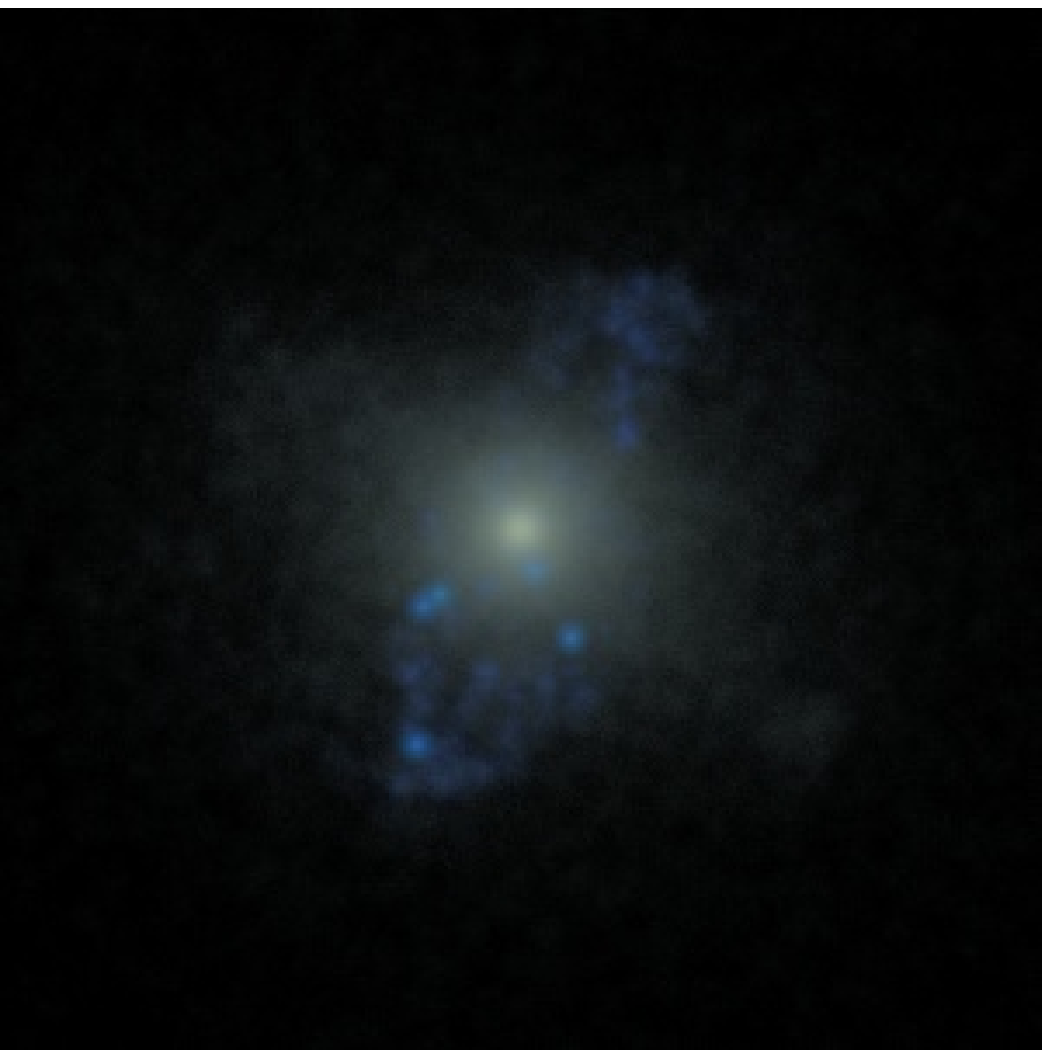}\hspace{0.1cm}\includegraphics[width=2.7cm]{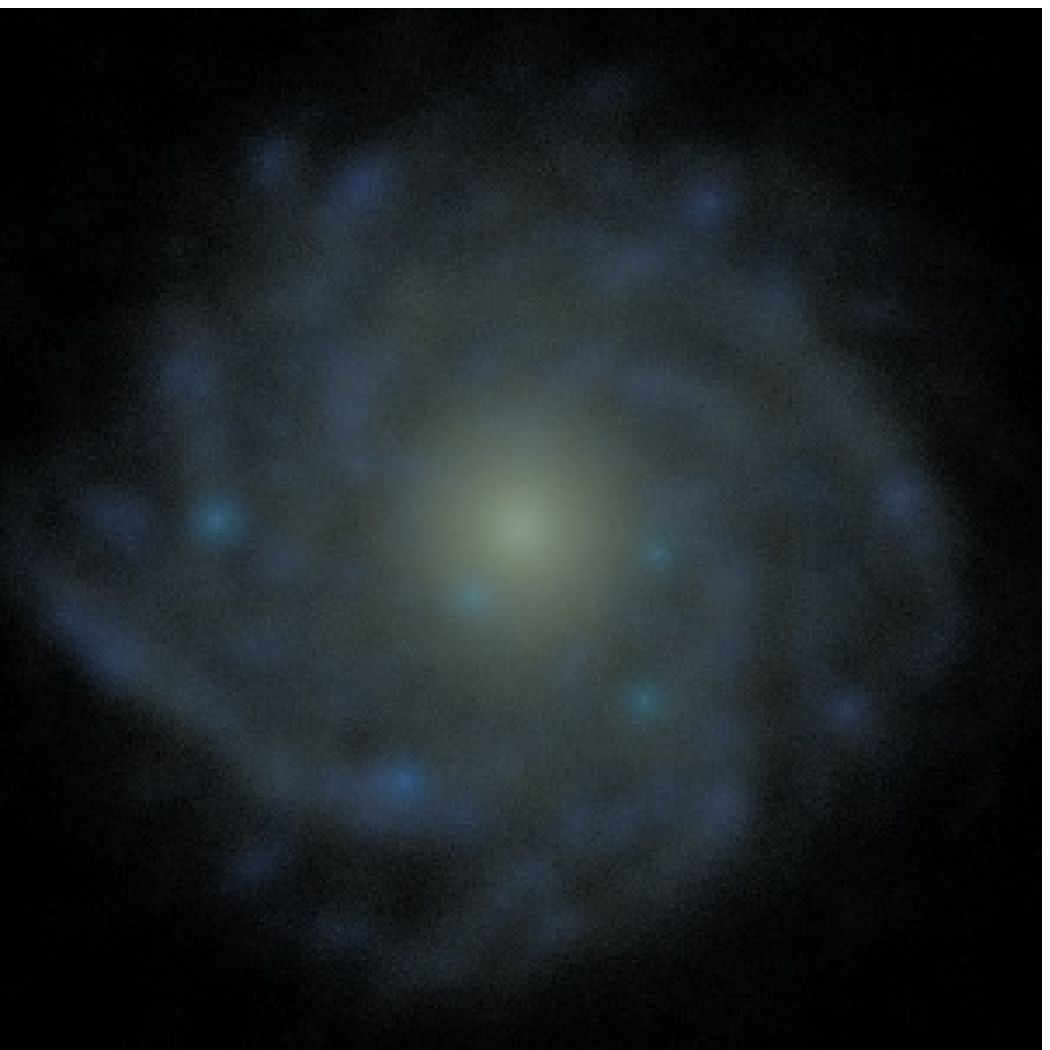}\includegraphics[width=2.7cm]{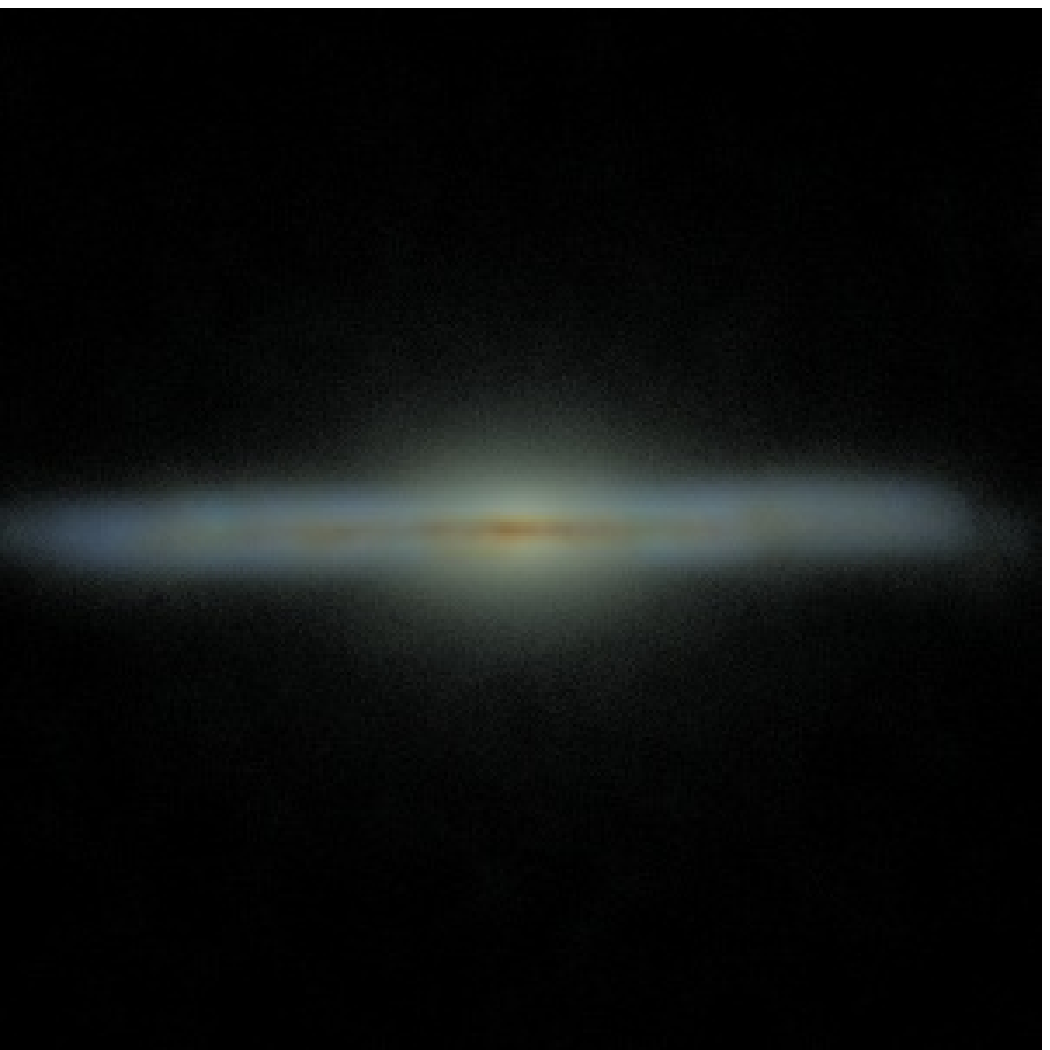}}

\vspace{0.2cm}

  {\textbf{C-CS\hspace{5.cm}C-CS$^+$\hspace{5.cm}C-MA}\par\medskip\vspace{-0.2cm}\includegraphics[width=2.7cm]{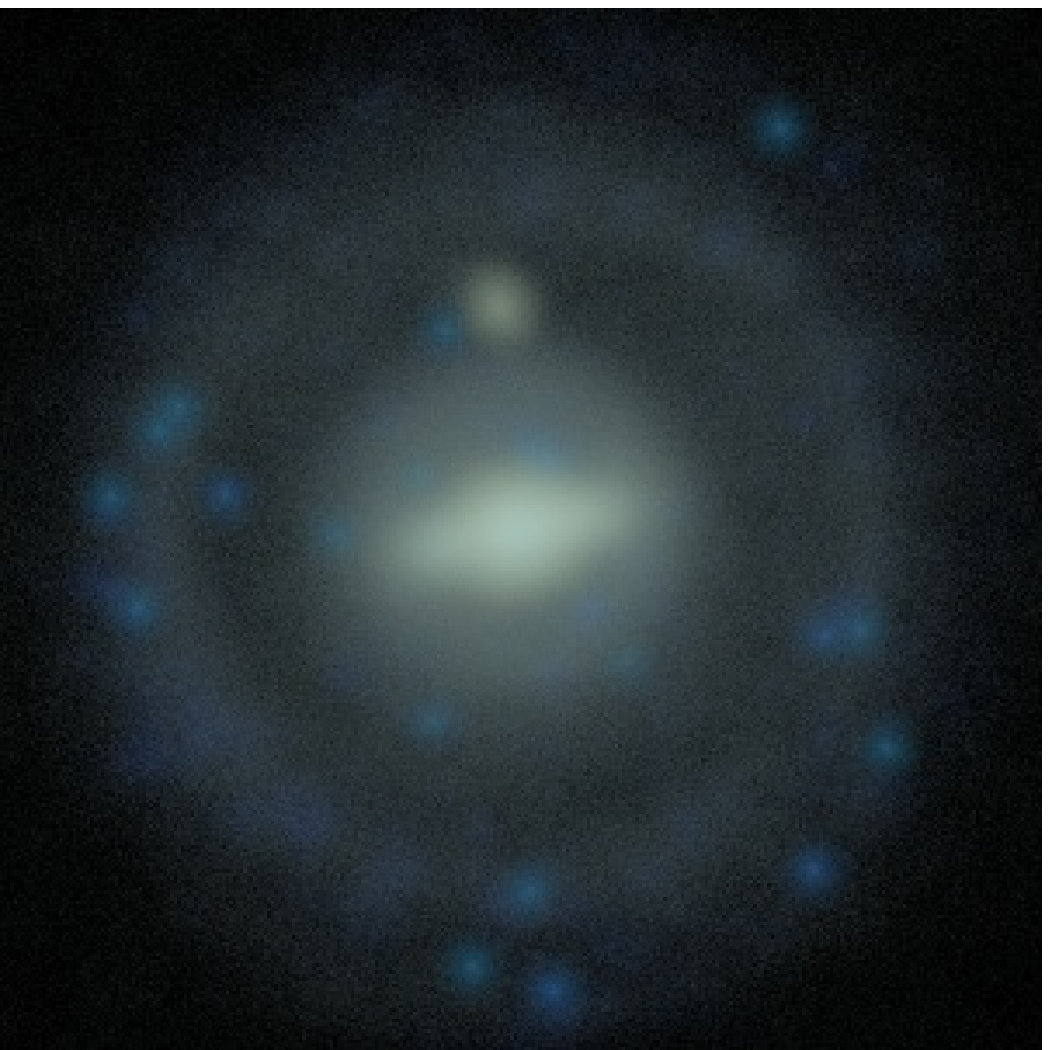}\includegraphics[width=2.7cm]{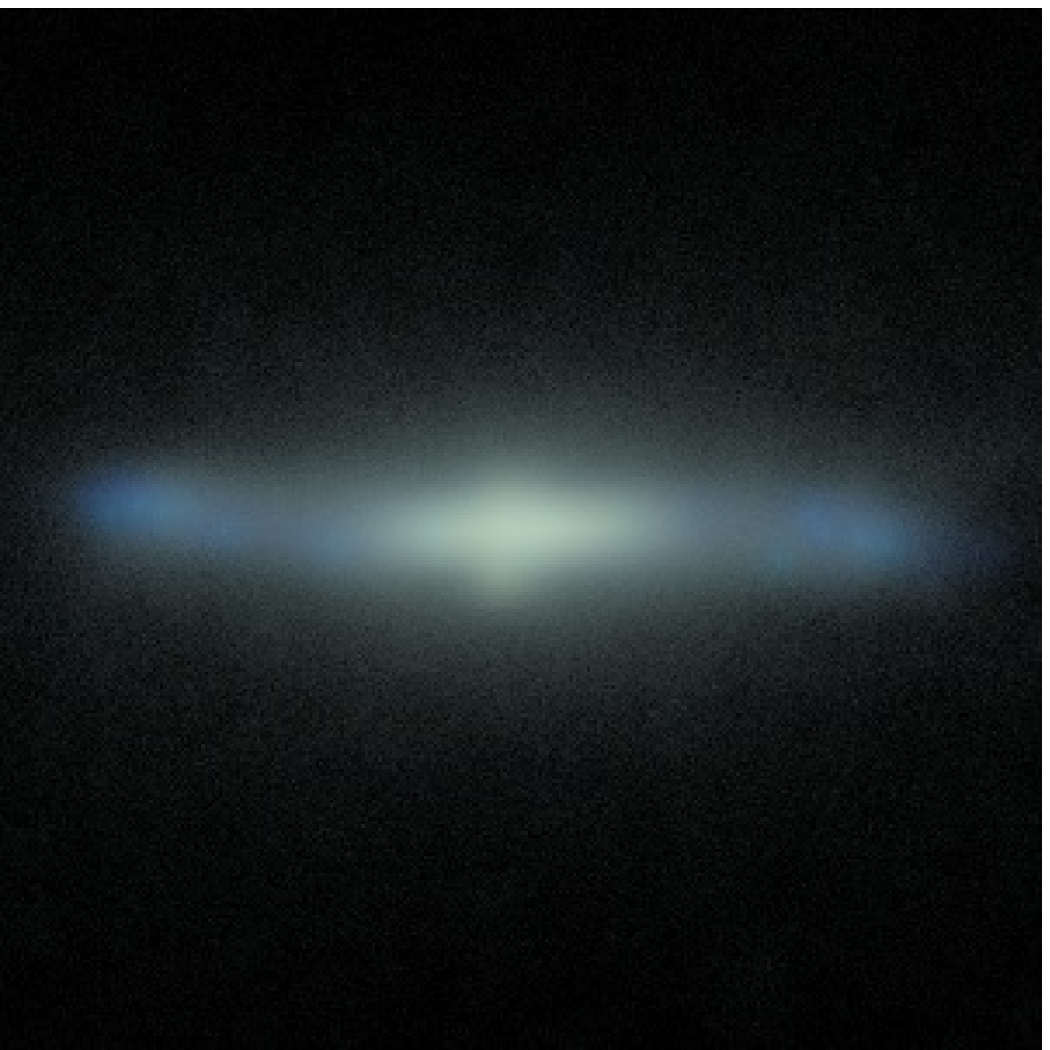}}\hspace{0.1cm}{\includegraphics[width=2.7cm]{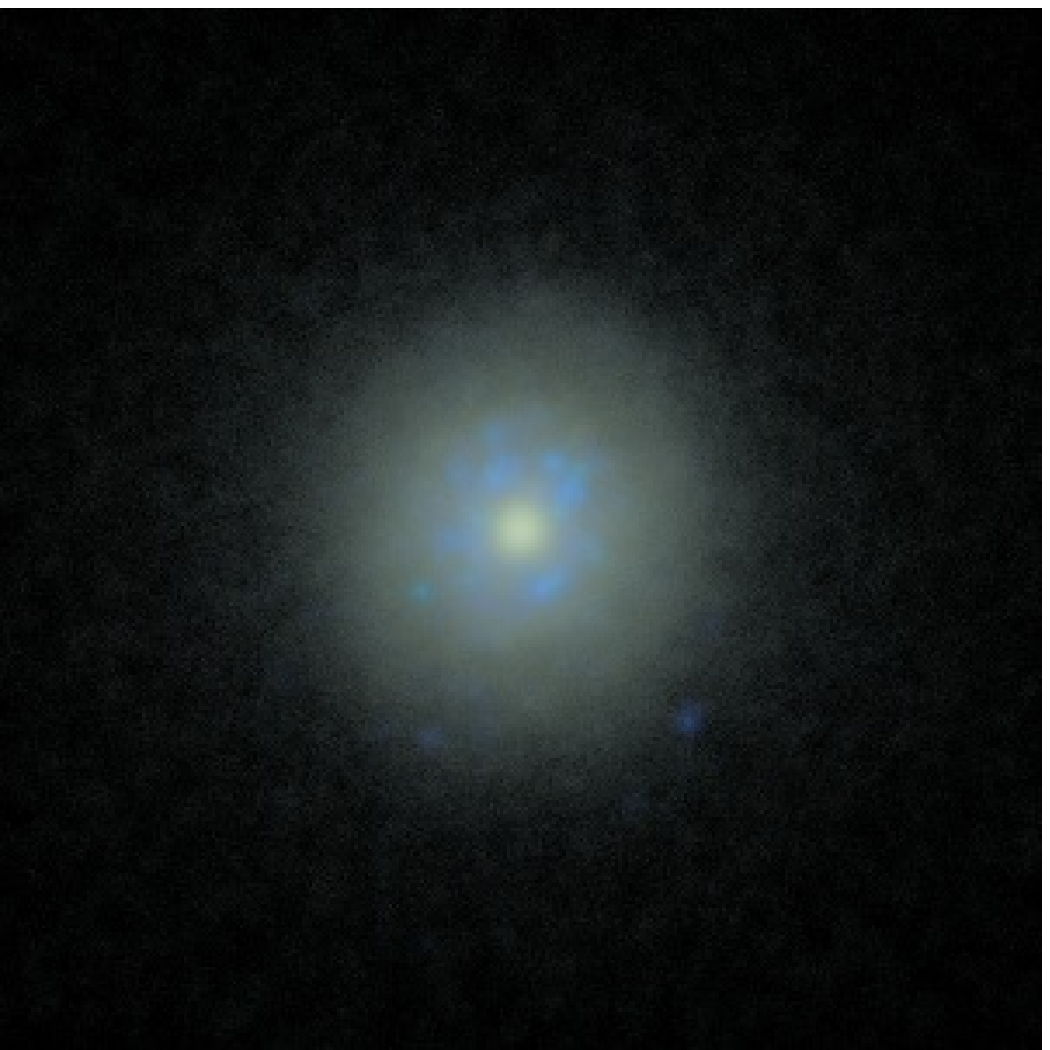}\includegraphics[width=2.7cm]{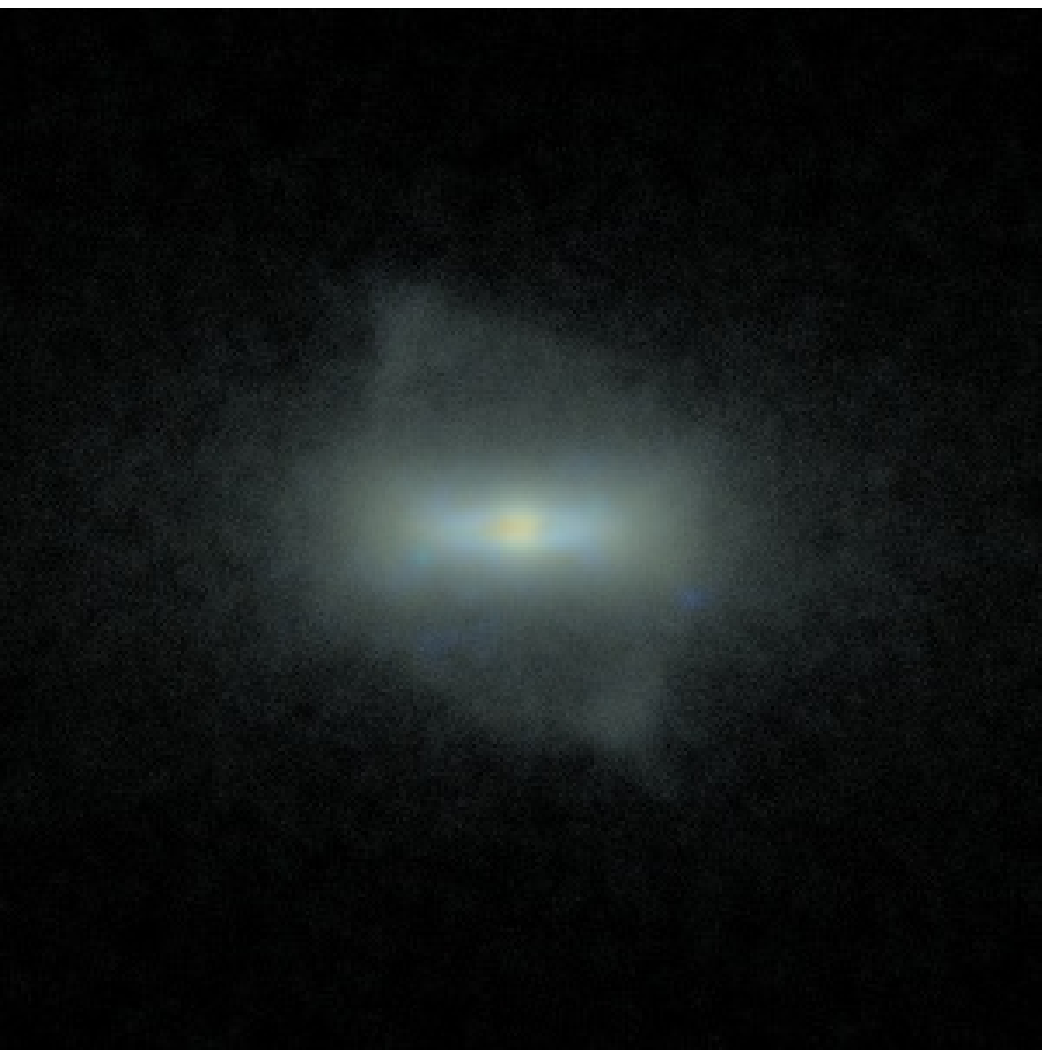}\hspace{0.1cm}\includegraphics[width=2.7cm]{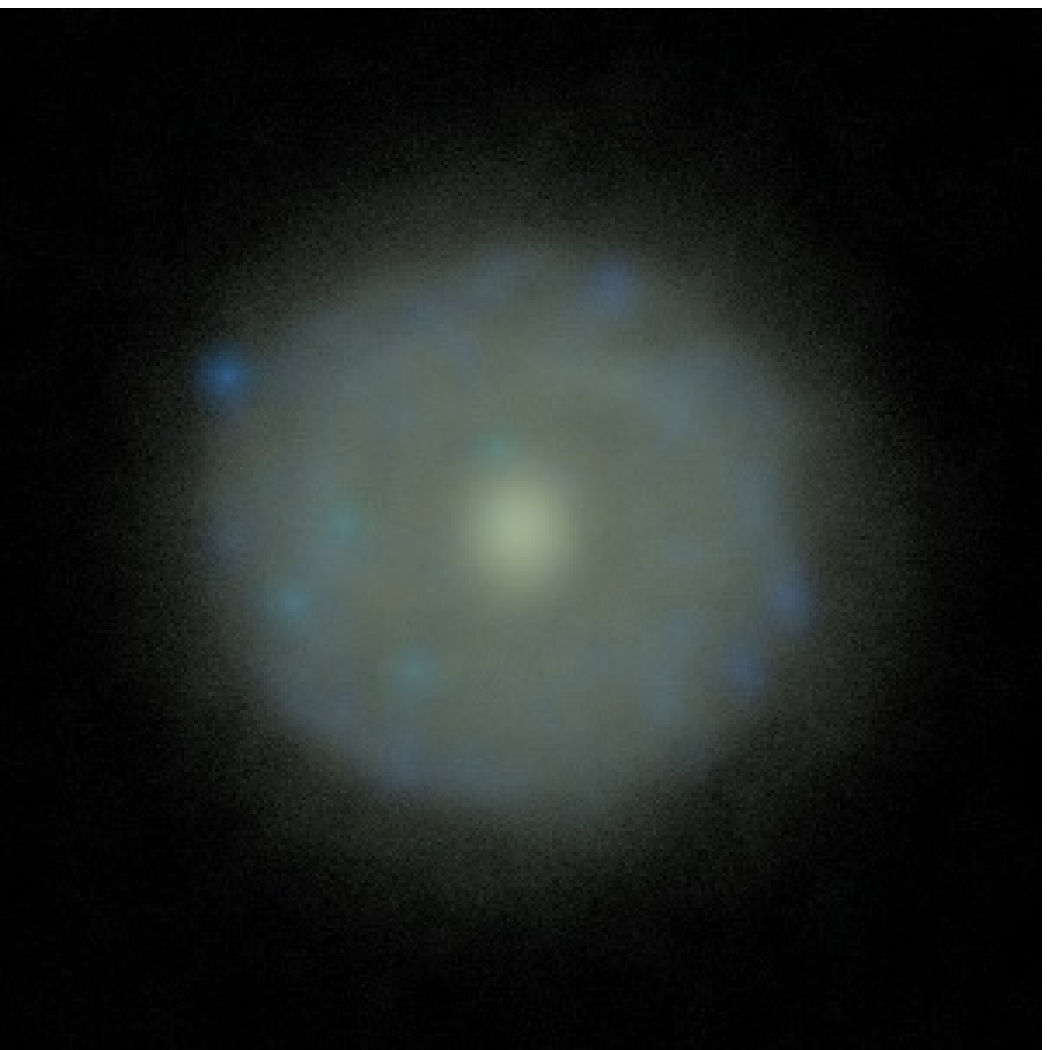}\includegraphics[width=2.7cm]{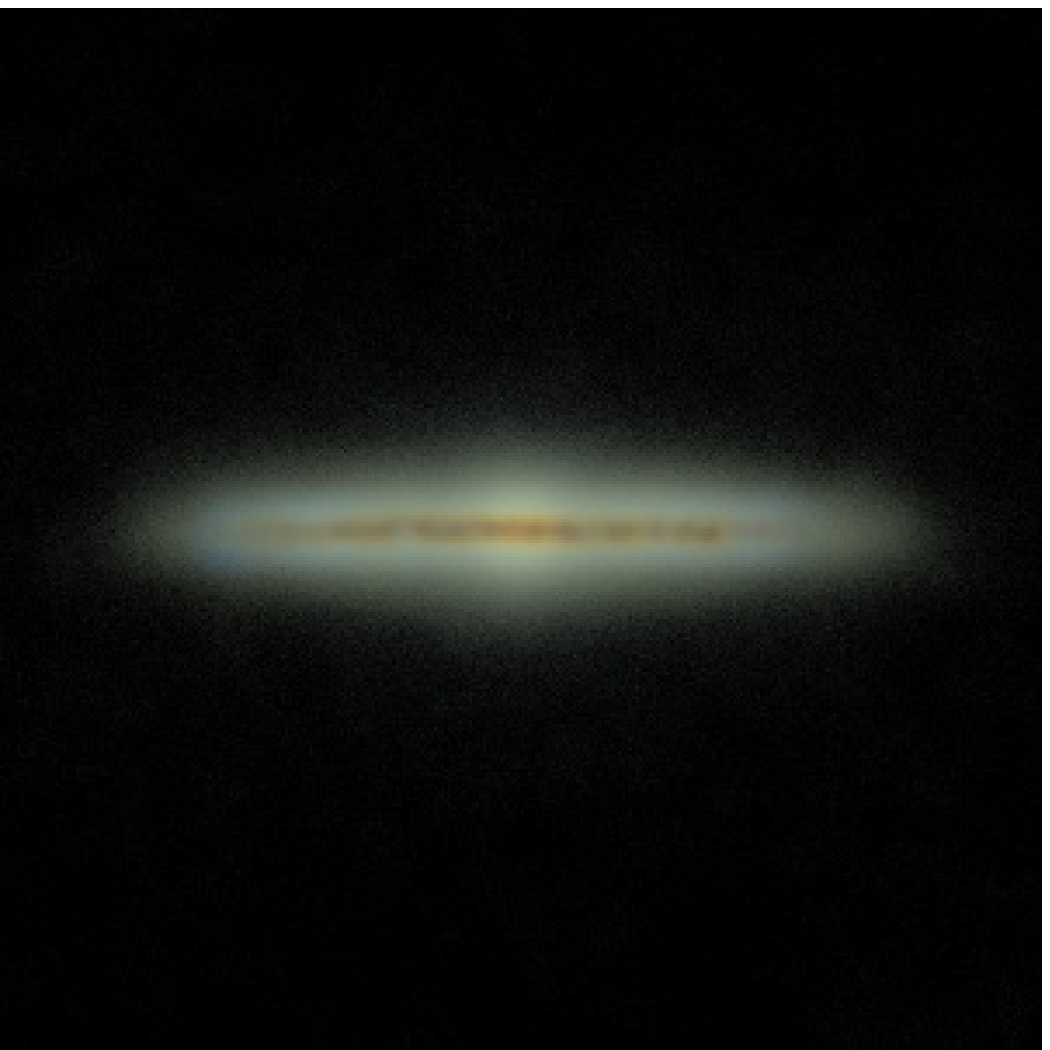}}

\vspace{0.2cm}

  {\textbf{D-CS\hspace{5.cm}D-CS$^+$\hspace{5.cm}D-MA}\par\medskip\vspace{-0.2cm}\includegraphics[width=2.7cm]{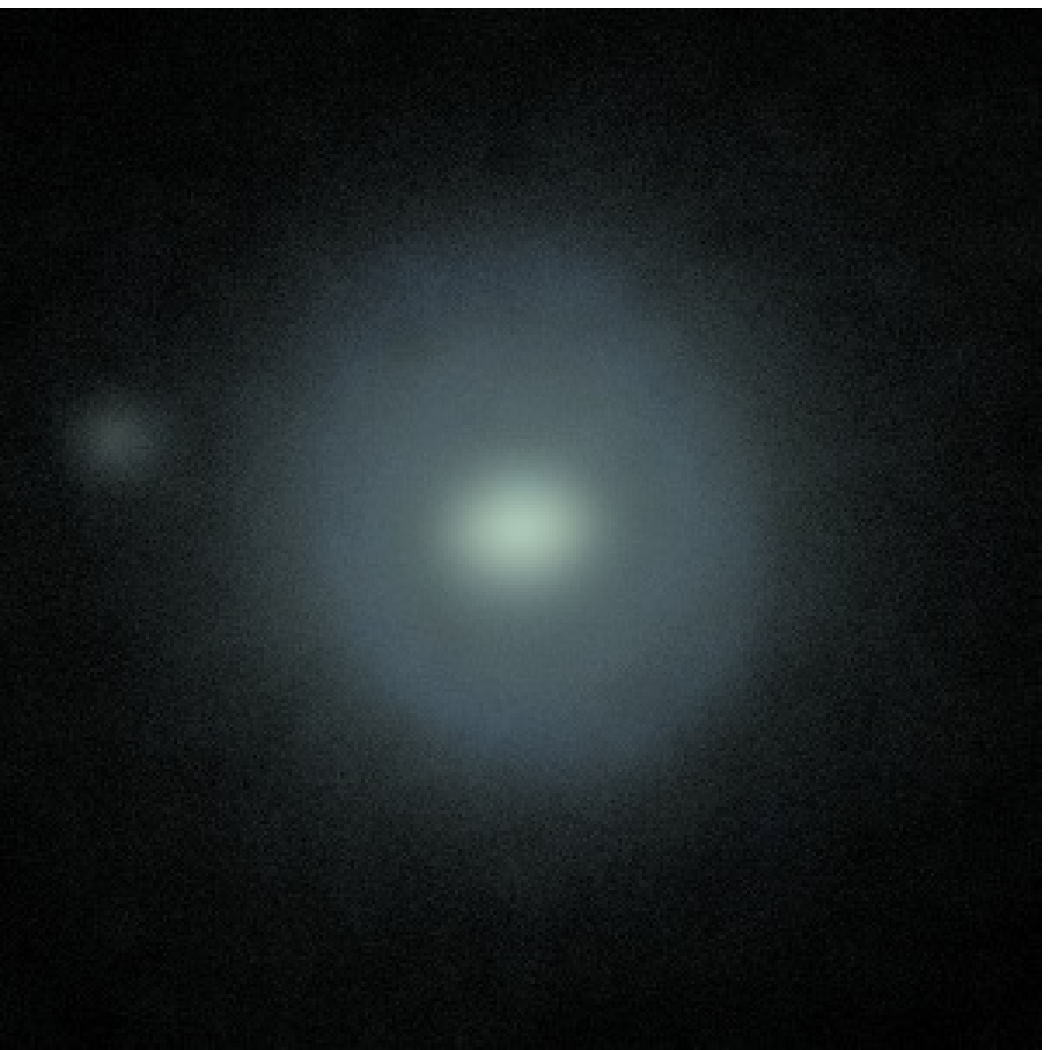}\includegraphics[width=2.7cm]{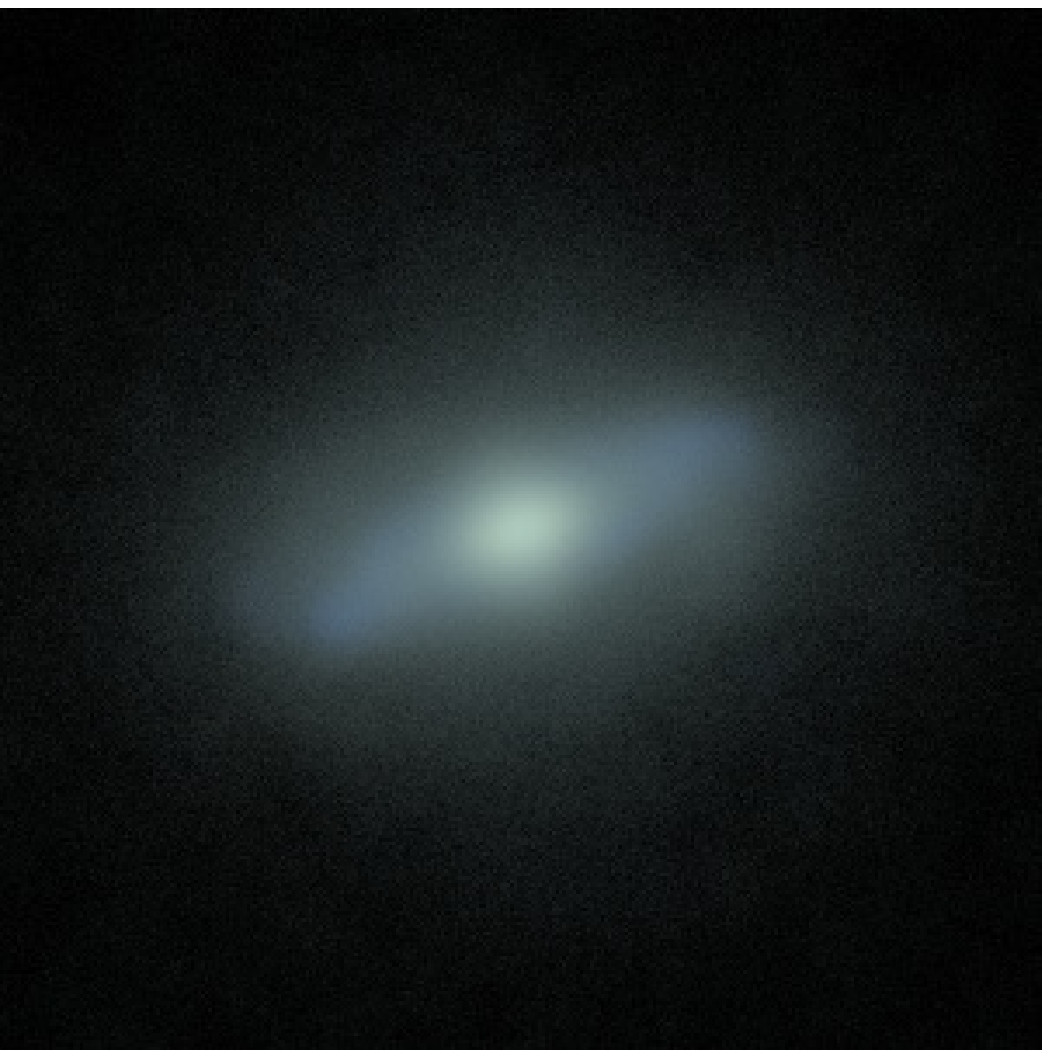}}\hspace{0.1cm}{\includegraphics[width=2.7cm]{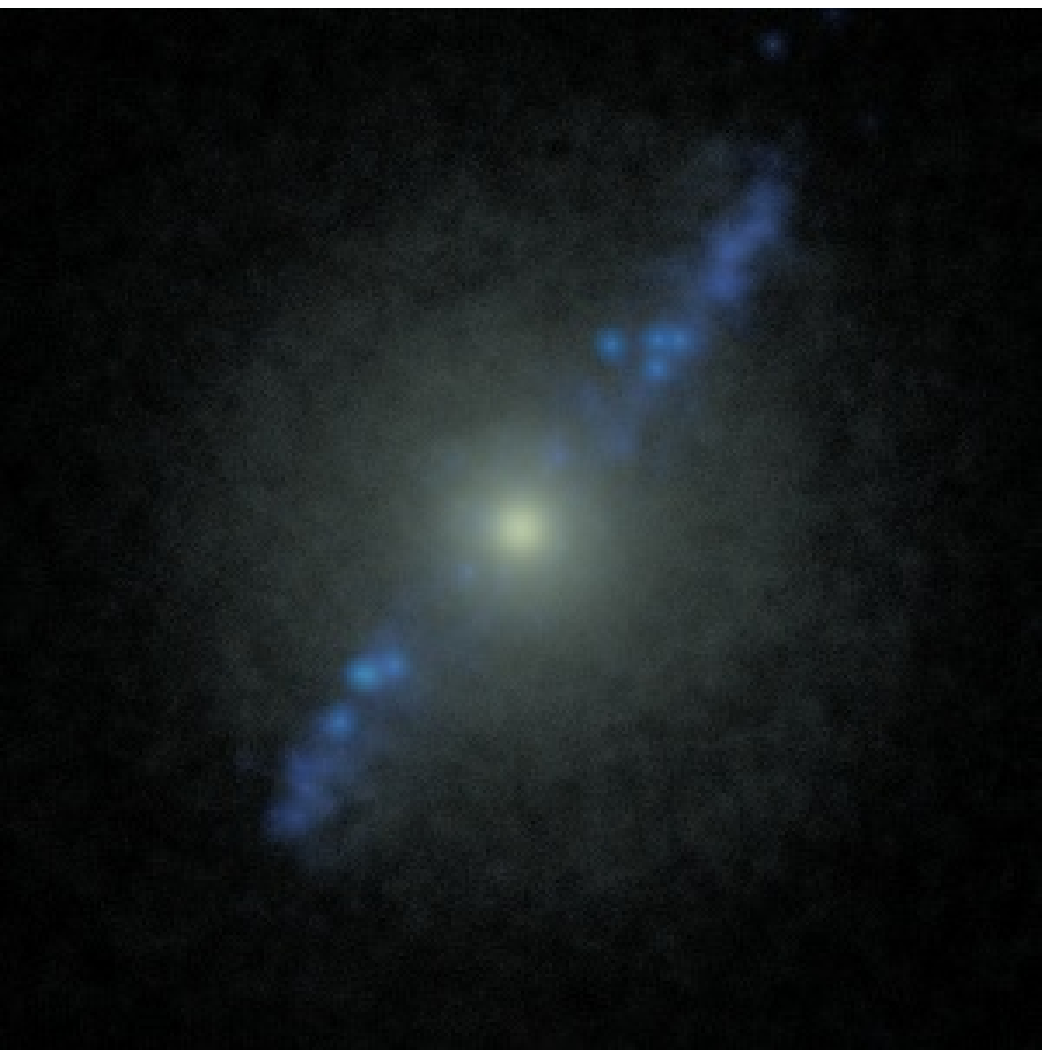}\includegraphics[width=2.7cm]{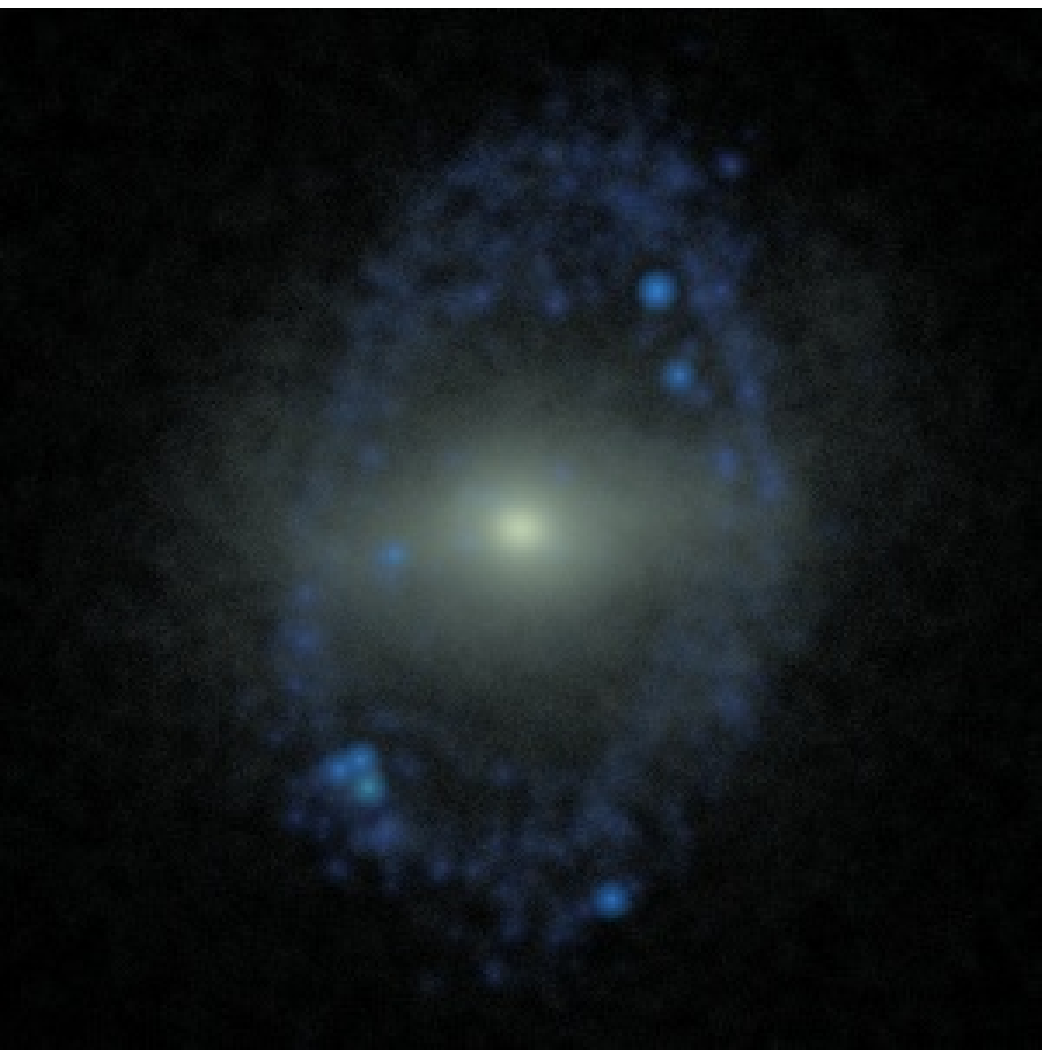}\hspace{0.1cm}\includegraphics[width=2.7cm]{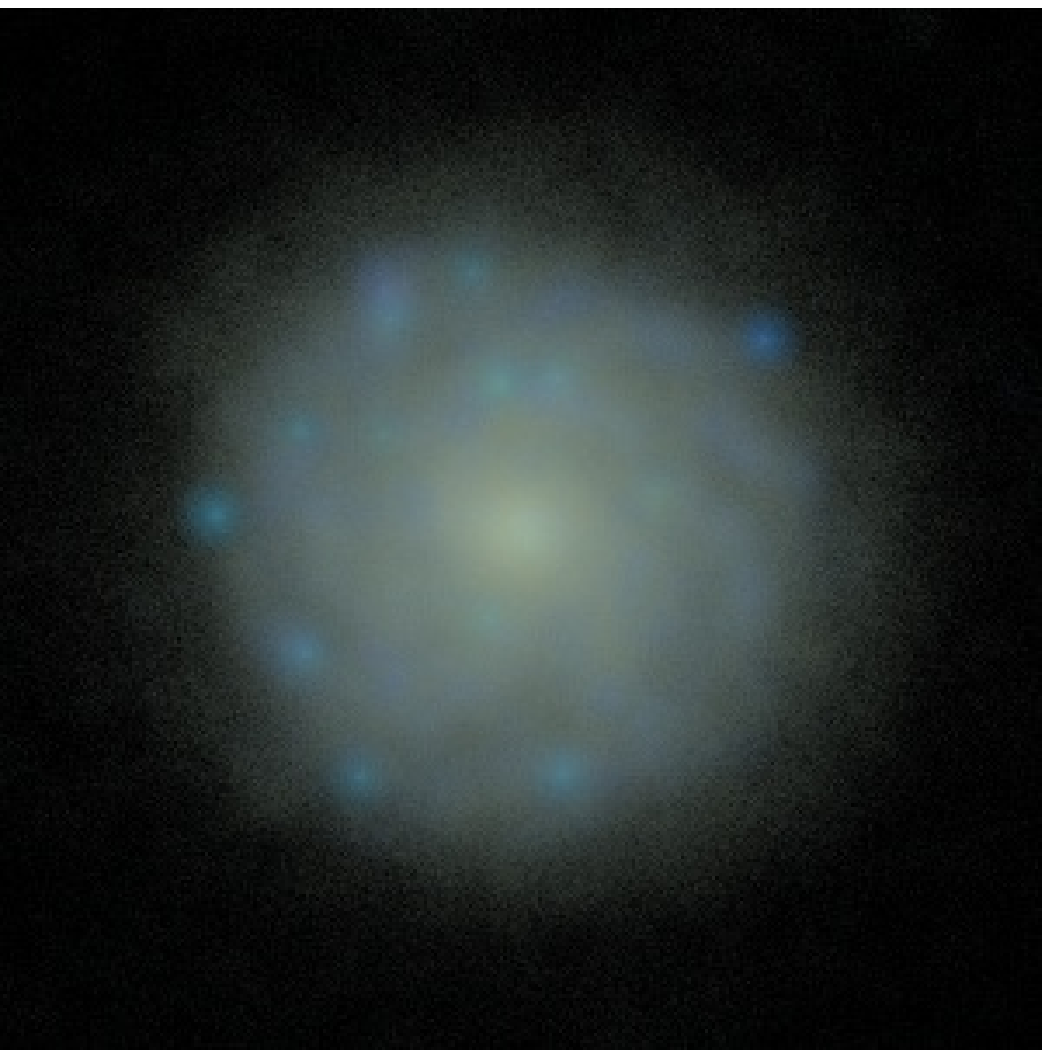}\includegraphics[width=2.7cm]{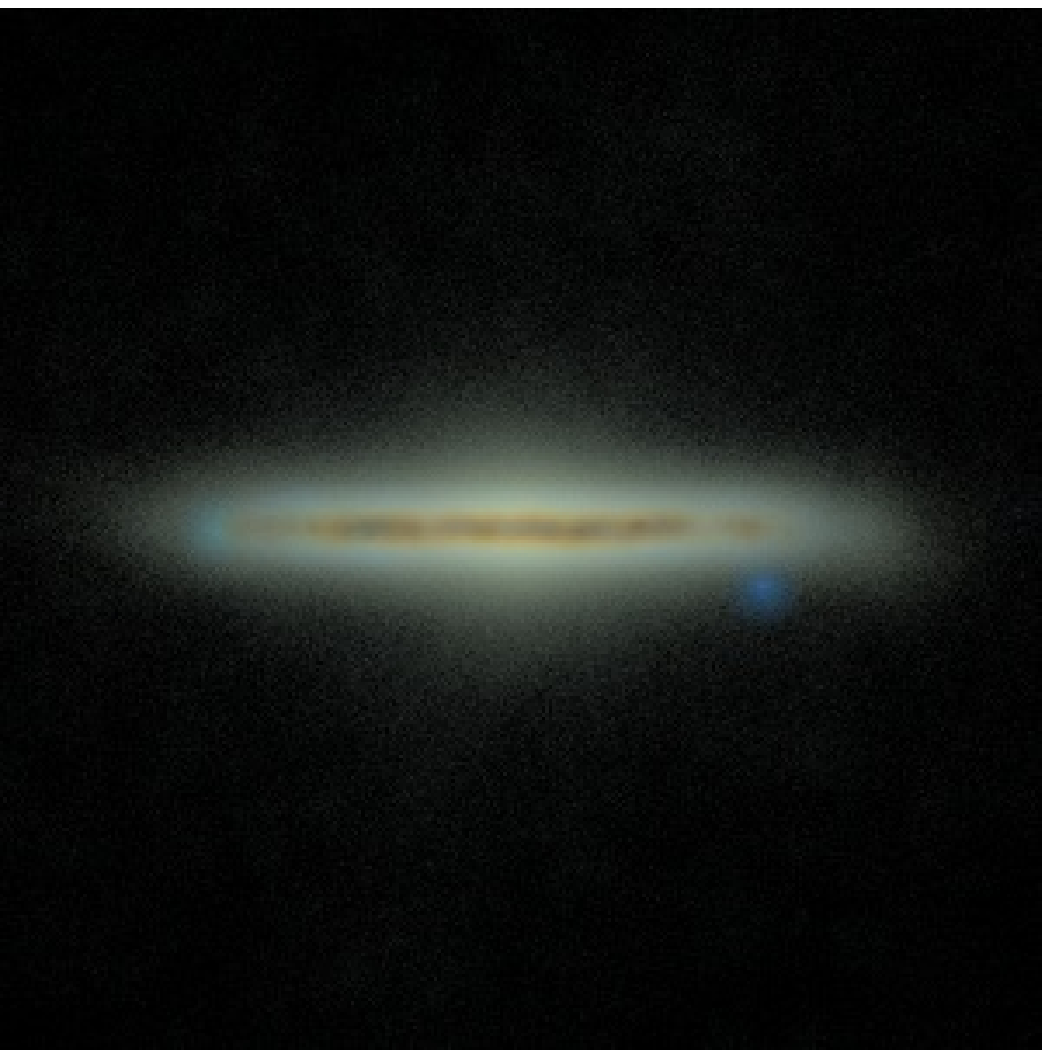}}

\vspace{0.2cm}

  {\textbf{E-CS\hspace{5.cm}E-CS$^+$\hspace{5.cm}E-MA}\par\medskip\vspace{-0.2cm}\includegraphics[width=2.7cm]{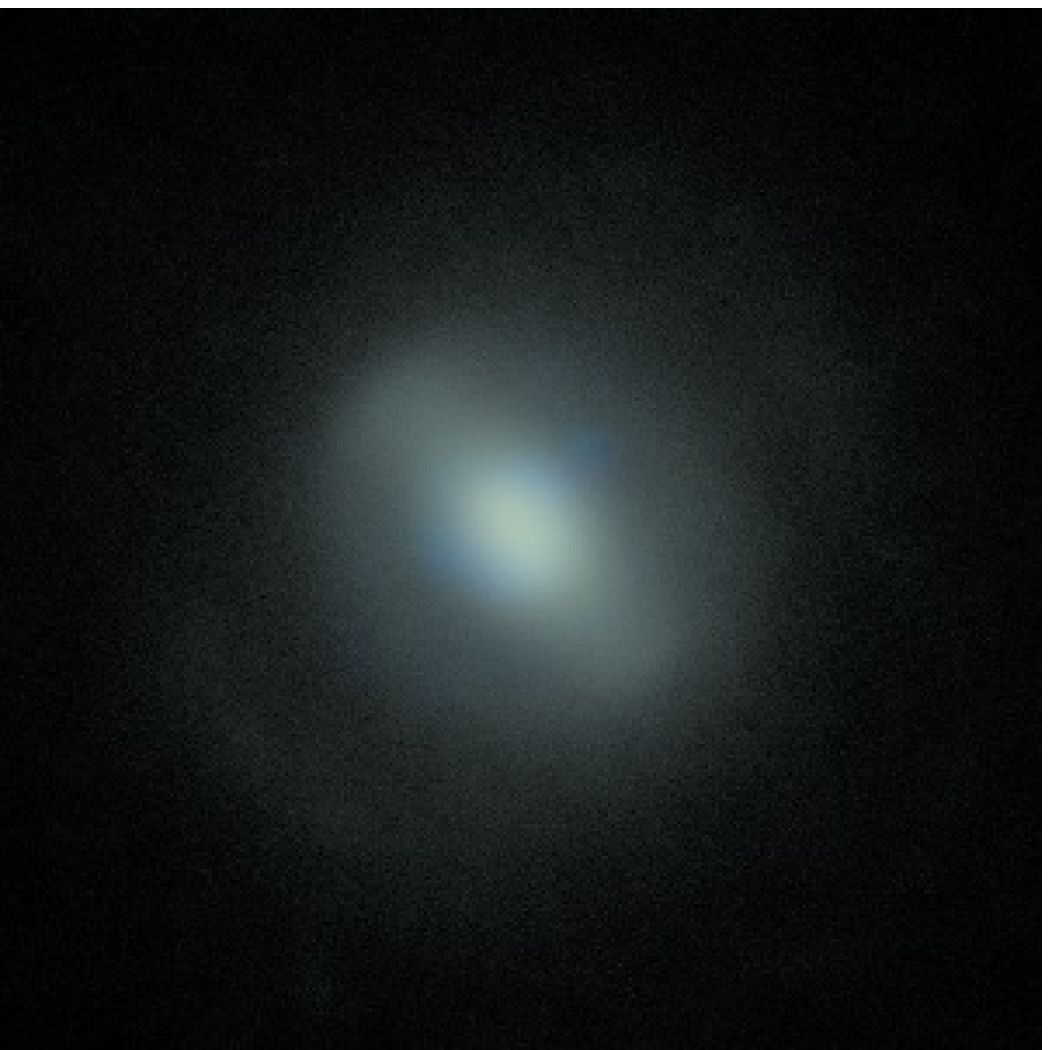}\includegraphics[width=2.7cm]{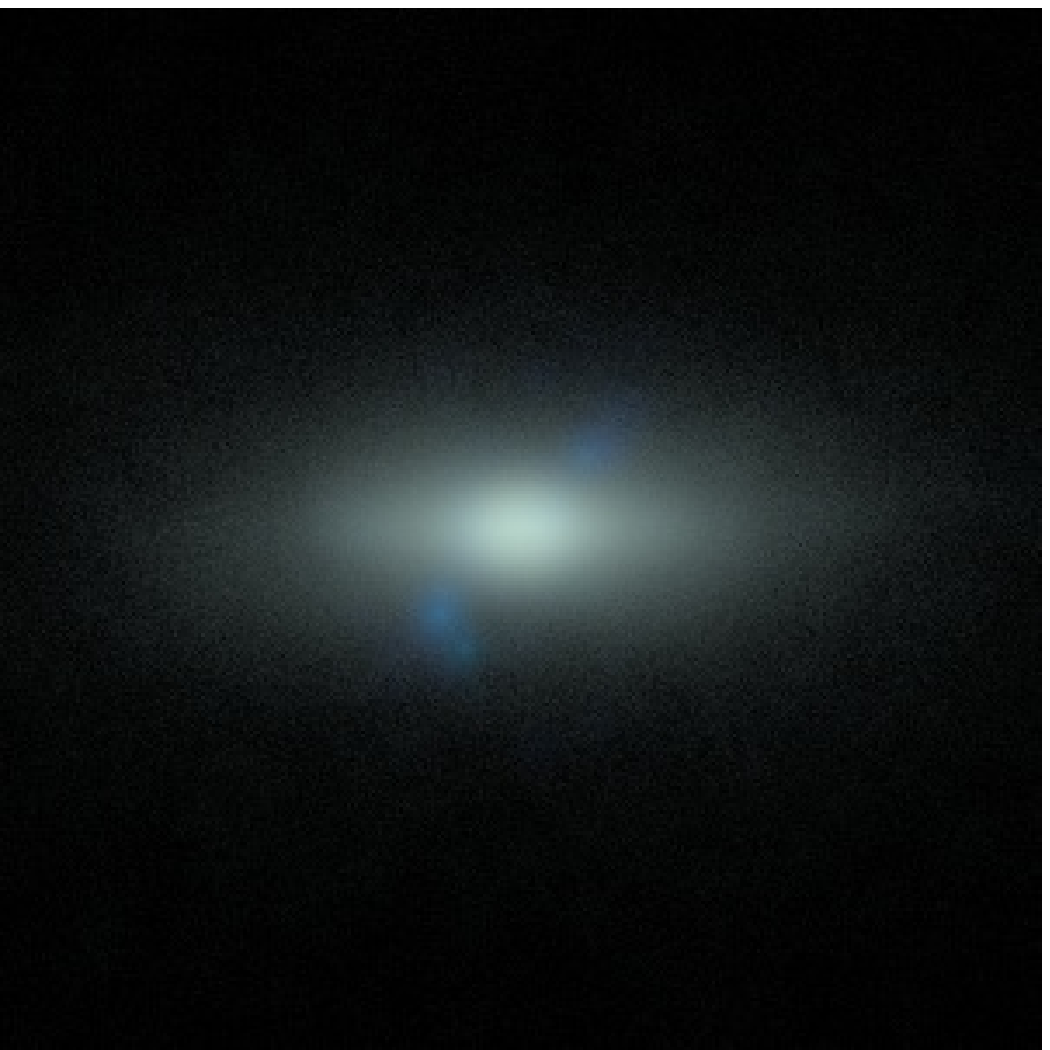}}\hspace{0.1cm}{\includegraphics[width=2.7cm]{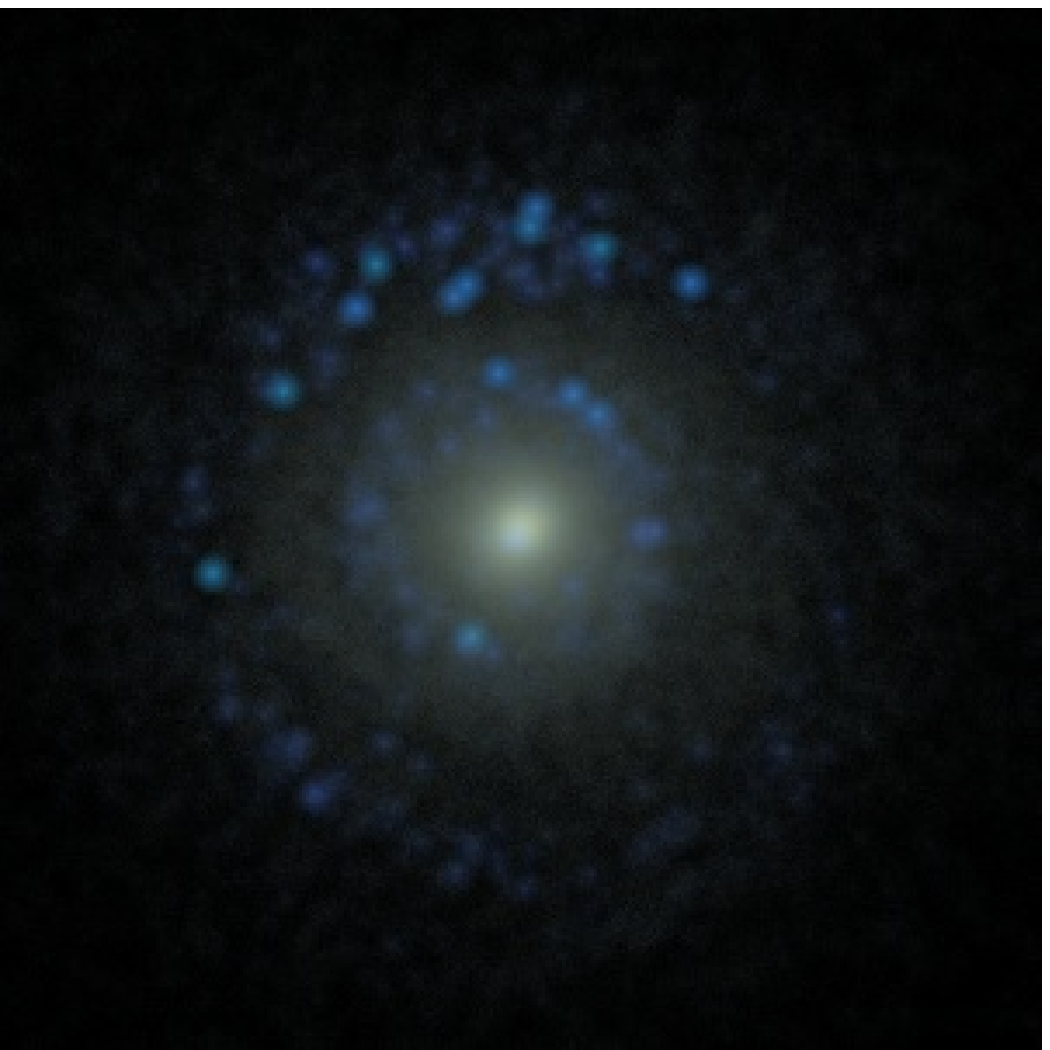}\includegraphics[width=2.7cm]{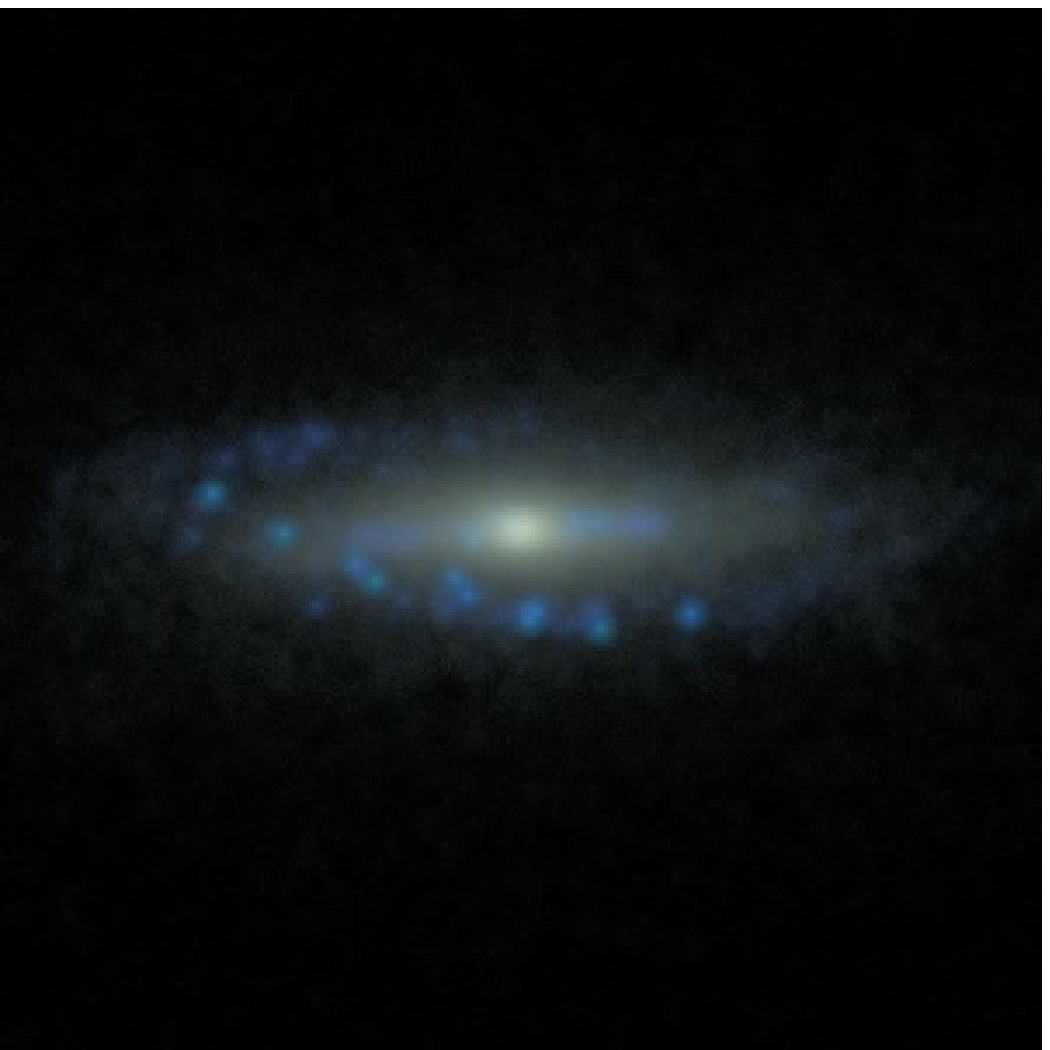}\hspace{0.1cm}\includegraphics[width=2.7cm]{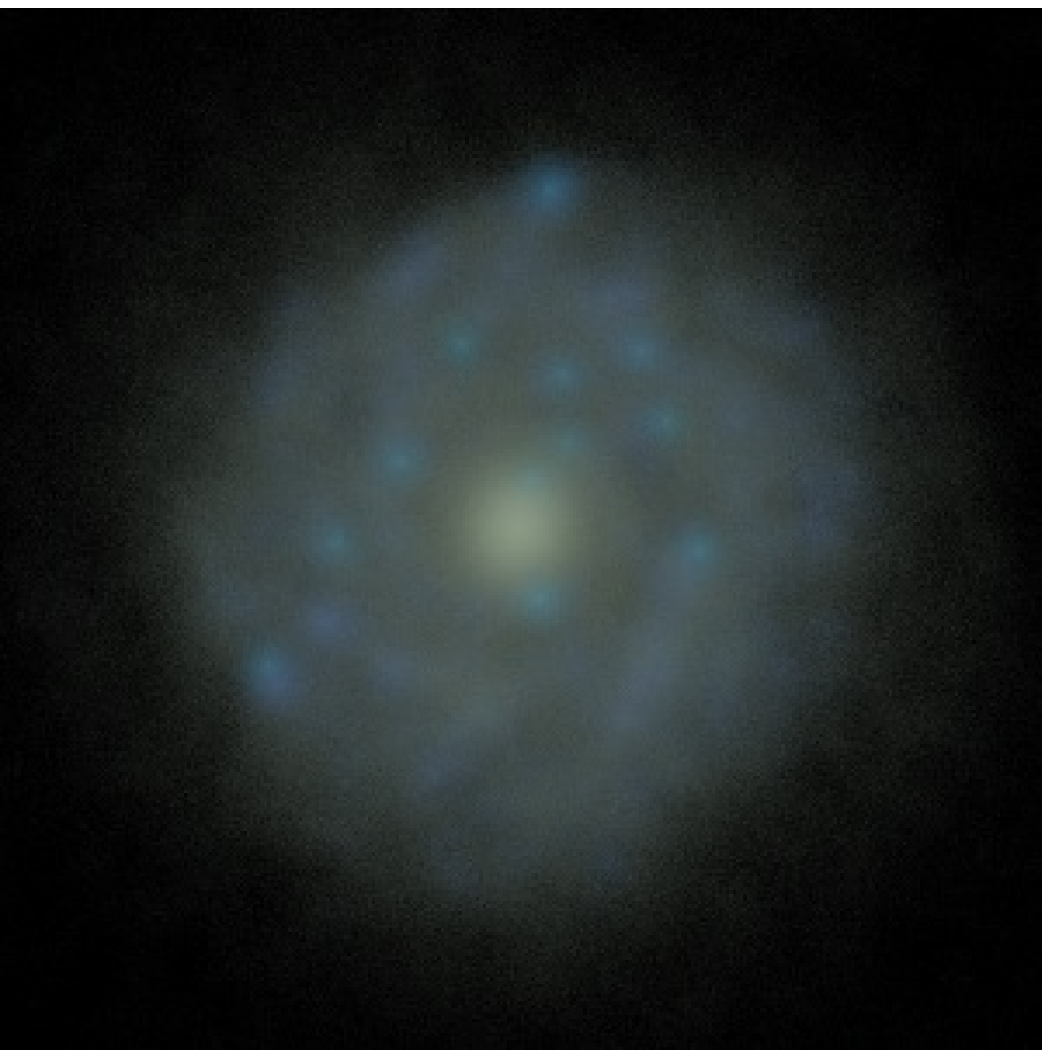}\includegraphics[width=2.7cm]{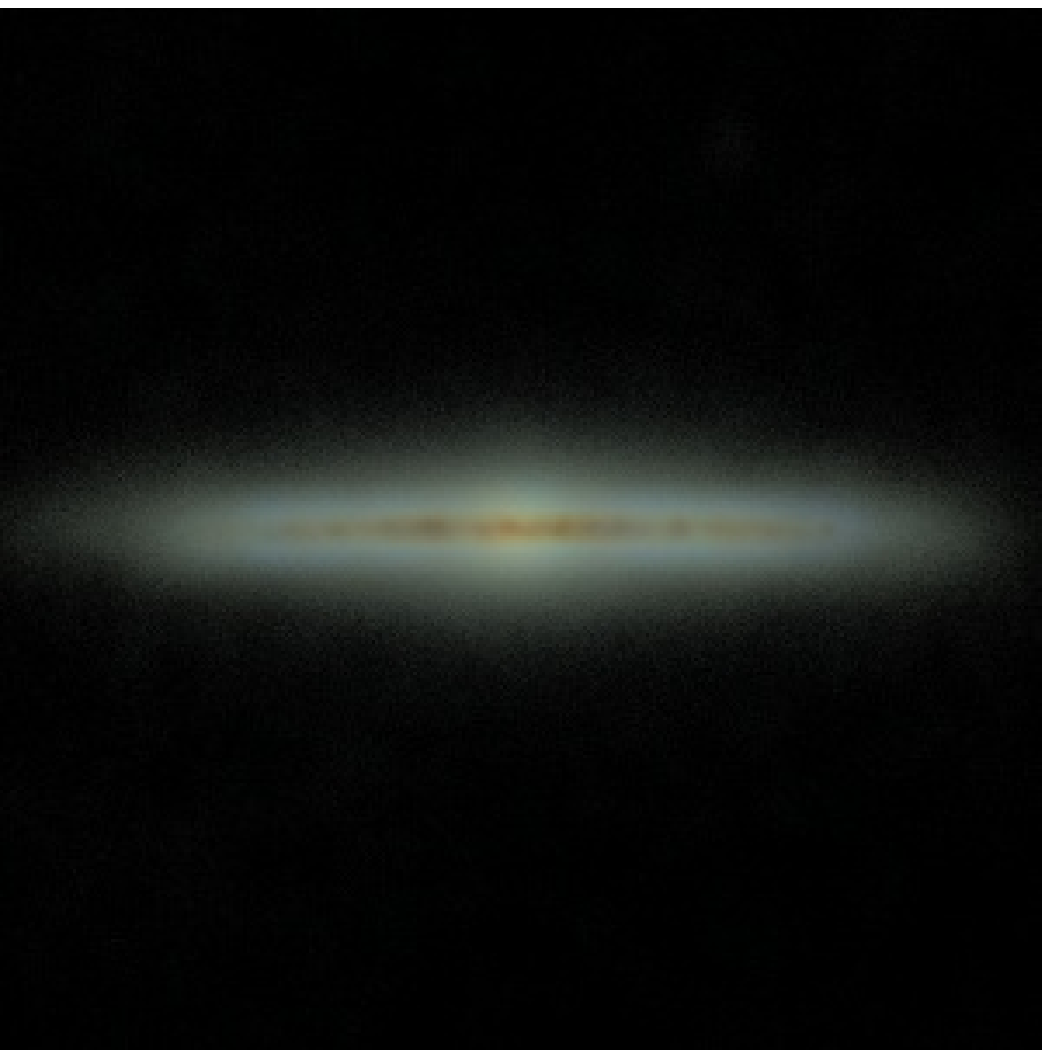}}

\caption{Multi-band {\bf{($u,r,z$)}} images of our fifteen galaxies, for face-on and edge-on views.}
\label{fig:galaxy_images}
\end{figure*}

Fig.~\ref{fig:galaxy_images} shows the color-composite images of the fifteen 
simulated galaxies, 
in face-on and edge-on views, obtained with the code {\sc sunrise} (see below).
The composite images in the ($u, r, z$)-bands are generated using the 
algorithm described in \citet{Lupton04}. 
The edge-on and face-on views are defined such
that the total angular momentum of the stars in the 
galaxy is aligned with the $z$-direction.

\section{Creating  synthetic spectra of the simulated galaxies}

To derive the observables of our simulated galaxies and investigate
the biases introduced in the process, we follow three approaches. First,
we use different Stellar Population Synthesis (SPS) models, which provide the 
resulting spectrum due to the emission of the stars, as well as different
informations about the stellar populations at a given metallicity and age 
(e.g. mass loss, ionizing photon flux, Lick indices, number of black holes 
and neutron stars).
Second, we add a simple dust model to the prediction of the SPSs, 
in order to better compare  the resulting spectra
from those of observed galaxies. 
Finally, we postprocess the simulations with the radiative transfer code 
{\sc sunrise} \citep{Jonsson06, Jonsson09}, which is computationally slower but 
more consistent with the underlying hydrodynamical simulation. 
This gives the full SED including stellar absorption features, nebular emission,
and extinction due to dust as light travels out through the interstellar medium.

The SPS models are commonly used as a postprocessing of hydrodynamical
simulations 
as they provides
a fast and easy way to estimate the light distribution of an ensemble of stars.
The use of radiative transfer codes has also been applied to
hydrodynamical simulations (e.g. \citealt{Governato09,Scannapieco10,Hayward13,Christensen14}). This
approach has the advantage that can consistently consider the distribution
of dust (traced by the metals)  although it also needs to introduce certain 
assumptions and simplifications.

In the following subsections, we describe the main aspects of these models 
and the assumptions that are relevant for our work, as well as discussing
the main
uncertainties and sources of biases in the creation of the synthetic SEDs.

\subsection{Stellar population synthesis Models}
\label{sec:SPS}

We use various SPS models to create synthetic spectra of our simulated
galaxies. 
The spectra are obtained assuming that each star 
particle represents a Simple Stellar Population (SSP) parametrized
by the age and metallicity of the star, and normalized by the mass of 
the particle.
In the simulations, the mass of a star particle changes
with time owing to the supernova ejecta and the mass loss
during the AGB phase;  the mass normalization should be made with
the mass of the stars {\it at the formation time}
(note that this is important in order to avoid double-counting
the effects of stellar mass loss).
The spectrum of the galaxy is  obtained by summing up the
spectra of individual stars. The galaxy spectra can be 
convolved with given photometric bands and integrated to get the 
magnitude of the galaxy in the bands.

We use 5 different SPS models to obtain spectra, magnitudes
and colors of our simulated galaxies, which allows us to
identify systematic effects due to their varying assumptions. 
Although moderate, some differences between these models are expected, as 
uncertainties in modelling the stellar evolution still exist 
(e.g. different treatment of convection, rotation, mass-loss, thermal 
pulses during AGB evolution, close binary interaction), as well as in the 
empirical and theoretical stellar spectral libraries (see the reviews 
by \citealt{Walcher11,Conroy13}).
The main characteristics of the SPS models are described in the following,
and
in Table~1 
we give a summary of the 
input parameters we chose, taken as homogeneous as possible to 
make the interpretation of our results more clear. 

\begin{itemize}
\item \textbf{BC03} \citep{Bruzual03}
 computes, using different evolutionary tracks, the spectral 
evolution of stellar populations at a resolution 
of $3 \, \mathring{A}$ FWHM in 
the optical and at lower resolution 
in the full wavelength range.

\item \textbf{SB99} ({\small ``STARBURST99"}, \citealt{Leitherer99,Vazquez05})  
is a web-based platform that allows users to run customized SPS 
models for a wide 
range of IMF, isochrones, model atmospheres, ages and 
metallicities. The spectral resolution reaches $\sim 1 \, \mathring{A}$ 
FWHM in the optical wavelength range for the fully theoretical spectra.

\item \textbf{PE} ({\small ``PEGASE"}, \citealt{Fioc97,Fioc99})  
uses an algorithm that can accurately follow the stellar tracks of very 
rapid evolutionary phases such as red super-giants or TP-AGB.
The stellar library includes also cold star parameters, and a simple model 
for nebular emission (continuum + lines) can be added to the stellar spectrum.
 
\item \textbf{FSPS} (``Flexible Stellar Population Synthesis 2.3", 
 \citealt{Conroy09}) is a flexible SPS package that can 
compute spectra at resolving power 
$\lambda/\Delta \lambda \approx 200-500 $. In addition to the choice of 
IMF, metallicities, ages, the user can select a variety of 
assumptions on Horizontal Branch (HB)
morphology, blue straggler population, location of the 
TP-AGB phase in the HR-diagram and post-AGB phase.

\item \textbf{M05} \citep{Maraston05}
is different from the other SPS models in the treatment of Post Main Sequence
stars, namely using the Fuel Consumption theorem \citep{Renzini86} to 
evaluate the energetics. 
It gives spectra at a resolution $5- 10  \, \mathring{A}$ in the visual region 
and at $20 -100  \, \mathring{A}$ from the NUV to the near-IR for either blue or red 
populations on the horizontal branch. 

\end{itemize}

\begin{table*}\label{SSP_parameters}
\begin{center}
\caption{Summary of the characteristics and choices of parameters for the 5 SPS models used in our work. The input stellar model in {\sc sunrise} is SB99 with the same parameters.}
\resizebox{\textwidth}{!}{
\begin{tabular}{lllllll}
\hline
Model & IMF & Age range (yr) & Metallicity range & Wavelength range & Stellar tracks & Stellar library \\
\hline \hline
\bf{BC03}  & Chabrier$^{(1)}$ & $10^5 - 2 \cdot 10^{10} $ &0.0001 - 0.05 & $91 \, \mathring{A} -160 \, \mu m$ & Padova 1994 & BaSeL3.1 / STELIB  \\
\bf{M05}  & Kroupa$^{(2)}$ & $10^3 - 1.5 \cdot 10^{10}$ &0.0001 - 0.07 & $91 \, \mathring{A} - 160 \, \mu m$ & Cassisi / Schaller & BaSeL3.1 / Lan\c{c}on \& Mouchine \\
\bf{SB99}  & Kroupa$^{(2)}$ & $10^4-1.5 \cdot 10^{10}$ & 0.0004 - 0.05 & $91 \, \mathring{A} - 160 \, \mu m$ & Padova 1994 & Pauldrach / Hillier \\
\bf{PE}  & Kroupa$^{(2)}$ & $0 - 2 \cdot 10^{10}$ & 0.0001 - 0.05 & $91 \, \mathring{A} - 160 \, \mu m$ & Padova 1994 & BaSeL3.1 / ELODIE \\
\bf{FSPS}  & Kroupa$^{(2)}$& $3 \cdot 10^5 -1.5 \cdot 10^{10}$ & 0.0002 - 0.03 & $91 \, \mathring{A} - 10 \, mm$ & Padova 2007 & BaSeL3.1 / Lan\c{c}on \& Mouchine \\
\hline
\end{tabular}
}
\end{center}
{\small {\sc notes:} (1) Mass range: $m =$ 0.1-100 M$_{\odot}$, $\alpha = 2.3$ for $m>1$ M$_\odot$;  
(2) $\alpha= 1.3$ for $m = $0.1-0.5 M$_{\odot}$,  $\alpha = 2.3$ for $m = $ 0.5-100 M$_{\odot}$.}
\end{table*}

\subsection{Dust}\label{sec:dust}

Dust extinction is an important ingredient in the estimation
of observables from the simulations, as in observed galaxies
dust effects can be large, especially for edge-on systems.
Different authors have modelled dust extinction curves of the Milky Way 
\citep[e.g.][]{Seaton79,Cardelli89}, of the Large and Small 
Magellanic clouds \citep[e.g.][]{Fritzpatrick86,Bouchet85},
and also of  external galaxies \citep[e.g.][]{Silva98, 
Charlot00,Calzetti00,Fischera05}, which are often used to 
correct for dust extinction.

In our calculations of the dust-corrected  magnitudes and colors,
we use the model of \citet[][CF00 hereafter]{Charlot00} 
in the slightly different formulation given in \citet[][dC08 hereafter]{DaCunha08}, 
that consistently extend the CF00 model to include dust emission, 
in order to allow the interpretation of the full UV-far infrared galaxy SEDs.

CF00 is an angle-averaged time-dependent model, with extinction curve depending 
on the wavelength, and on the Stellar Population (SP) age.
This model has been proven to work reasonably well for a wide class of
galaxies, and it has been already implemented in SPS models
\citep{Bruzual03}. In CF00, stars are assumed to born in 
``birth clouds'' that disperse after 
a certain amount of time \citep[see also][]{Silva98}; the transmission 
function of the SP is the product of the transmission function in the 
birth cloud (which depends on the SP age $t$) and in the ISM:

\begin{displaymath}
T_{\lambda}^{\text{BC}}(t) = \left\{
  \begin{array}{l l}
    \exp\left[-(1-\mu) \, \tilde{\tau}_V\left(\frac{\lambda}{5500 \, \text{A}}\right)^{-N_{\text{BC}}}\right] & \quad \text{for $t \le t_{\text{BC}}$}\\
    1 & \quad \text{for $t > t_{\text{BC}}$}
  \end{array} \right.
\end{displaymath}

\begin{flalign*}
\;\;\;& T_{\lambda}^{\text{ISM}} = \exp\left[- \mu \, \tilde{\tau}_V \left(\frac{\lambda}{5500 \, \text{A}}\right)^{-N_{\text{ISM}}}\right]& 
\end{flalign*}
where $t_{\text{BC}}$ is the birth cloud life-time (see CF00 and dC08 for details).
The resulting attenuated luminosity $L_{\lambda}$ given the intrinsic luminosity $S_{\lambda}$ is then
$$
L_{\lambda}(t) = S_{\lambda} \; T_\lambda^{BC}(t) T_\lambda^{ISM}.
$$
We set the free parameters to the values given in dC08  
for normal star-forming galaxies, namely: $t_{\text{BC}} = 10$ Myr, 
$\tilde{\tau}_V = 1.5$, $\mu = 0.3$, $N_{\text{BC}} = 1.3$ and $N_{\text{ISM}} = 0.7$.

\subsection{Radiative Transfer}
 \label{sec:sunrise}

To calculate the full far-UV to submillimeter SED of our 
simulated galaxies we use the Monte Carlo Radiative 
Transfer code {\sc sunrise} in the post-processing phase.
{\sc sunrise} \citep{Jonsson06, Jonsson09} is a 3D adaptive grid 
polychromatic Monte Carlo radiative transfer code, suited to process 
hydrodynamical simulations. 
{\sc sunrise} assigns a spectrum to each star particle in the simulation,
and then propagates photon ``packets'' from these sources 
through the dusty ISM using a Monte Carlo approach, assuming a constant 
dust-to-metals mass ratio, which we fixed to 0.4 according 
to \citet{Dwek98}. 

In the standard {\sc sunrise} implementation, each stellar particle older than 
10 Myr is assigned a spectrum corresponding to its age and metallicity from the
input SPS model, in our case SB99 (see Table~1).
Star particles younger than 10 Myr are assumed to be located in 
their birth clouds of molecular gas, and are given a modified 
spectrum which accounts for the effects of HII and photo-dissociation 
regions (PDRs). The evolution of HII regions and PDRs are described 
by the photo-ionization code {\sc mappings III} \citep{Groves04,Groves08}, 
which is used to calculate the propagation of the 
source spectrum through its nebula. The HII regions absorb 
effectively all ionizing radiation and are the sources of hydrogen
recombination lines, as well as hot-dust emission.  
The only {\sc mappings III} parameter not constrained by the hydrodynamical 
simulation is the PDR clearing timescale; in {\sc sunrise} this free 
parameter has been changed to the time-averaged fraction of stellar 
cluster solid angle covered by the PDR ($f_{\text{PDR}}$), for which we use 
the fiducidal value of $f_{\text{PDR}} = 0.2$ as in \citet{Jonsson09}. 

Dust extinction in {\sc sunrise} is described by a Milky Way-like 
extinction curve normalized to $R_V =3.1$ \citep{Cardelli89,Draine03}
 which includes also the $2175 \AA$ bump observed in 
our galaxy \citep{Czerny07}. 
Model cameras are placed around the simulated galaxies, to sample a range of 
viewing angles. The emergent flux is determined by the number of photons
that flow from the galaxy unscattered in a given camera's direction, as well 
as those scattered into the line of sight or reemitted in the infrared 
by dust into the camera. In the calculations presented here we use two cameras, 
one looking face-on and the other edge-on to the galaxy, to take into account the two
extreme cases in terms of optical depth.
Finally, the flux in the cameras is convolved with the bandpass 
filters 
and integrated to get the broadband images and magnitudes in the chosen
photometric bands.

\begin{figure*}
  \centering
  \includegraphics[width=16cm, height=7cm]{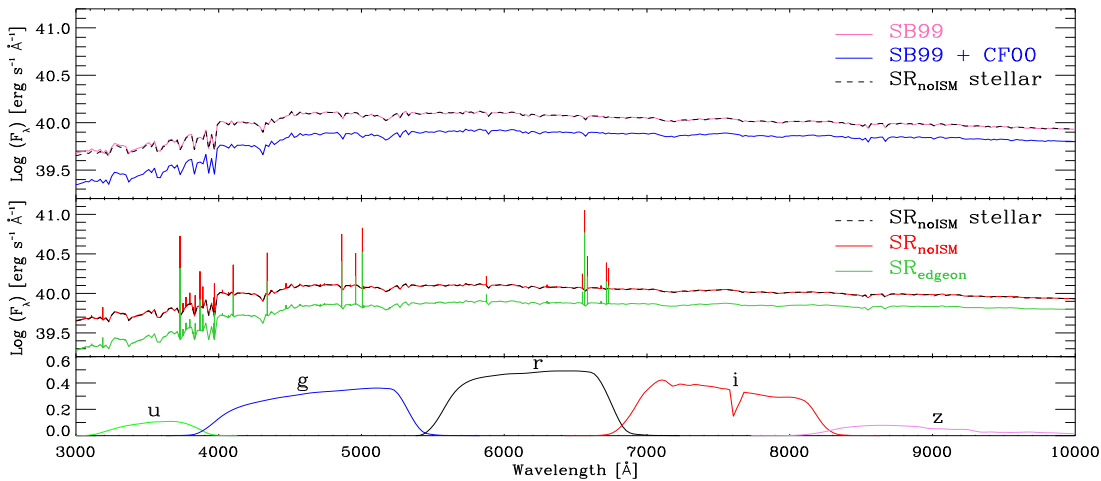}
  \hspace{5mm}
  \caption{Examples of the SEDs obtained for one of our simulated galaxies
(C-MA) using different methods, including the stellar spectrum only
(SB99), the stellar spectrum and the CF00 simple dust model (SB99 + CF00), 
the stellar {\sc sunrise} dust-free spectrum (SR$_{\rm noISM}$ stellar) and 
total {\sc sunrise}
spectra in the absence (SR$_{\rm noISM}$) and presence
of dust, for the edge-on view (SR$_{\rm edgeon}$).
The lower panel shows the SDSS filter transmission functions for
the $u, g, r, i$ and $z$-bands.}
\label{fig:sample_spectra}
\end{figure*}

Fig.~\ref{fig:sample_spectra} is an example of the spectra 
obtained 
for one of our simulated galaxies, C-MA, showing in the upper panel
the SB99 spectrum both without dust (pink line) and corrected with the CF00 model
(blue line), as well as the dust-free stellar spectrum with {\sc sunrise}
(black dashed line).
The lower panel shows again the {\sc sunrise} stellar spectrum without 
dust, the full nebular+stellar spectrum without 
dust (red), 
and the full spectrum with dust from the edge-on camera (green).

\subsection{Uncertainties and biases}\label{sec:biases}

In addition to the different assumptions of the SPS models and
the treatment of dust and radiative transfer, there are a number of
choices on the process of constructing the synthetic SEDs of simulated galaxies.
In some cases, these constitute important sources of biases
that need to be kept in mind when results are interpreted, while
others do not strongly affect the final SEDs.
When appropriate we will test and quantify
these effects, which we describe below.

\subsubsection{Aperture bias}
\label{sec:fiber_bias}

\begin{figure*}
  \centering
  {\includegraphics[width=18cm,height=5cm]{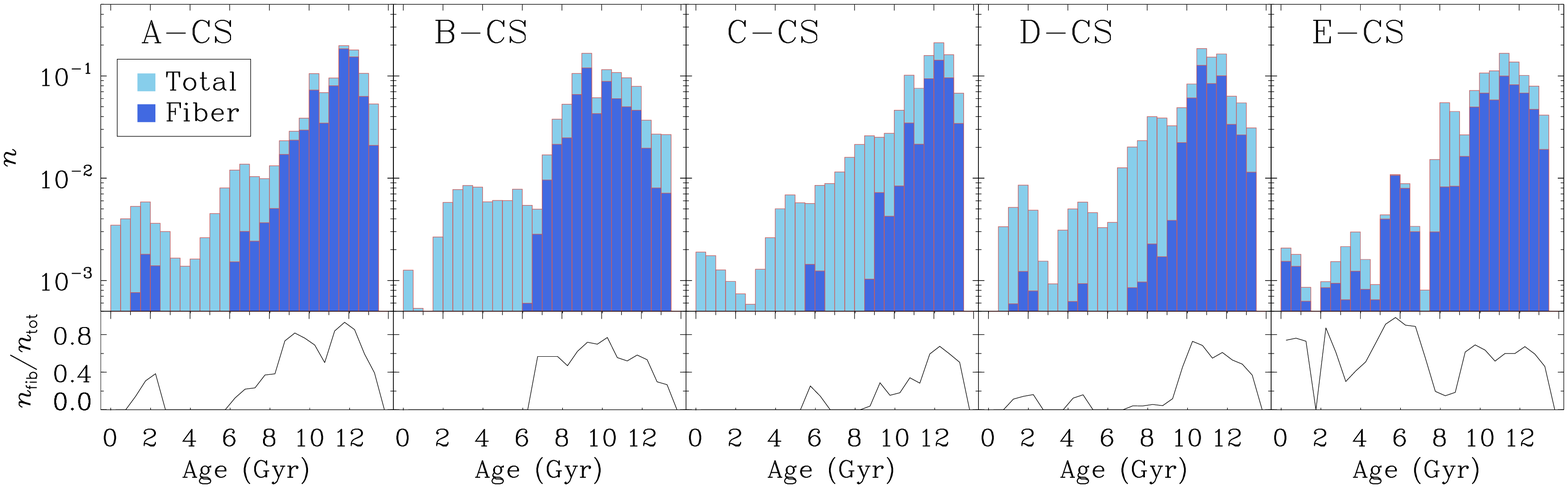}}
  {\includegraphics[width=18cm,height=5cm]{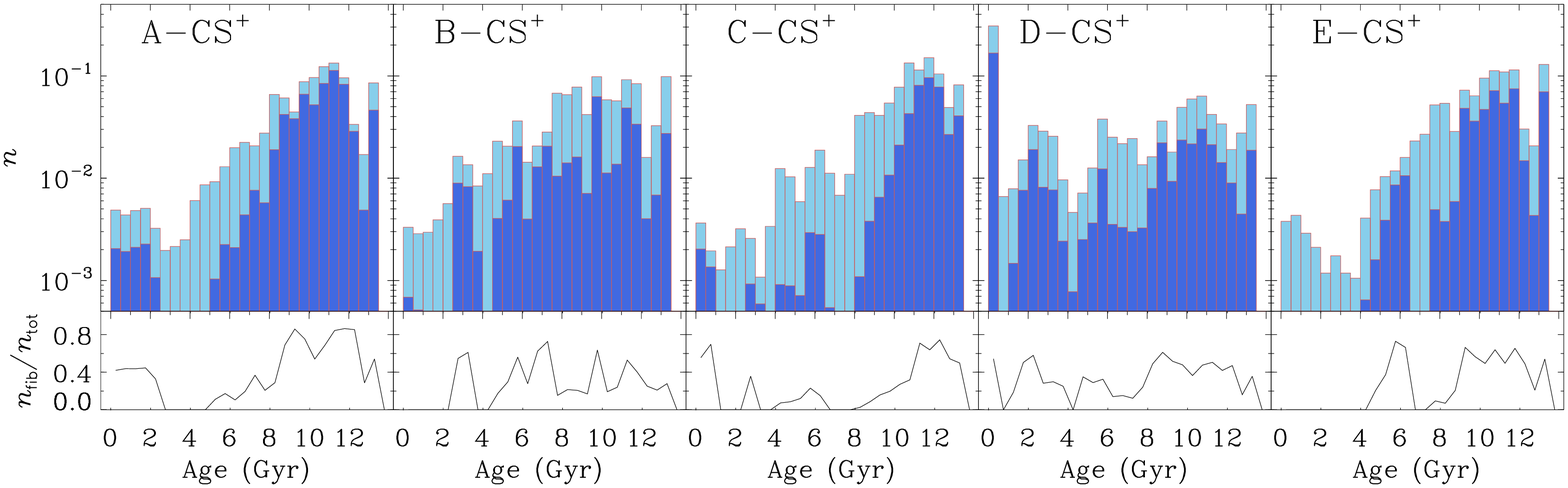}}
  {\includegraphics[width=18cm,height=5cm]{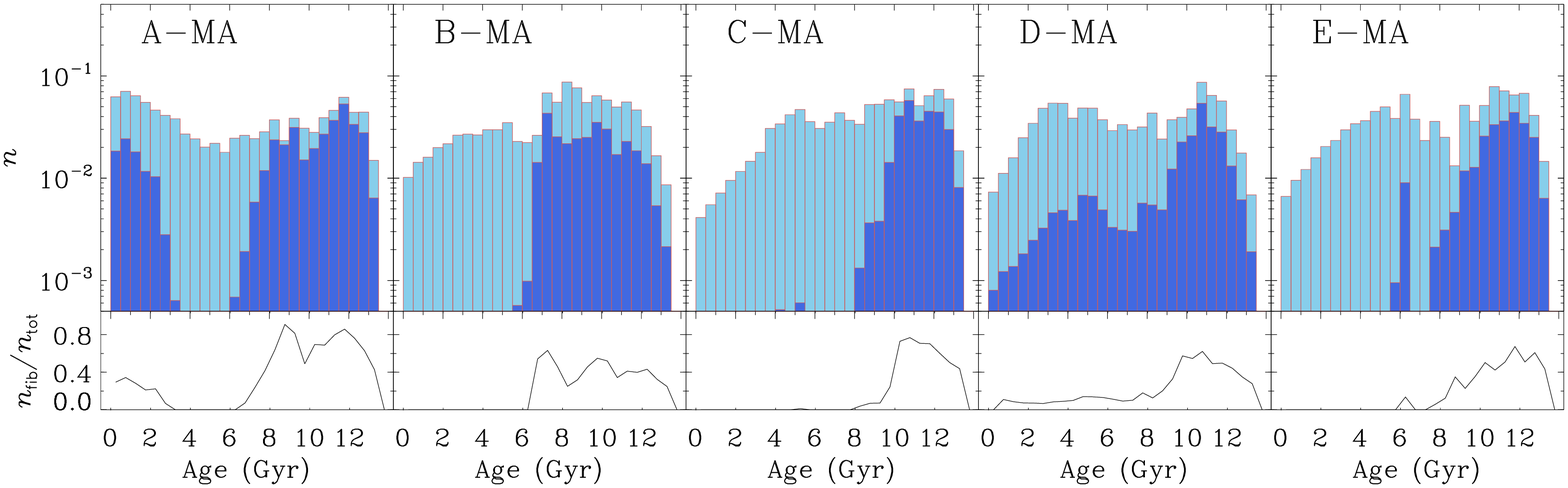}}

\caption{Number-density histograms of stellar ages for the 15 simulated galaxies, considering all star particles in the galaxies as well as only those within the fiber.
For each galaxy, we also show the fraction between the
number of stars within the fiber to the total number of stars in each bin, 
$n_{\rm fib}/n_{\rm tot}$, that we refer to as the sampling function.}
\label{fig:hist_stellar_age}
\end{figure*}

\begin{figure*}
  \centering
  {\includegraphics[width=18cm,height=5cm]{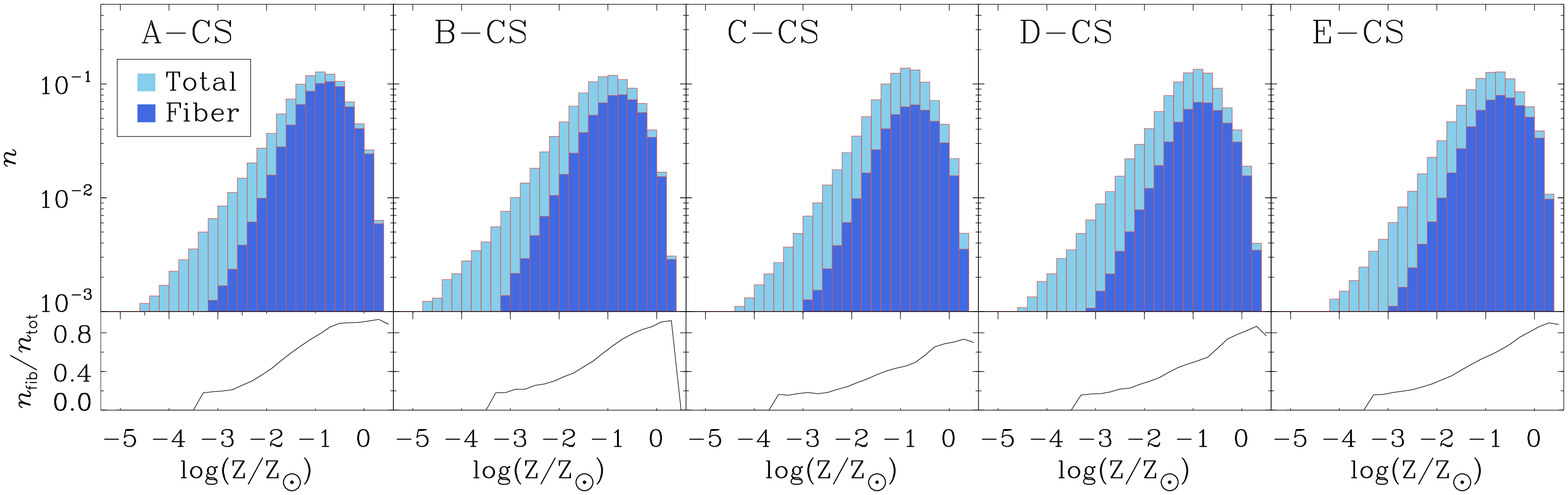}}
  {\includegraphics[width=18cm,height=5cm]{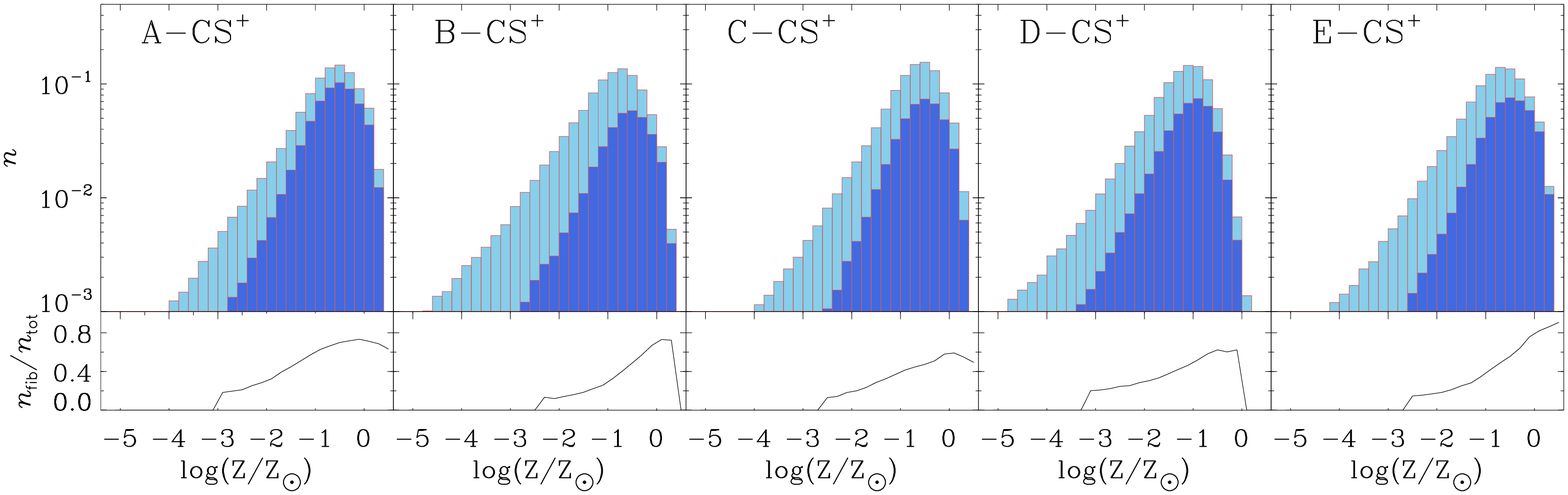}}
  {\includegraphics[width=18cm,height=5cm]{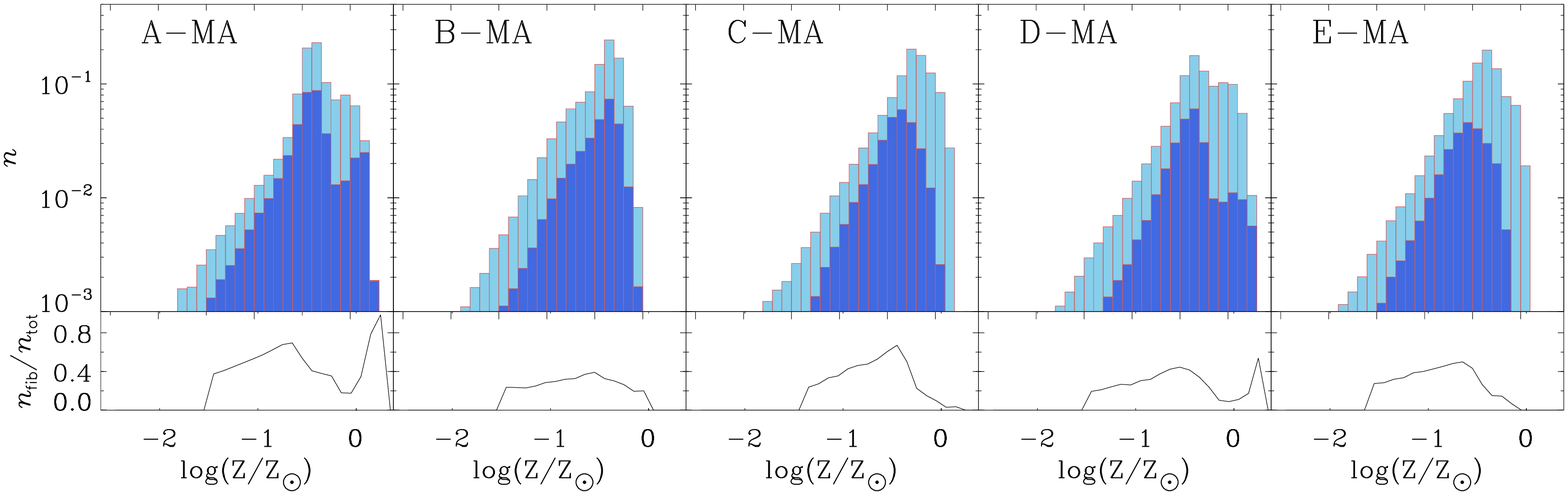}}

\caption{Number-density histograms of stellar metallicity (in solar
units, assuming Z$_\odot = 0.02$), both for all
star particles and inside the fiber field of view. For each galaxy, we also show the fraction between the
number of stars within the fiber to the total number of stars in each bin, 
$n_{\rm fib}/n_{\rm tot}$, that we refer to as the sampling function. }
\label{fig:hist_stellar_met}
\end{figure*}

\begin{figure*}
  \centering
 {\includegraphics[width=18cm,height=5cm]{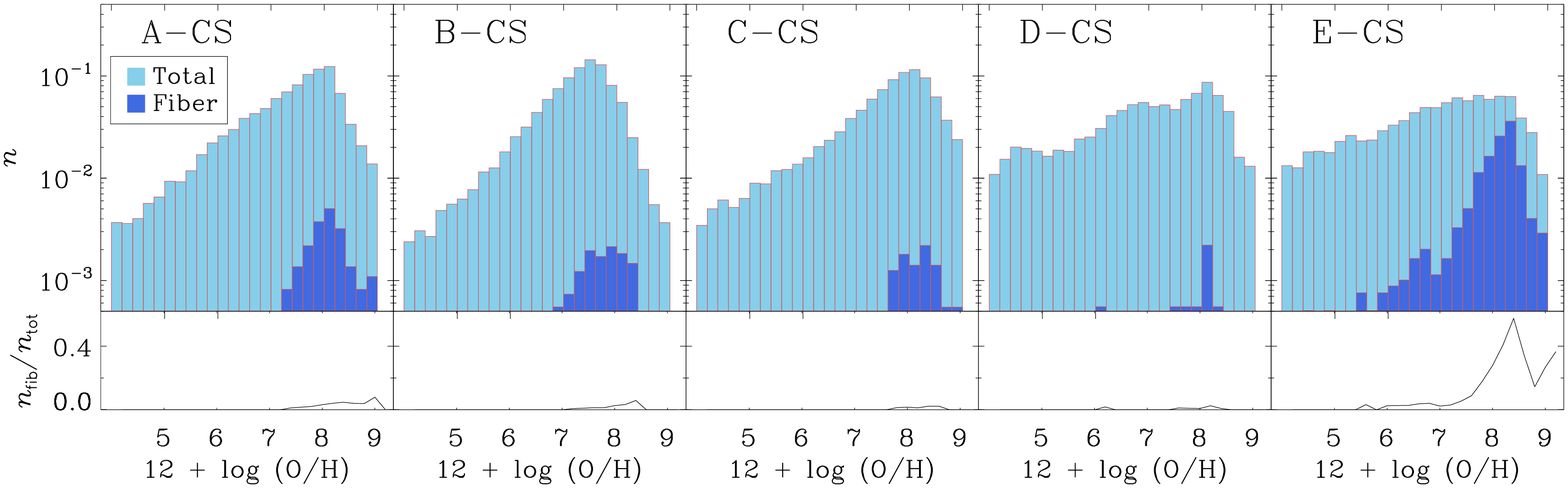}}
 {\includegraphics[width=18cm,height=5cm]{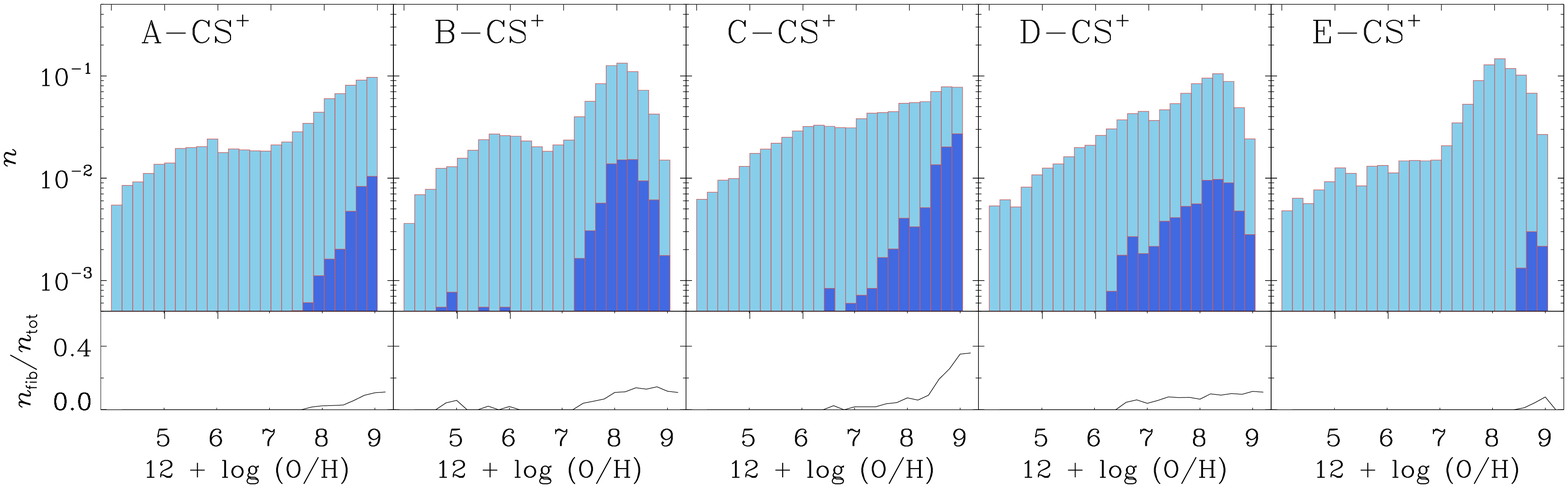}}
 {\includegraphics[width=18cm,height=5cm]{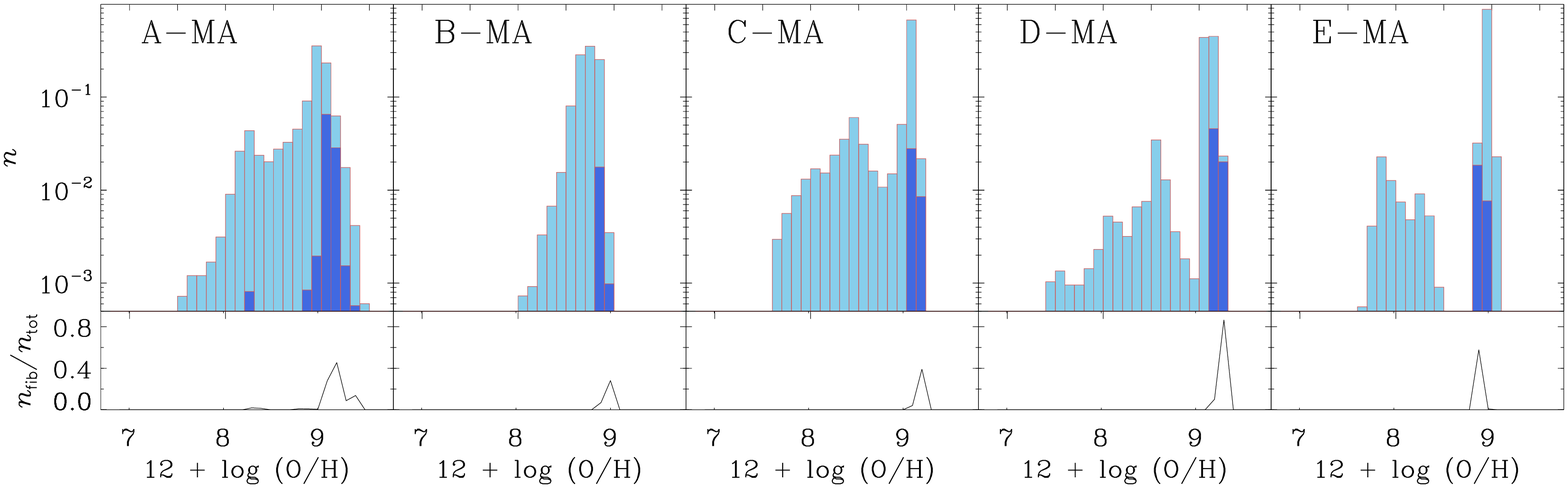}}
\caption{Number-density histograms of gas metallicity, both for all star particles and inside the fiber field of view. For each galaxy, we also show the fraction between the
number of stars within the fiber to the total number of stars in each bin, 
$n_{\rm fib}/n_{\rm tot}$, that we refer to as the sampling function. 
Note the different x-range plotted for the CS/CS$^+$ and MA samples. }
\label{fig:hist_gas_met}
\end{figure*}

While simulations give the full information on
the phase space of particles that constitute
a galaxy, observations usually have limitations
related to the region that can be observed/measured, 
and the different techniques used to define this region in observations
and simulations can introduce 
some systematics \citep{Stevens14}.
In SDSS, the properties derived from spectra (e.g. stellar age and metallicity, 
gas metallicity, SFRs)  are affected by the small aperture (3'' arcsec) of the 
SDSS 
spectrograph, which samples only the inner part of the target galaxies. 
The presence of metallicity gradients in galaxy properties and colors 
\citep{Bell00, Pilkington12, Welikala12, Sanchez15} can therefore 
lead to substantial uncertainties in the
observed measurements when one wants to estimate the total 
quantities. 
The implications of this bias have been discussed by several authors 
\citep[][e.g.]
{Bell00, Kochanek00, Baldry02, Gomez03, Brinchmann04, MacArthur04};
in some cases methods for aperture correction which exploit the spatially-resolved color 
information can be used  to extrapolate to global quantities 
(e.g. SFRs, \citealt{Brinchmann04}, \citealt{Salim07}).

We used our simulations to test the
effects of using a single fiber in the derivation
of properties such as metallicities and
stellar ages, by including in the calculations both 
all particles in a 60x60 kpc field of 
view (FoV) in the face-on projection (full FoV),  
or only those within a small region that mimics the size
sampled by the SDSS fiber spectrograph, i.e. a
circular region of 4 kpc radius with the galaxy in the center observed 
face-on (fiber FoV)\footnote{In the z-direction we count all stars 
in the halo as identified with {\sc sunfind} as bound to the halo.}, 
which according 
to the cosmology adopted in our simulations corresponds to redshift $z \sim 0.15$ for
the SDSS instrument aperture. 
Note that we are limited to further decrease the size of the fiber field of 
view by having enough particles to reach a good statistical sample,
in particular in the case of gas particles.

Figs.~\ref{fig:hist_stellar_age},
\ref{fig:hist_stellar_met} and \ref{fig:hist_gas_met} show 
the number-density histograms of stellar 
ages and stellar/gas metallicities  for the 15 simulated galaxies,
 when we consider all particles in the galaxy
and only those within the fiber.
For each galaxy, we also show the fraction between the
number of star/gas particles within the fiber to the total number of 
star/gas particles in each bin, $n_{\rm fib}/n_{\rm tot}$, that we refer
to as the sampling function.
In the case of stellar ages we find that, 
in general, fiber quantities preferentially
sample the older populations, 
with sampling functions typically higher than $50\%$.
In contrast, the young populations are sampled
in various ways, depending on the age profile of the
galaxy.
In some cases, young populations are not at all sampled and will
be completely  missed in the calculations using particles within the fiber.
It is clear, in any case, that the {\it sampling within the fiber
is not constant or even similar for all galaxies, which means
that it is not
possible to reliable estimate the mean stellar age of a galaxy
from the fiber quantities, or even estimate the fiber bias without 
having additional information (e.g. importance of age/metallicity profiles)
of the regions outside the fiber.
}
We note that
the relative contribution of old and young populations in a galaxy will in general
reflect the relative importance of bulges and disks, but with
fiber quantities this information might not be a reliable reflection
of the real relative contribution of the different stellar components
of a galaxy.

Similar considerations can be made in the case of 
the stellar metallicities,
shown in Fig.~\ref{fig:hist_stellar_met}.
The metallicity distributions of the galaxies in the
CS  sample is broader compared to those in the MA sample,
with the CS$^+$ galaxies lying in between.
This results from the different implementation of chemical enrichment
(chemical yields increase from the CS galaxies, to CS$^+$ and then
to MA), chemical
diffusion (only included in the MA simulations) and feedback (from
weaker in CS to stronger in MA, with CS$^+$ in between). 
In particular,  the MA galaxies have  more peaked distributions,
and have less strong metallicity gradients (see also \citealt{Aumer13}) compared
to the CS and CS$^+$ galaxies  (see
also \citealt{Tissera12}).
Unlike for
the stellar ages, {\it for the stellar metallicities we find that
the sampling shows much less variation from galaxy to
galaxy, as in all cases there is a preferential sampling of the 
more  metal-rich populations.}

The distributions of gas metallicities are more complex,
and show important
differences
depending  on the details of the implementation of chemical enrichment
(IMF, chemical yields, etc.) and on the
absence/presence of chemical diffusion,
which is included only in the MA sample
and leads to a much less broad distribution of gas metallicities
(note the different x-scales in the lower panels). 
From Fig.~\ref{fig:hist_gas_met}, it is clear that
the sampling of gas metallicities varies significantly
from galaxy to galaxy. While in general there is a preferential
sampling of the metal-rich regions, gas particles with
intermediate metallicities can also contribute
significantly to the final metallicity when fiber quantities
are considered. For the MA sample, we find that, within
the fiber, only a narrow range of gas metallicities
are sampled, which reflects the effects of metal diffusion
that tends to give a smoother metallicity distribution.
In the case of the D-CS galaxy, we also find that although
the metallicity distribution is broad, only a very narrow
range of metallicities are sampled within the fiber. 
According to our findings, {\it for estimating the gas metallicy
of a galaxy it is of primary importance that the fiber
bias can be quantified; otherwise it is not possible
to derive the mean gas metallicity of the whole galaxy
reliably.}

In summary, our results show that {\it the bias due to the fiber 
is in general  reflected in the tendency to sample older and more 
metal-rich stellar populations and metal-enriched regions
of the ISM, although the shape
of the sampling function  highly depends on the 
significance of gradients in ages/metallicities} (which also somewhat 
depend on the modelling of 
physical processes in the simulations), {\it affecting in particular 
galaxies with stronger gradients}.

\subsubsection{Variation of the assumed Initial Mass Function}

The choice of the Initial Mass Function (IMF) can strongly bias the 
derivation of many observables of galaxies \citep[see][]{Salpeter55, 
Miller79, Scalo86, Kroupa02},
for example, the assumed IMF has a direct influence on the 
$M/L$ ratio depending on the age, composition and past star formation 
history \citep{Baldry03,Chabrier03}, which in turn
influences the stellar mass derivation. 
The IMF is measured in many environments, still there is
not yet consensus on a number of important aspects,
such as whether it
is universal or not \citep[e.g.][]{Hoversten08, Bastian10}.

We assume in general a Kroupa IMF with slope $\alpha = 1.3$ 
for masses 0.1-0.5 $M_{\odot}$ and $\alpha = 2.3$ in the mass range
0.5-100 $M_{\odot}$ when we apply the different SPS models,
except in the case of  BC03, where we instead use
a Chabrier IMF  (Table~1).
The difference between these two IMFs is moderate, however,  
due to their very similar shape, particularly at the
high-mass end (see Fig.~17 
in \citealt{Ellis08}). 
Note that our three galaxy samples use different IMFs
to calculate feedback and chemical enrichment:
Salpeter (CS), Chabrier (CS$^+$) and Kroupa (MA).


\section{Results}
\label{sec:results}

In this section we describe the methods we use to derive
the galaxy properties -- magnitudes and colors, stellar masses,
stellar ages, stellar and gas metallicities and star formation
rates -- from their synthetic spectra, and make a detailed
comparison between results obtained with the different
methods, in order to identify biases and
systematics. For each property, different techniques are
used in the derivation, the selection of which has been
made taking into account the questions we want to answer:
($i$) what is the range of variation
in observationally-derived quantities when different
techniques are used, ($ii$) do they agree with the 'real'
quantity;  and
($iii$) how do observational definitions of galaxy properties effect their 
measured values.
In the case of observations, we particularly focus on the techniques
employed in SDSS, which will be used in PaperII to compare simulated
and observed galaxies.

In order to avoid additional biases, we always use 
the same field of view defined as 
a 60 kpc$\times$60 kpc region with the galaxy in the center
both when we derive the properties directly from the simulations
and from the SPS/{\sc sunrise} spectra.


\subsection{Magnitudes and colors}
\label{subsec:magnitudes}

We first compare the colors and the 
absolute magnitudes of our simulated galaxies,
in the 5 SDSS photometric
bands ($u, g, r, i, z$, see Fig.~\ref{fig:sample_spectra}),
obtained using different methods, namely:
\begin{itemize}
\item {\bf BC03, SB99, FSPS, PE, M05}: these refer to the
magnitudes calculated by applying the five different SPS models described in
Section~\ref{sec:SPS}.
As explained in the previous section, for each star particle 
the magnitudes are obtained via a linear 
interpolation of the SPS tables according to its age and 
metallicity, normalizing with the particle mass  {\it at the formation time}.

\item {\bf{\small CF00}}: we include dust effects to BC03
using the model of CF00/dC08 described
in Section~\ref{sec:dust}.

\item  {\bf {\small SR$_{\rm noISM}$, SR$_{\rm faceon}$, SR$_{\rm edgeon}$}}:
results from the radiative transfer code {\sc sunrise} as explained
in Section~\ref{sec:sunrise}, in the absence of dust\footnote{
We note, however, that dust is included through the sub-grid
dust model around young stars in MAPPINGS.}, and for
the face-on and edge-on views including dust, respectively. 
Note that the input SPS model
for
{\sc sunrise} is SB99.

\end{itemize}

\begin{figure*}
  \centering
  \includegraphics[height=9.7cm]{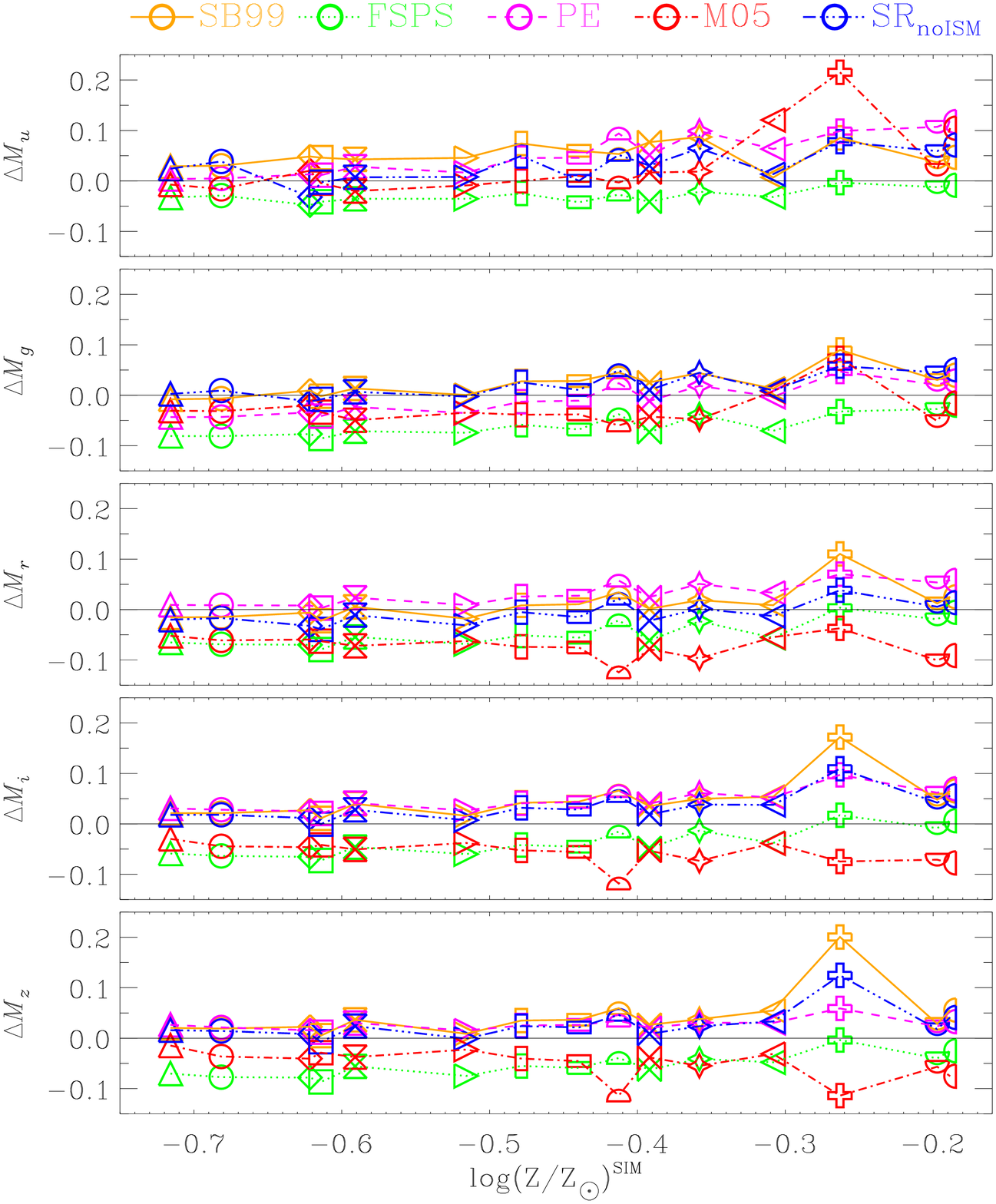}
  \includegraphics[height=9.7cm]{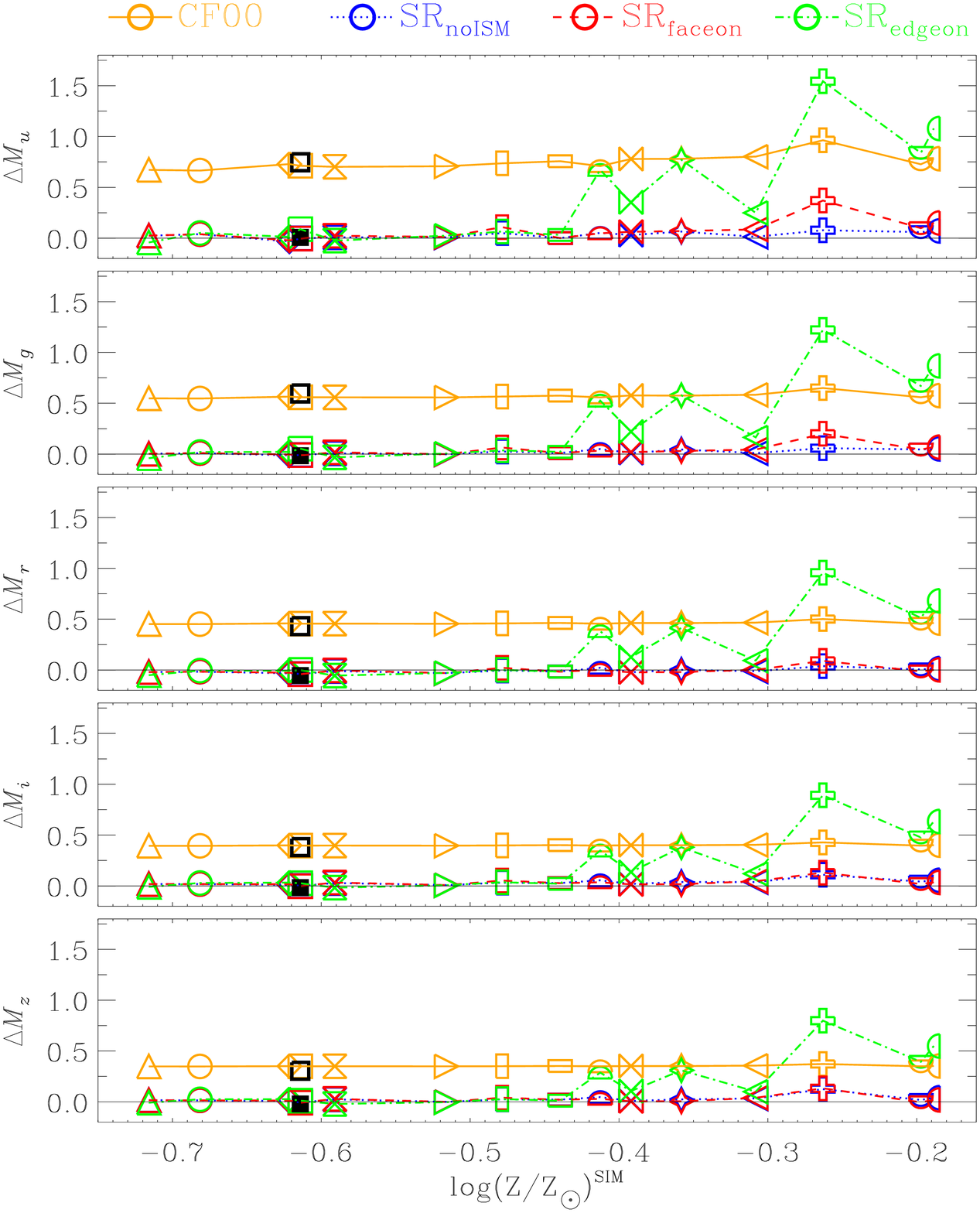}
  \includegraphics[height=9.7cm]{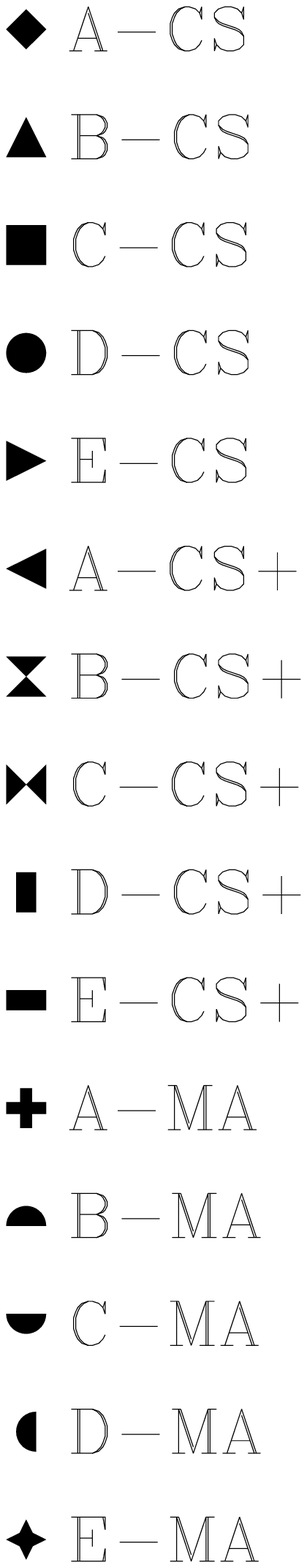}
\caption{Difference between the absolute magnitudes of the 
simulated galaxies obtained using different methods and those predicted
by the dust-free BC03 model, in the 5 SDSS bands and
as a function of stellar metallicity.  The {\it  left}-hand panel shows
results for dust-free models and the {\it right}-hand panel compares
results of the CF00 simple dust model and of the {\sc sunrise} code.
The open/filled black squares are results for C-CS, when a higher
dust-to-metals ratio is assumed in the {\sc sunrise} calculations
in edge-on and face-on views, respectively.
}
\label{fig:magnitudes}
\end{figure*}

The left-hand panel of Fig.~\ref{fig:magnitudes}
compares the magnitudes obtained using the different SPS methods and
 {\sc sunrise}, when we ignore the effects of dust. 
We show the differences with respect to BC03\footnote{We
note that, even if {\small BC03} assumes a different IMF
(Chabrier) compared to the other SPS models, 
this does not introduce
any significant systematic effect in the derived
magnitudes/colors, since for the more
luminous stars ($M_* > 1 M_{\odot}$) 
the IMF slope is the same as Kroupa (see Table~1).}
as a function
 of the stellar metallicity of the galaxy\footnote{This 
is calculated as the average mass-weighted metallicity over 
all stellar particles in
the simulated galaxy (method {\small SIM} in Section~\ref{sec:stellar_ages}).} 
(in solar units, assuming Z$_\odot = 0.02$), as
systematics 
in SPS models are expected to increase
both at  low/high metallicity 
(as well as at younger stellar ages, e.g., \citealt{Conroy09,Conroy10}).  
The  different SPS models show in general
very good agreement, 
with magnitude differences of $\lesssim 0.1$. 
We detect however some systematics:  models FSPS and M05 
usually predict lower magnitudes (i.e. brighter galaxies) 
compared to BC03, while models SB99 and
PE give systematically higher magnitudes than BC03. 
As expected, differences are somewhat larger (but
still moderate) for the most metal-rich galaxies,
which are also those that exhibit a higher fraction of young (ages $< 10$ Myr)
and intermediate  (ages in the range [$0.1-2$] Gyr) stellar populations.
For these ages  the uncertainties in the treatment
of young stars, post red-giant branch and TP-AGB phase stars 
are larger. 
The galaxy which exhibits the largest
$\Delta M$ (in all bands) is A-MA,
which has $\log(Z/Z_{\odot}) \approx -0.26$ 
the most extreme SFR (Section~\ref{subsec:star_formation_rate}) and the
youngest mean stellar age in our sample (see next sections).

\begin{figure*}
  \centering
   \includegraphics[height=9.7cm]{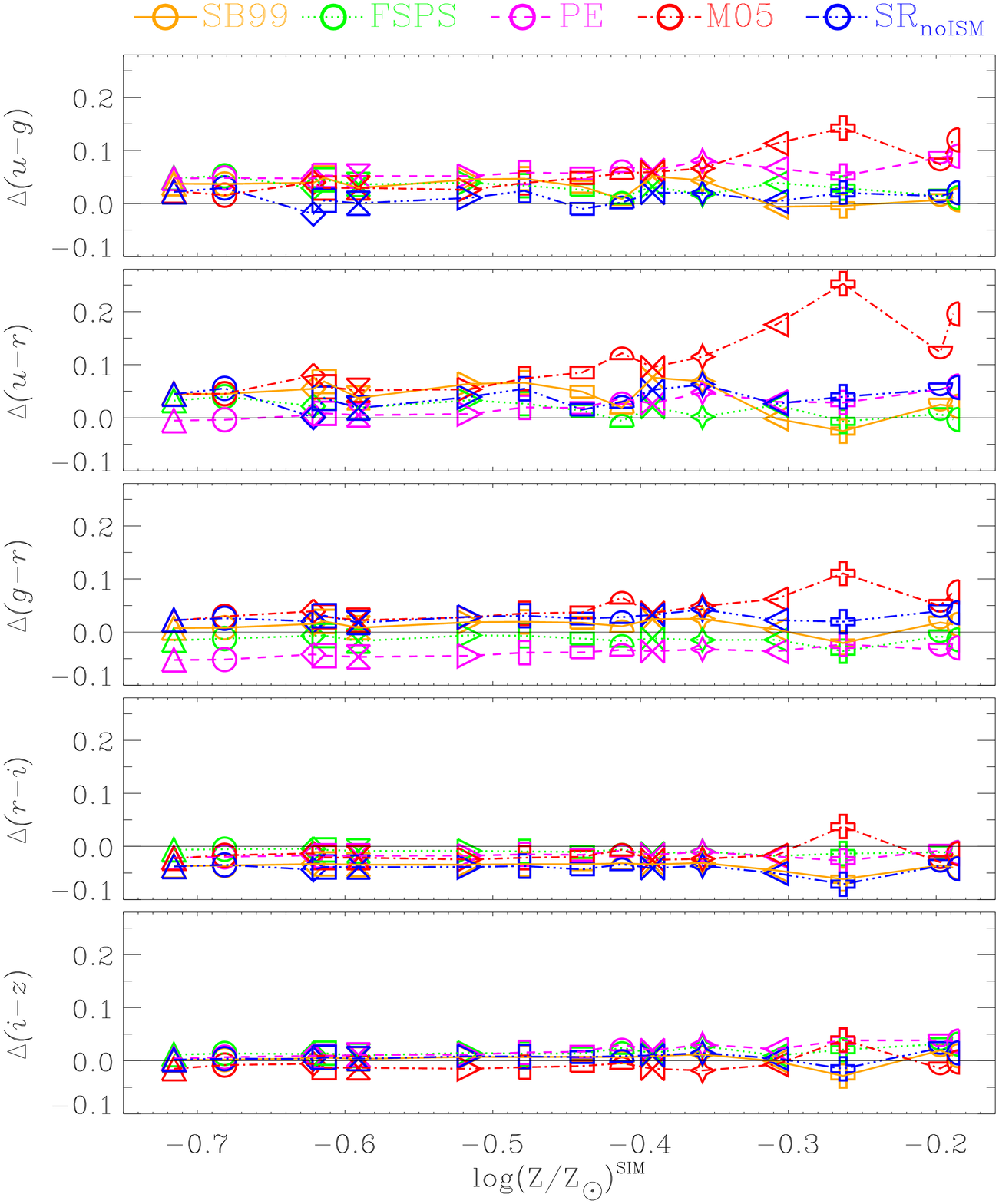}
  \includegraphics[height=9.7cm]{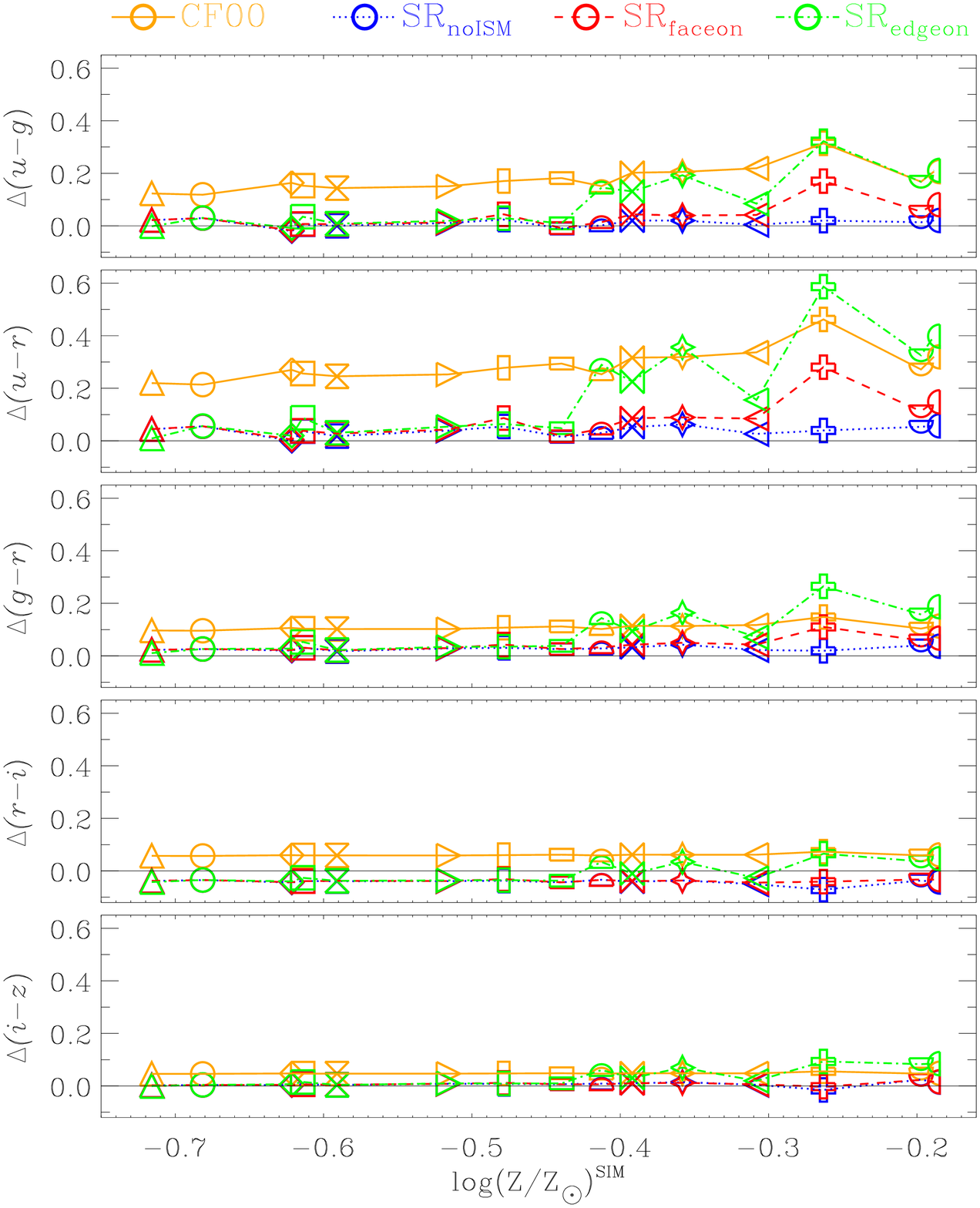}
  \includegraphics[height=9.7cm]{figures/magnitudes_label_vertical.eps}
\caption{Difference between the colors of the 
simulated galaxies obtained using different methods and those predicted
by the dust-free BC03 model, 
as a function of stellar metallicity.  The {\it  left}-hand panel shows
results for dust-free models and the {\it right}-hand panel compares
results of the CF00 dust model and of the {\sc sunrise} code.}
\label{fig:colors}
\end{figure*}

The results of  {\small SR$_{\rm noISM}$} also agree
well with the rest of the models, particularly
with those of SB99, as SB99 is the input SPS model of {\sc sunrise}.
The remaining differences are because in 
{\sc sunrise} the spectrum for young
stars is calculated with {\sc mappings III},
which includes the contribution from 
nebular continuum and emission 
as well as dust absorption and IR emission
that is modeled as sub-grid physics. 
The largest differences 
are found for A-MA, the galaxy with the highest number 
of young star particles. In fact,
on one side the ionizing photons are reprocessed 
as nebular emission lines and continuum increasing the total flux in the 
optical and, on the other hand, 
the young particles are also more extincted due to the
sub-grid treatment of dust
absorption in {\sc mappings III}, giving total magnitudes in 
general  lower (brighter) by $ < 0.1$ mag than modelling the SED 
including only stellar emission.

The right-hand panel of Fig.~\ref{fig:magnitudes} shows
our results in the case of including the effects
of dust, 
either with the CF00/dC08 correction or, more consistently,  
with the {\sc sunrise} code. 
For comparison, the magnitude differences are calculated
with respect to the BC03 model.
When the CF00 simple dust model is included we find, 
 as expected, fainter galaxies, with
differences between $\sim 0.6-0.8$ mag for the $u$ and $g$-bands
and of $\sim 0.4-0.5$ mag for $i, r$ and $z$. 
As the CF00 model is angle-averaged, there is no difference
between face-on and edge-on views.
When we consider radiative transfer effects
in {\sc sunrise}, we detect almost no
difference if we see the galaxies face-on
(SR$_{\rm faceon}$) or if dust is ignored
(SR$_{\rm noISM}$ method, included here for comparison).

As expected, larger differences are detected
when galaxies are seen edge-on, as dust effects are
maximal in this case. However, we find 
significant differences only for metal-rich galaxies,
while for galaxies
that are metal-poor ($\log(Z/Z_{\odot}) \lesssim$-0.45) dust effects
are unimportant. This is because the amount of dust
in {\sc sunrise} is directly proportional
to the gas metallicity (galaxies with low stellar metallicity
also have low gas metallicity).
Observations also confirm that the
dust-to-metals ratio is nearly constant over a large range of 
metallicities and redshifts, and dust effects are small in metal-poor 
galaxies (\citealt{Zafar13}, \citealt{Mattsson14}).
For the metal-rich galaxies, the inclusion of dust
makes them fainter by factors
of $\sim 0.8 -1$ dex in the $u$-band and of $\sim 0.3-0.5$ dex
for the $z$-band.
Interestingly, for the most metal-rich galaxies 
the magnitudes derived from the SR$_{\rm edgeon}$ and {\small CF00} methods are
very similar.
Note also that the CS$^+$ galaxies have a systematically lower 
reddenning compared to the MA ones, even at high metallicity. This is 
explained by the lower amount of metals in
the ISM, and therefore lower amount of dust of the galaxies
in the CS$^+$ sample compared to those in MA.

To explore the 
dependence of dust effects on metallicity for the metal-poor sample,
we rerun {\sc sunrise} for one galaxy (C-CS), using a higher dust-to-metals 
ratio of $4$ ($10$ times larger than the standard value of $0.4$ assumed
in our {\sc sunrise} calculations, \citealt{Dwek98}).
The results are shown as open and filled squares, respectively for the edge-on
and face-on views. In this case, we indeed find 
little effects in the case of the face-on view and  larger differences
in the edge-on case, with results very similar to
those of CF00.


The differences in magnitudes, when different methods are applied,
are translated into differences in the colors, as shown in 
Fig.~\ref{fig:colors}. 
Both in the absence (left-hand panel) and in the presence
(right-hand panel) of dust, differences are larger
for the bluer colors ($u-g$, $u-r$ and $g-r$), and for
the most metal-rich galaxies.


Our results show that  
{\it applying different SPS to simulations gives visual magnitudes of galaxies
with spread $0.05-0.1$ dex for metal-poor
galaxies and of $0.1-0.2$ for metal-rich ones}, depending on the model,
{\it with BC03 appearing as intermediate among the 5 SPS models tested. }
Furthermore, {\it while for face-on galaxies or
edge-on galaxies with low metallicities 
{\rm(}$\log(Z/Z_{\odot}) \lesssim -0.45${\rm )} the 
effects of dust can be in first approximation ignored, when galaxies are 
seen edge-on and have
a significant amount of dust (i.e. are metal-rich),
errors in the magnitudes can be up to $0.7-1.5$ dex if dust is ignored}, 
depending on the band. This means that for edge-on galaxies, particularly 
metal-rich ones, 
reliable magnitudes will not be possible to obtain without
a proper modelling of the dust extinction. 
Similar considerations can be made for the colors, that
can not be calculated with precision better than $0.1-0.5$ dex.


\begin{figure*}
  \centering
\includegraphics[width=8.cm]{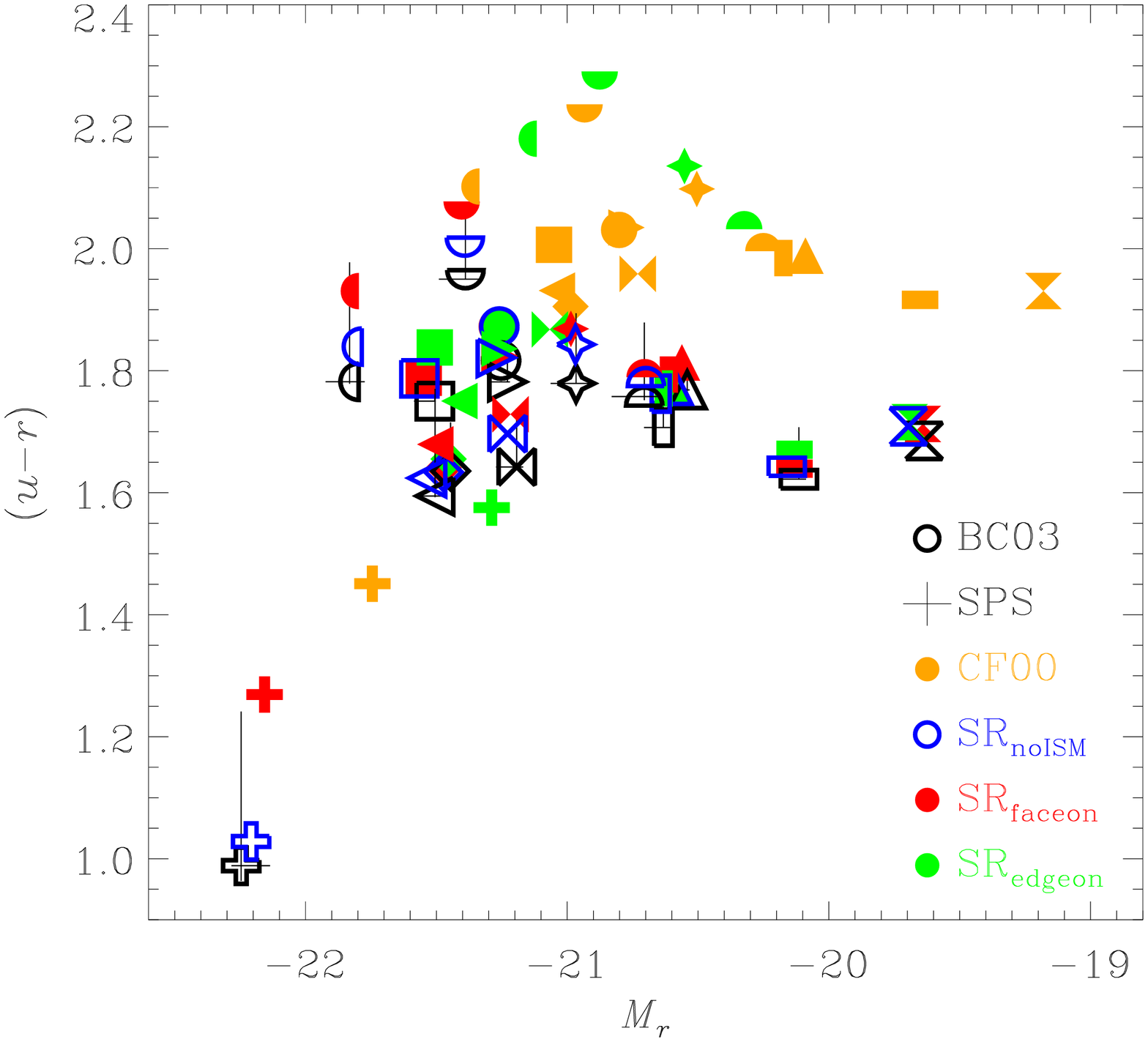}
\includegraphics[width=8.cm]{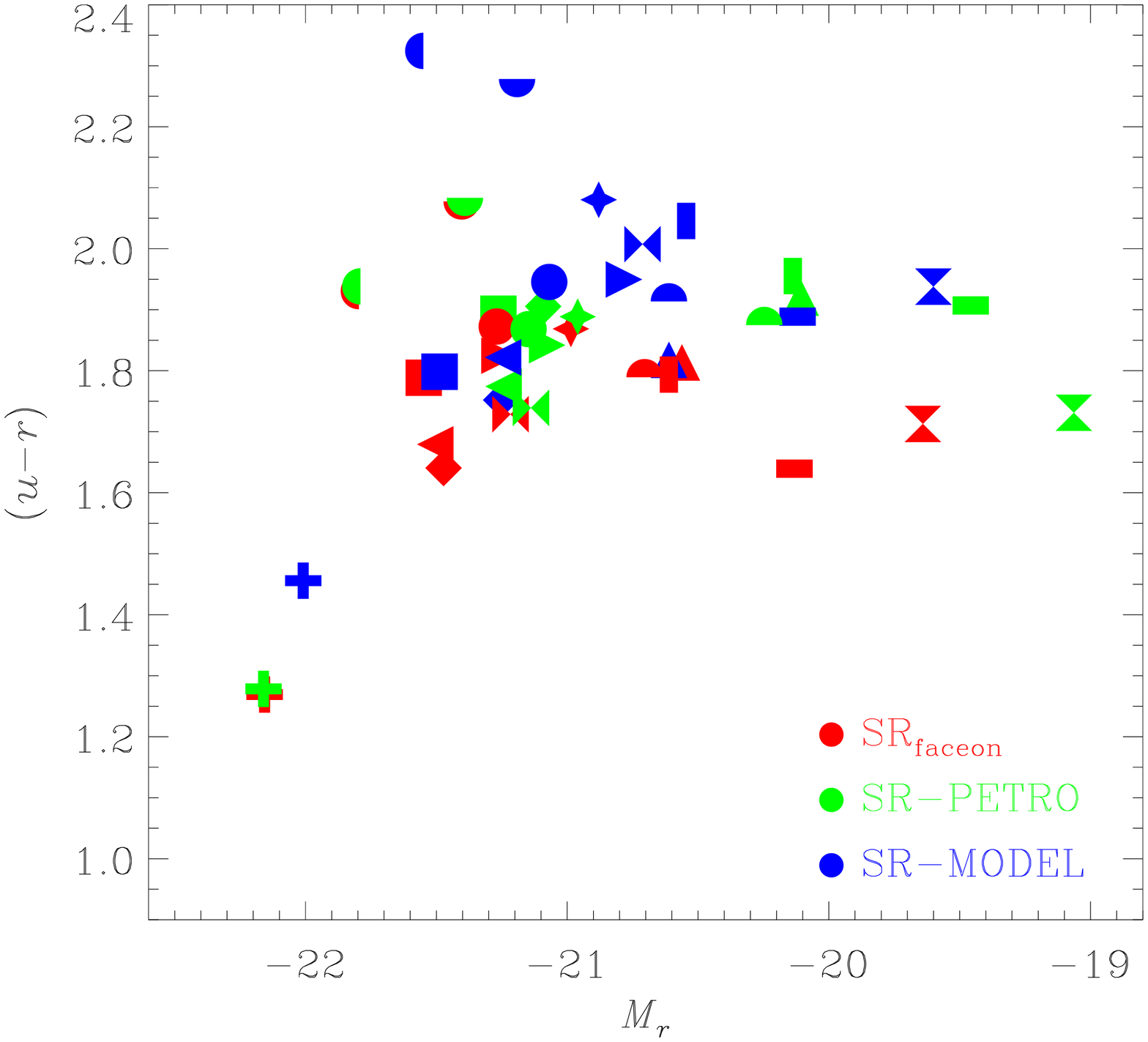}
\includegraphics[width=1.05cm]{figures/magnitudes_label_vertical.eps}
\caption{Color-magnitude diagram of the simulated galaxies, obtained
using different methods to derive the synthetic spectra. The left-hand
panel show results for the SPS, CF00 and {\sc sunrise} methods
and the right-hand panel compares results obtained following
the SDSS techniques.}
\label{fig:col_mag_SPS}
\end{figure*}

The previous figures showed the dependence of the galaxies'
magnitudes on different models and assumptions. In simulation
studies, it is common to use SPS models 
to convert masses into luminosities, and to compare
models with observational data (e.g. light profiles, 
magnitudes, etc). We have shown that the use of different
SPS models introduces some changes in the predicted magnitudes, that
are however moderate. On the other hand, when dust effects are
included or a proper radiative transfer treatment is
considered, larger differences might appear.
In the left-hand panel of Fig.~\ref{fig:col_mag_SPS} we show the position of
our simulated galaxies  in the color-magnitude
diagram; in this case in the $M_r$ vs ($u-r$) plane. We show
the results for {\small BC03}, the spread found for
all SPS models  in both
plotted quantities, and the results for the BC03 model
with the CF00 correction. We also show results
for SR$_{\rm noISM}$,  SR$_{\rm faceon}$ and  SR$_{\rm edgeon}$.
These
cover the commonly used ways to post-process
simulation results, and allow to understand what
are the possible offsets expected when more sophisticated
methods are used to calculate the magnitudes.
Most of our simulated galaxies have $M_r$ between $-22$ and
$-19$, and ($u-r$) color between $\sim 1.5$ (except
for A-MA) and $\sim 2.3$ (see also PaperII for a comparison
with SDSS data).
The  A-MA galaxy  lies at a different position compared to the other galaxies,
particularly when dust effects are ignored or when it is included but
the galaxy is seen face-on. 
Note that this is a young galaxy with strong emission lines, and
the H$\alpha$ 
line falls in the $r$-filter at $z=0$.
A-MA however  moves photometrically closer to 
the other galaxies if dust extinction is included.

In observations, measuring the total magnitudes of galaxies can be 
plagued by different observational problems (e.g. low signal-to-noise, 
sky brightness, bad calibrations and sky subtraction), which
affect particularly the outer part of the profiles where the S/N is lower.
For this reason, definitions on how to measure the total light are used
in all galaxy surveys, although they can differ substantially
among authors and collaborations (e.g. \citealt{Kron80}, \citealt{Blanton01}, 
\citealt{Norberg02}).  
We have calculated the magnitudes and colors of our
simulated galaxies following the techniques used
in SDSS, in particular calculating the 
so-called Petrosian and Model magnitudes
  \citep{Petrosian76,Blanton01,Yasuda01,Kauffmann03,Salim07}.
These were derived from the images 
obtained with {\sc sunrise}, including dust and
in the face-on view  (SR$_{\rm faceon}$), as follows 
 \citep{Blanton01}:
\begin{itemize}
\item {\bf{\small PETRO}}: 
we calculate 
the Petrosian Radius \citep{Petrosian76} in the $r$-band, and we derive the Petrosian 
magnitudes in all bands taking the flux inside $N_P = 2$ Petrosian radii;
as in SDSS, we assume a Petrosian Ratio $R_P = 0.2$. Note
that, in general, this method samples only part of the total flux, 
the fraction depending on the luminosity profile (see \citealt{Graham05} 
for detailed calculations).
\item {\bf{\small MODEL}}: in this case, each image is matched to 
different 
luminosity profiles, and the magnitudes are calculated from the profile
which gives the best fit.
We used the code {\sc galfit}  to perform the fit \citep{Peng10}, assuming 
arbitrary axis ratio and position angle, and used two different profiles:
Exponential and DeVaucouleur. Depending on the galaxy, one of these
profiles provided the best fit: for 5 galaxies this was an exponential, for 10
a DeVaucouleur profile.
\end{itemize}

The right-hand panel of Fig.~\ref{fig:col_mag_SPS} compares
the results
obtained with the {\small PETRO} and {\small MODEL}  methods to those 
of SR$_{\rm faceon}$.
We find in general similar values of magnitudes 
for the {\small PETRO} and
SR$_{\rm faceon}$ methods, with {\small PETRO} magnitudes
always higher (i.e. fainter) than those of SR$_{\rm faceon}$
as expected. 
In the case of the {\small MODEL} method,
we also find  fainter galaxies compared to the results of SR$_{\rm faceon}$.
In both cases, differences are lower than $0.3$ dex for 9/15 galaxies,
with a maximum of $0.6$ dex for the remaining 6 systems.
Colors are similar in the three models, with
differences in $(u-r)$
$\lesssim 0.1-0.2$ dex.
Our results show that, for our simulated galaxies 
the {\it {\small PETRO} and {\small MODEL} methods
can not recover the real magnitudes/colors of galaxies
with a precision better than $0.2-0.3$ dex.}


\subsection{Stellar mass}
\label{subsec:stellar_mass}

The stellar mass of galaxies is an important proxy of how the galaxy 
populations evolved over cosmic time \citep[e.g.][]{Bell03}. 
In this section, we compare the stellar mass of our simulated galaxies,
 obtained  in different ways, including the simple
sum over the mass of stellar particles (as done in simulation studies)
and 
those obtained using different post-processing techniques that mimic 
observations, in particular the ones used in  SDSS.
These are described below:
\begin{itemize}
\item {\bf {\small {SIM}}}: the total mass of star particles in the simulated 
galaxy. We include all stars in the 60 kpc$\times$60 kpc field of view 
(see sec.~\ref{sec:results}), i.e., the same field of view of
{\sc sunrise}.
\item {\bf{\small BC03}}: the final mass of star particles is calculated with
the BC03 model, which considers the mass lost by a stellar population since
it was formed, and until the present time. 
The BC03 final mass of each star particle is obtained normalizing
with its initial mass. 
The total stellar mass of the galaxy is the sum over the final
masses of the star particles within the field of view.
\item {\bf{\small SED}}: the total stellar mass is estimated by fitting 
the {\sc sunrise} face-on 
optical galaxy spectrum (SR$_{\rm faceon}$ method) over the range $3800 - 9000$ $\: \mathring{A}$ 
using  the {\sc starlight} code \citep{CidFernandes09} with the BC03 SPS model. The 
spectra include nebular emission (masked during the fit) and are dust-extincted.
\item {\bf{\small PHOTO}}: we fit the photometric $(u,g,r,i,z)$-band 
magnitudes obtained from the SR$_{\rm faceon}$ spectrum
 to a grid of models 
as updated from BC03 in 2007 (CB07, Charlot \& Bruzual, priv. comm.); the grid spans a large 
range in galaxy star formation histories, ages and metallicities
 (see \citealt{Walcher08} for a 
description of the code).
To remove the nebular contribution from the broad-band magnitudes
(the fitted model considers only stellar light),  we 
calculate the relative contribution of nebular emission within the 
fiber in each photometric band fitting the fiber spectrum with the {\sc starlight} 
code, and 
assume that the relative contribution of nebular emission for the total
galaxy is the same as in the fiber. This procedure allows one to
mimic the stellar mass estimation of the
Garching SDSS DR7 (which we will use in PaperII to
compare simulated and observed galaxies). 
\item {\bf{\small PETRO}}: Petrosian stellar masses are derived
as in {\small PHOTO-MASS}, but using the Petrosian magnitudes (Section~\ref{subsec:magnitudes}) as input for the fit (i.e. within 2 times the Petrosian radius).
The same procedure to remove nebular emission as in {\small PHOTO-MASS}
 was applied in this case.
\item {\bf{\small MODEL}}: The same as {\small PETRO-MASS}, but using the
Model magnitudes (Section~\ref{subsec:magnitudes}).
\end{itemize}

\begin{figure}
  \centering
  \includegraphics[width=8.5cm]{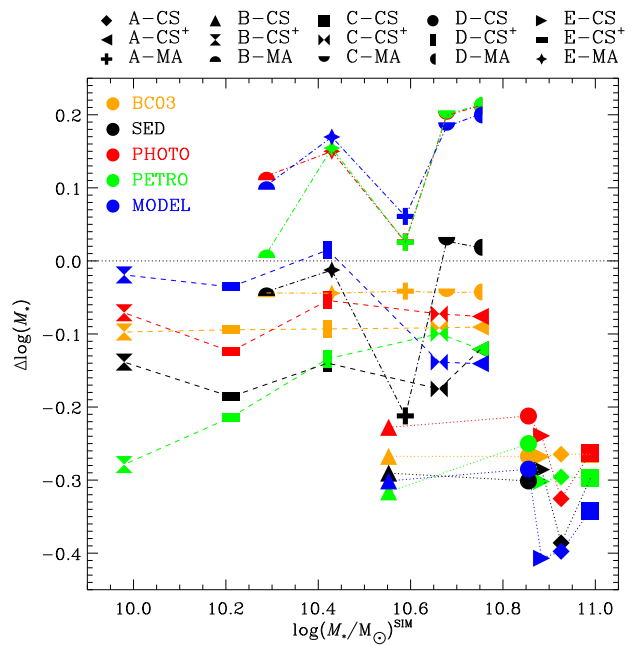}
  \caption{Comparison between different estimators
of the stellar mass of galaxies, with respect to the
real stellar mass, calculated as the
simple sum of the mass of star particles ($M_*^{\rm SIM}$).
Dashed and dot-dashed lines indicate simulations with a proper
accounting of stellar mass loss, while dotted lines are for simulations
where mass loss is not consistently followed.}
\label{fig:stellar_mass}
\end{figure}

In order to assess the effects of observational biases
on the determination of the stellar mass of galaxies, it
is important to consider the effects of mass loss of stellar particles
in the simulations, as
for all observationally-derived quantities, 
the fitting of spectra/magnitudes is done using the SPS models
that include a mass loss prescription.
Our 15 simulations cover the cases where mass
loss is not followed, and where it is properly treated,
allowing one to assess how important this effect can be. 
Moreover, simulations assume sometimes choices for the IMF that
are different than those used in the SPS models.
It is for these reasons that we calculated the stellar mass
of simulated galaxies using the {\small BC03} method additionally
to the direct result of the simulation ({\small SIM} method).

In Fig.~\ref{fig:stellar_mass} we compare the stellar masses
of simulated galaxies, obtained with the different methods,
to the stellar mass obtained directly from the simulations ({\small SIM}).
Also shown is the 1-to-1 relation (solid black line).
In the lower panel, dotted, dashed and dot-dashed lines indicate 
respectively galaxies
in the CS/CS$^+$/MA samples, that differ in the treatment
of mass loss of stars. 
We find
that for galaxies where mass loss was
not properly treated (CS), the
stellar masses obtained with all methods
are systematically lower compared to the  direct result of the simulations.
Furthermore, the offset is very similar for all galaxies, of the
order of $0.3$ dex in log($M_*$).
The results of the {\small PHOTO}, {\small PETRO}
and {\small MODEL} methods are similar, with the former
giving systematically higher stellar masses, and the latter
predicting the lowest values of stellar masses.

In contrast, for galaxies in the simulations with a proper mass loss treatment
(CS$^+$/MA), the stellar masses obtained with the {\small BC03} 
shows better agreement with the direct result ({\small SIM}),
with differences up to $\sim 0.1$ dex.
The {\small SED} method predicts slightly lower stellar masses,
with typical differences of the order of $0.05-0.2$ dex.
In the case of the
{\small PHOTO}, {\small PETRO}
and {\small MODEL} methods, stellar masses have some scatter 
up to $\sim 0.1-0.2$ dex in log($M_*$), and 
are systematically lower for CS$^+$ (dashed lines) and higher 
for MA (dot-dashed lines) compared to the direct result of simulations.

The results of this section show that {\it
the observational estimators are able to recover the mass up
to 0.1-0.2 dex in logarithmic scale, only in simulations where
the mass loss of stars is consistently included. In contrast, simulations
with no treatment of mass loss can still be compared with observations
in a meaningful manner after a fixed correction (at least for the
mass range of our sample) is applied.}


\subsection{Stellar ages and metallicities}
\label{sec:stellar_ages}

The determination of stellar metallicities and ages of galaxies
from observational data is usually accomplished 
by fitting either selected sensitive absorption line 
indices (e.g. Lick indices, \citealt{Worthey94}, \citealt{Trager98}) 
derived from 
high S/N spectra to a grid of SPS models (see 
\citealt{Trager00a,Trager00b,Gallazzi05,Gallazzi06}) 
or the full available spectrum ($\lambda$-to-$\lambda$ 
method, e.g. \citealt{CidFernandes05,Tojeiro07,Tojeiro09,Chen10}), 
which also requires high quality 
and S/N spectra for a reliable fit. Some authors also exploit colors 
to 
 estimate ages and metallicities, especially at high 
redshift \citep[e.g.][]{Lee10,Pforr12,Li13}.

In this Section, we compare the stellar ages and metallicities of the 
simulated galaxies obtained in different ways.
In order to assess the effects of the fiber bias, for all
methods we consider  stars within the same field of view than in the 
previous section, i.e.
in a projected area of $(60\text{kpc})^2$, and also only those within the
fiber. The different methods are as follows:
\begin{itemize}
\item {\bf{\small SIM}}: the direct result of the simulations, i.e., 
the mean stellar age/metallicity of a galaxy
is calculated as the corresponding mass-weighted mean over the stellar particles. 
This requires no post-processing or additional assumptions, and is the
most common way to derive ages/metallicities from simulations.
\item {\bf{\small SIM-LUM}}: 
we weight with the luminosity of stellar particles to 
calculate the average ages/metallicities,
which more closely reflects what 
would be obtained in an observation;
we use the $r$-band particle luminosity for stellar ages and the
particles' luminosity in all visible bands for the metallicities (both
obtained with the BC03 model) to mimic the SDSS analysis 
(\citealt{Gallazzi05}).
\item {\bf{\small SED-FIT}}: we fit the spectra of simulated galaxies obtained with 
{\sc sunrise} using the {\sc starlight} code 
\citep{CidFernandes05}, selecting the  best mixture of $\sim 300$ 
instantaneous-burst SSPs with different ages and metallicities. The 
fitted  spectra correspond to the face-on views and
include dust and nebular emission (SR$_{\rm faceon}$ method). 
From the fitted mixture of SSPs we calculate the mean (optical) 
light-weighted ages and metallicities.
\item {\bf{\small LICK-IND}}: the mean ages and metallicities were 
computed by Anna
Gallazzi (priv. comm.)  for a subsample of 10/15 galaxies 
using the method described in \citet{Gallazzi05}.
The method is based 
on fitting five 
absorption features (Lick indices D$4000n$, H$\beta$, H$\delta_A$+H$\gamma_A$, 
[$\text{Mg}_2$Fe], [MgFe]) to libraries constructed with the BC03 model 
with different star formation history, metallicity and
velocity dispersion, deriving the 
(optical luminosity-weighted) 
stellar metallicities  and (r-band weighted) mean ages. 
As the spectra generated with SUNRISE have the resolution of the input
stellar model (SB99) with  
spacing 20 $\mathring{A}$ (which is too low to compute 
meaningful absorption indices), we thus explore an alternative way to 
measure the Lick indices of the simulated galaxies directly from 
the BC03 tables, after interpolating with 
age and metallicity for each star particle and averaging weighting with 
the particle mass. However, we note that these may 
differ from the indices measured directly from the spectrum, as done in 
observations and in the model library of Gallazzi et al. used to interpret 
observations. For this reason, this method is not fully consistent
with the one used in SDSS. 
We will explore this point in subsequent analysis (PaperII) with 
higher-resolution spectra.
\end{itemize}

\begin{figure}
  \centering
  \includegraphics[width=8.5cm]{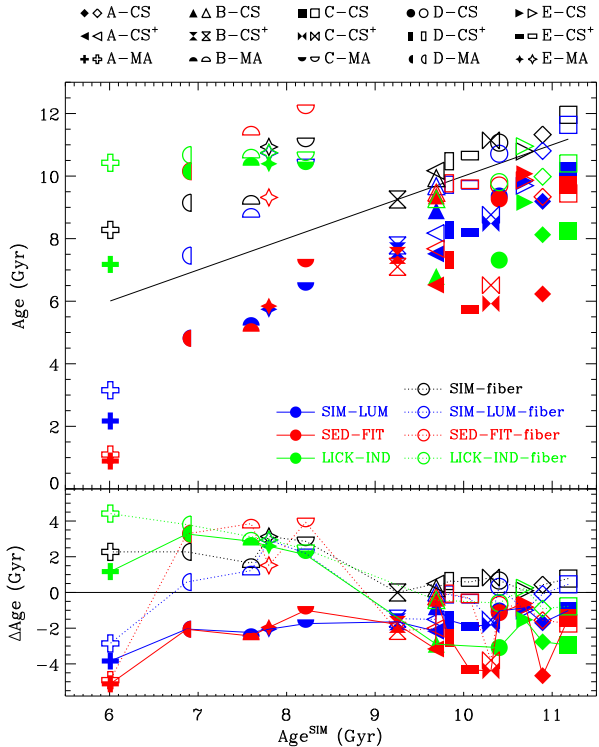}
  \caption{Difference between the mean mass-weighted stellar age in 
simulations and the ages calculated with the observational methods. The galaxies are sorted by increasing mean age to the right.}
\label{fig:stellar_age_diff}
\end{figure}

\begin{figure}
  \centering
  \includegraphics[width=8.5cm]{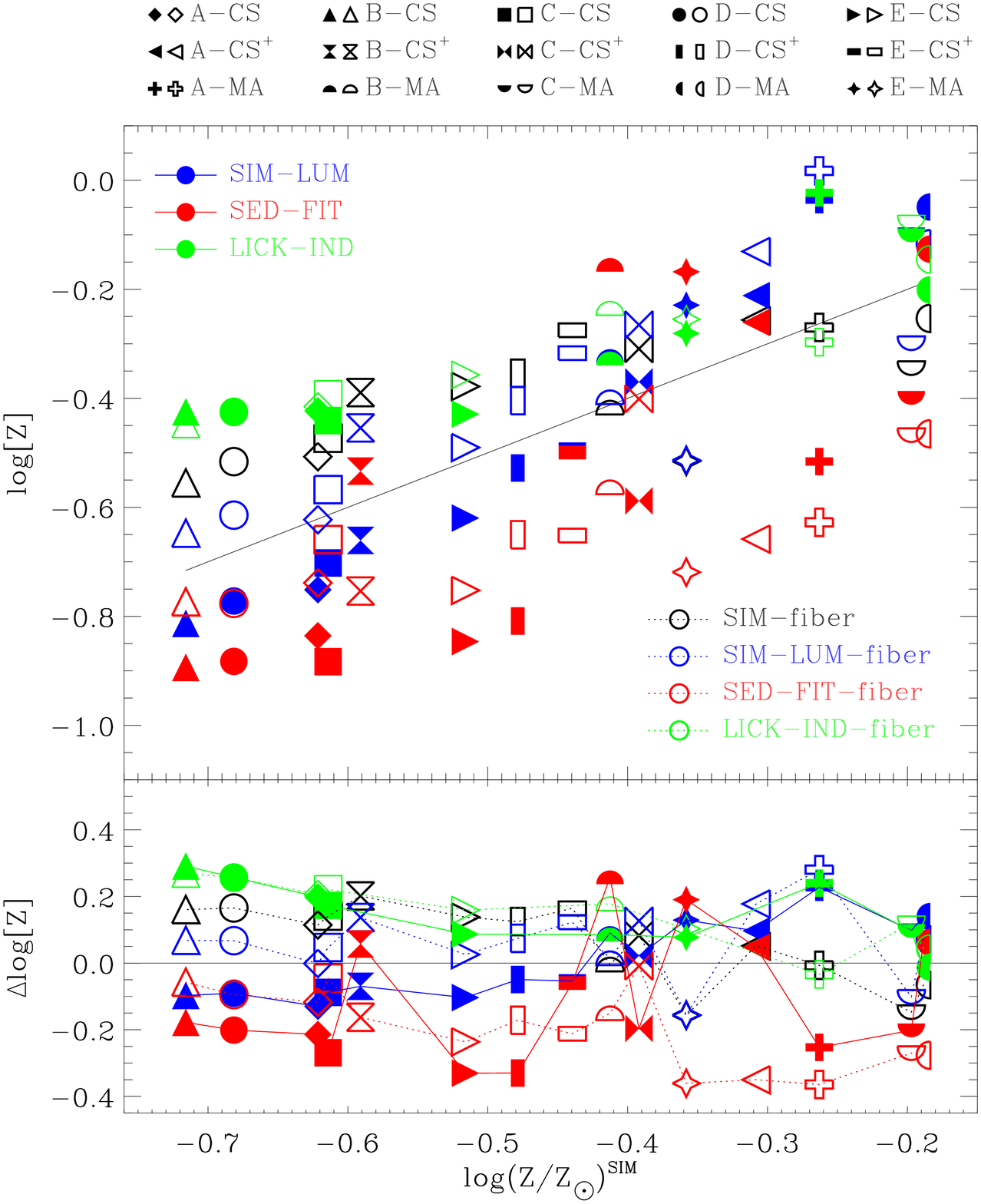}
  \caption{ Difference between the mean mass-weighted stellar metallicity and 
the observational values, ordering the galaxies by increasing stellar 
metallicity to the right.}
\label{fig:stellar_met_diff}
\end{figure}

The upper panel of Fig.~\ref{fig:stellar_age_diff}  shows
the galaxies' stellar ages obtained with the various methods described
above as a function of the age obtained directly from the simulations
(i.e. with the {\small SIM} method, Age$^{\rm SIM}$),
together with the 1:1 relation, while the lower panel shows the
corresponding differences. 
Results are shown in the cases of considering all stars in the field of
view and stars within the fiber.

In general, when all stars in the field of view are
considered (filled symbols and solid lines), 
the ages obtained using the {\small SIM-LUM} and {\small SED-FIT}
methods are similar, and are systematically lower than the simple average of 
the age
of stellar particles. This results from the fact that
young stellar populations emit more light than old ones,
therefore have a comparatively higher weight in {\small SIM-LUM} and 
{\small SED-FIT}. 
Typical variations are of the order of $\lesssim 2$ Gyr, but
the discrepancies are larger for some systems, 
with a maximum of about $\sim 5$ Gyr for the youngest galaxy.
The estimation with the (mass-weighted) Lick indices gives systematically 
older ages 
for young galaxies  (Age$^{\rm SIM}<$ 8 Gyr)  compared to SIM-LUM and SED-FIT, 
and slightly 
older ages compared to SIM, while it is in the range of the other observational 
estimators for the older galaxies.

Contrary to what we found for the ages estimated
from all stars, in the case of the fiber quantities (open symbols and dotted lines)
we find
systematically higher ages compare to
Age$^{\rm SIM}$, with differences up to $\sim 4$ Gyr. 
Note that the differences depend sensitively on the age but also on the presence/absence of
age gradients which, as shown in Fig.~\ref{fig:hist_stellar_age},
can vary significantly from galaxy to galaxy. 
For old galaxies, all methods, in general,  agree better with each other.
We note that, for quantities within the fiber, the two sources of bias 
(preferential
sampling of inner populations and dependence on age gradients) can
lead to both positive and negative differences with respect
to the direct result of the simulation. 
The LICK-IND method taking only particles in
the fiber (LICK-IND fiber) gives in general slightly older ages compared 
to LICK-IND.


We also applied the three methods and calculated total and fiber quantities
for the stellar metallicities. 
Fig.~\ref{fig:stellar_met_diff} shows a comparison of results.
Compared to the direct result of the simulation, the {\small SIM-LUM} and {\small
SED-FIT} methods give in general 
 lower metallicities for metal-poor (old) galaxies and
 higher metallicities for more metal-rich (younger) ones.
This is again explained by the different relative weight of old and
young stars to the average metallicity (note that the spread
in stellar metallicities is much smaller than that for stellar ages). 
Differences are however moderate, always of the order of $\pm 0.3$ dex. 

For quantities within the fiber, we also find moderate differences; in this
case, metallicities tend to be higher compared to total quantities,
as the contribution of very low metallicity stellar populations gets smaller
(Fig.~\ref{fig:hist_stellar_met}).
When we apply the LICK-IND and LICK-IND fiber methods, we find, for the 
low-metal sample
($\log(Z/Z_{\odot})\lesssim -0.45$), systematically higher metallicities, 
while for metal-rich galaxies we find better agreement with
the other indicators. 
We detect small differences between LICK-IND and LICK-IND fiber, the 
latter giving in general
slightly higher metallicities.
It is notable that for the metal-poor galaxies,
most observational methods disagree with the direct result of the
simulations by an approximately constant factor, instead of
scaling with the metallicity.

Our results show that in our simulations
it is not possible to estimate the mean stellar age
with accuracy better than $\sim 3$ Gyr, and
the stellar metallicities better than $\sim 0.3$ dex, depending on 
the method and on the way averaged quantities are calculated. The fiber 
bias has a strong effect on the derived ages/metallicities, due
to the preferential sampling of old and metal-rich stars;
our 15 galaxies show wide variety of gradients, and
therefore we find that the fiber bias when applying
different methods can significantly change from galaxy to galaxy.


\subsection{Gas metallicity}
\label{subsec:gas_metallicity}

In this section, we compare the gas oxygen abundance of our simulated
galaxies, using different estimators.
Observationally, the determination of the
chemical composition of the gas in galaxies is based
on metallicity-sensitive emission line ratios.
A number of different metallicity calibrations are used 
to estimate the O/H ratio in nebulae and galaxies, and comparisons
among the results of various calibrations reveal
large discrepancy with systematic offsets 
that can reach $\sim 0.7$ dex (\citealt{Pilyugin01}, see also 
Appendix~\ref{app:gas_metallicity}).

As in the previous sections,
we calculate the gas chemical composition 
following the most common methods from simulations and observations,
and, when appropriate, considering all gas particles and only those within the fiber (labelled
as ``fiber''). 
When we use (face-on) {\sc sunrise} spectra 
for the calculation ({\small T04}, {\small KK04}, {\small $T_e$}), 
we first Balmer-correct 
the emission line ratios for dust extinction using the Calzetti law \citep{Calzetti94}.
The methods are described in the following.
\begin{itemize} 
\item {\bf{\small SIM}}: the gas metallicity 
of a galaxy is calculated as the mean (O/H) of gas particles, weighted by their
mass. 
\item {\bf{\small HII}}: is the mean mass-weighted O/H 
of gas particles, but only considering particles 
around stars younger than 20 Myr and inside 1 kpc radius. In this
way, we mimic the preferential bias to young stellar populations
in nebulae-based measurements. Note that although the HII regions have 
sizes which can vary in a wide range of $ 0.1 -200$ 
pc from ultra-compact to giant extragalactic HII regions, 
we are limited to define smaller sizes of the HII regions by the 
resolution of the simulations, which is of
the order of $\sim 1$ kpc. 
\item {\bf{\small T04}} \citep{Tremonti04}: the gas metallicity is 
computed using the calibration of the 
$R_{23}$-upper branch given in \citet{Tremonti04}. According to \citet{Kewley08} 
we use [NII]/[OII] to 
remove the degeneracy in the $R_{23}$-metallicity relation, and we define 
the upper branch as $\text{Log ([NII]/[OII])} \geq -1.2$ (see also 
Appendix~\ref{app:gas_metallicity}).
\item {\bf{\small KK04}} \citep{Kobulnicky04}: is a widely used metallicity calibration,
based on an iterative method that allows to solve both the oxygen abundance 
and ionization parameter using the $O_{32}$ and $R_{23}$ line ratios.
\item {\small $\mathbf{T_e}$}: the electron-temperature calibration 
(also referred to as ``direct'' method) is based
on the ratio between the auroral line [OIII] $\lambda 4363$ and [OIII] 
$\lambda 4959, 5007$; it is commonly used to determine gas metallicities 
when the weak auroral line can be detected. Here we follow the procedure 
outlined in \citet{Izotov06}.
\end{itemize}

\begin{figure}
  \centering
\includegraphics[width=8.5cm]{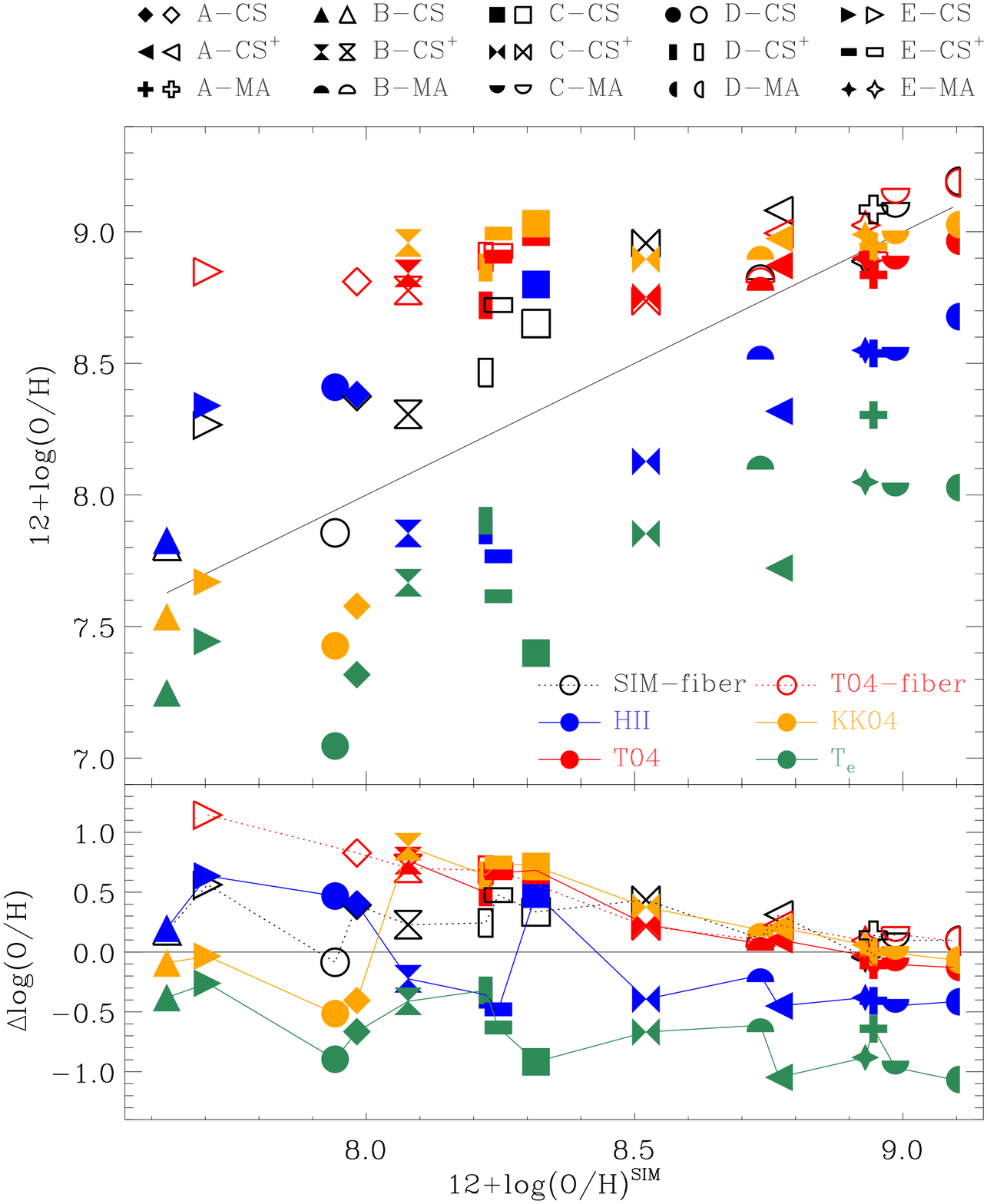}
\caption{Comparison between the different estimators for the
the galaxies' oxygen gas metallicities, as a function of that
obtained with the {\small SIM} method. For reference, in the upper panel we
include the 1:1 relation as a solid line. As explained in the text, 
T04 calibration is valid only for a subsample of the galaxies.}
\label{fig:gas_met_diff}
\end{figure}

Fig.~\ref{fig:gas_met_diff} shows our results for the oxygen abundances
of the simulated galaxies obtained using the different methods. 
We show results as a function of the real gas metallicity,
i.e. 12+log(O/H)$^{\rm SIM}$, the differences are also calculated
with respect to this estimator.
As expected, the metallicities predicted using the {\small SIM-fiber} 
and {\small T04-fiber} methods are systematically
higher  compared to {\small SIM} as the most metal-enriched 
regions in the bulge are preferentially sampled in these cases. 
This happens however only for the metal-poor galaxies, while
for the metal-rich ones these estimators agree very well.
Note that, in our simulations, the galaxies with low/moderate metal content (CS and CS$^+$ samples)
are also the ones with stronger metallicity gradients, 
while galaxies in the MA sample do not show significant
metallicity gradients, 
masking a possible bias due to preferential sampling (see also Fig.~\ref{fig:hist_gas_met}). 
The HII estimator, which samples 
the gas near young star particles, shows  $\pm 0.5$ dex scatter 
around the {\small SIM} value, in general higher for the galaxies
with $12+$log(O/H)$\lesssim$ 8.3 and lower for the more oxygen-rich ones.

The gas metallicities obtained
with the emission-line calibration of  {\small T04} are systematically
higher, with
large discrepancies only for galaxies with low metal content.
However, note that only for 11/15 galaxies this calculation 
is possible\footnote{Note that, when the {\small T04} 
calibration is used inside the fiber, the estimation is possible for 
12 out of the 15 galaxies. }. 
Also for the {\small KK04} method, we find good agreement
with {\small SIM} for the metal-rich systems,
with some offsets for metal-poor galaxies (i.e. 
$12+\log(\text{O/H}) \lesssim 8.3$).

The largest discrepancies in gas metallicities, both with respect to
the real value and to the other observational estimators,
are found for the  $T_e$ method, which 
gives the lowest metallicity values. Unlike the other methods,
the systematically lower metallicities are found even for the most metal rich 
galaxies where the rest of the methods agree well. The largest
differences are of the order of
 $\sim 1$ dex, and only a weak trend with the metallicity is detected.

It is worth noting that the emission line intensities
in the spectra calculated from {\sc sunrise} rely on the 
{\sc mappings III} photoionization code.
The line ratios are then affected by uncertainties, assumptions and 
approximations in the model that describe the photodissociation regions, 
e.g. on-the-spot approximation \citep{Stasinska02, Stasinska07}, 
assumptions on the geometry of the nebula, dust grain composition, shocks,
etc. 
\citep{Groves04}, which will in turn affect the derivation of 
the gas metallicities.
In any case, our results show that wide differences in metallicity 
can appear due to the use of various calibrations even on the same 
spectrum modeled with a given photoionization code.

In summary, our analysis shows that {\it the fiber bias has a strong 
effect on the mean gas metallicity of galaxies, particularly
for metal-poor systems and for systems with strong metallicity gradients,  
with metallicities systematically 
higher, up to  $\sim 1$ dex, compared to the total mean metallicities.
Deriving metallicities from the emission-line calibrations shows also
large offsets, that can reach $0.5-0.8$ dex compared to the value in the
simulations, but in this case, the results are more diverse,
with both positive and negative differences.
Finally, the
calibration based on electron temperature predicts systematically 
lower metallicities, by $0.7 - 1$ dex, compared to the rest of the methods
and to the metallicities derived directly from the simulations.
}


\subsection{Star formation rate}
\label{subsec:star_formation_rate}

In observations, the SFR of galaxies is estimated using different
methods at different redshifts, e.g. the luminosity of 
the H$\alpha$ line, of the  $[OII]\lambda3727$ line and the UV continuum 
for  low, intermediate and high redshifts, respectively. Each SFR indicator 
is affected by theoretical biases and observational uncertainties 
(e.g. different timescale, sparse wavelength sampling, contribution 
from old stars, AGN contamination) which can influence the final 
SFR value \citep[][]{Calzetti08}. Furthermore, the use of different 
SFR proxies at different redshifts may introduce systematic errors that 
can bias the comparison between simulations and observations 
\citep{Rosa-Gonzalez02}.

In this section, we calculate the SFRs of our simulated galaxies 
with different methods. We use the calibrations given in \citet{Kennicutt98}, applying 
the correction factor $f_{\rm IMF}=1.5$ \citep{Calzetti09}
to account for the use of a different IMF (\citealt{Kennicutt98} assumes 
a Salpeter IMF while we assume a Kroupa IMF in {\sc sunrise} and 
Chabrier in BC03). 
We note that, in SDSS, the fiber-derived SFRs \citep{Brinchmann04} are corrected to 
total SFR with the \citet{Salim07} method. For this reason,
we use the full face-on spectra
 obtained with {\sc sunrise}, dust-corrected with the Calzetti law, 
 to derive SFRs, instead of the spectra
 within the fiber.
The various methods to derive SFRs are described below. 

\begin{itemize}
\item {\bf{\small SIM}}: the real SFR  extracted directly from the simulations
(stellar mass formed over time interval),  averaged over the past 0.2 Gyr.
\item {\bf{\small BC03}}: we convert the rate of ionizing photons calculated
with BC03 into SFR:
\begin{equation*}
{\rm SFR}({\rm M}_\odot \, {\rm yr}^{-1}) = 1.08 \, f_{\rm IMF}^{-1} \times 10^{-53} Q(H^0) \, ({\rm s}^{-1}). 
\end{equation*} 
\item {\bf{H$\alpha$}}: we extract the H$\alpha$-luminosity 
from the  {\sc sunrise} spectrum and convert into SFR according to:
\begin{equation*}
{\rm SFR}({\rm M}_\odot \, {\rm yr}^{-1}) = 7.9 \, f_{\rm IMF}^{-1}  \times 10^{-42} L(H\alpha) \, ({\rm erg\ s}^{-1}) .
\end{equation*} 
\item {\bf{\small UV}}: we calculate the flux from {\sc sunrise} spectra 
of the nearly-flat region between $1500 - 2800 \: \mathring{A} $ and 
calculate the SFR as: 
\begin{equation*}
{\rm SFR}({\rm M}_\odot \, {\rm yr}^{-1}) = 1.4 \, f_{\rm IMF}^{-1} \times 10^{-28} L_\nu \, ({\rm erg\ s}^{-1} {\rm Hz^{-1}}). 
\end{equation*} 
\item {\bf{\small [OII]}}: we estimate the SFR using the luminosity of
the forbidden-line doublet [OII]$\lambda 3726,3729$ in {\sc sunrise}:
\begin{equation*}
{\rm SFR}({\rm M}_\odot \, {\rm yr}^{-1}) = 1.4 \, f_{\rm IMF}^{-1} \times 10^{-41} L({\rm OII}) \, ({\rm erg}\ s^{-1}). 
\end{equation*} 
\item {\bf{\small FIR}}: the SFR is estimated from the
{\sc sunrise}  FIR luminosity integrated over 
the range $8-1000 \: \mu \text{m}$ as: 
\begin{equation*}
{\rm SFR}({\rm M}_\odot \, {\rm yr}^{-1}) = 4.5 \, f_{\rm IMF}^{-1} \times 10^{-44} L_{FIR} \, ({\rm erg}\ s^{-1}).
\end{equation*} 
\end{itemize}

\begin{figure}
  \centering
\includegraphics[width=8.5cm]{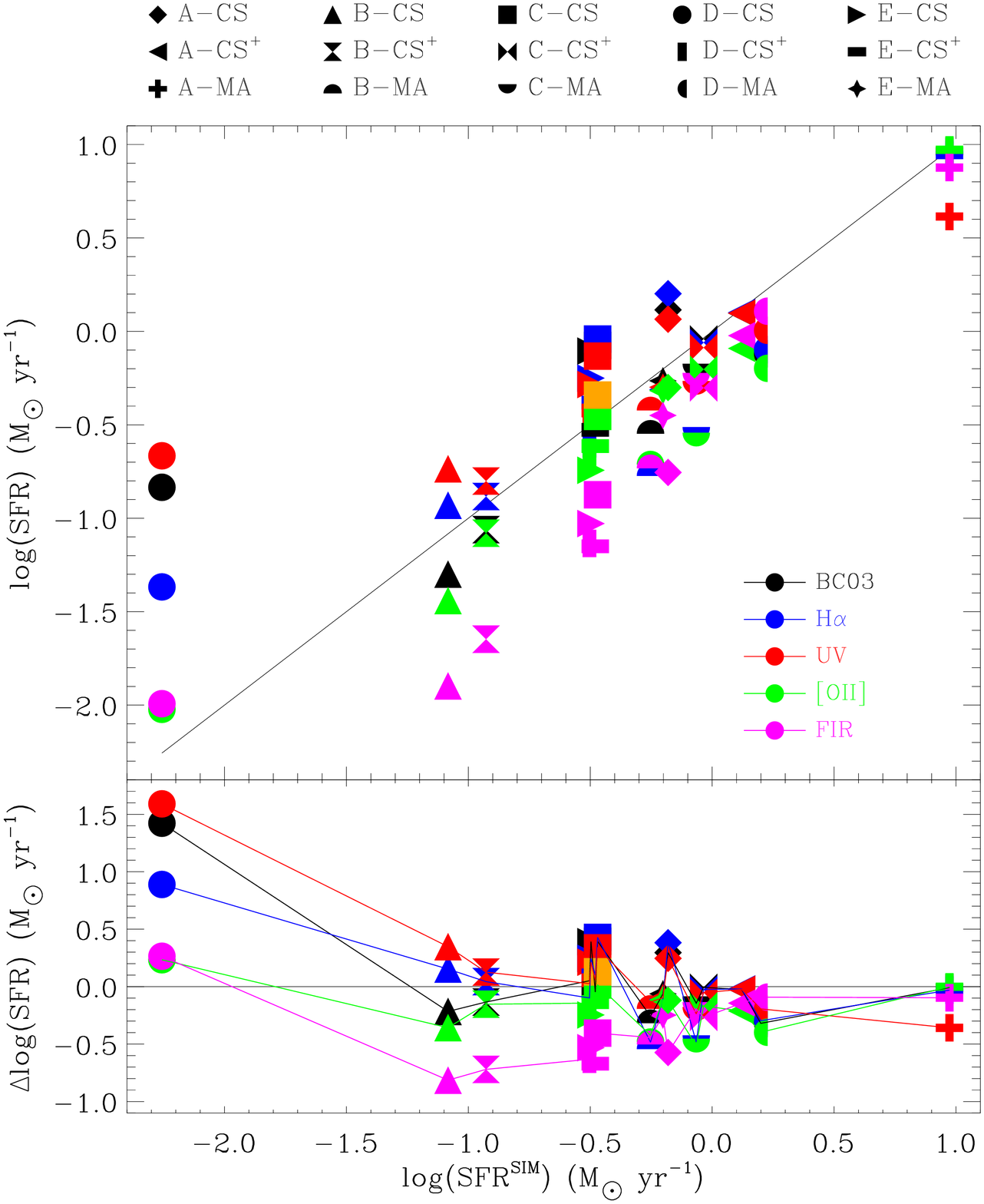}
\caption
{Comparison between the SFRs obtained using various methods
as a function of the real SFR of the simulation (method {\small SIM}).
In the upper box, we also include the 1:1 relation for reference.
}
\label{fig:sfr_diff}
\end{figure}

All observational 
methods are sensitive to the emission from young massive stars,
although the UV indicator has 10 times longer timescale 
($\sim 100$ Myr), because massive stars stay luminous for longer
time in the UV with small production of ionizing photons \citep{Calzetti08}.

Fig.~\ref{fig:sfr_diff} shows the 
differences between the SFR obtained with the different estimators
and that obtained directly from the simulations (SFR$^{\rm SIM}$).
Except for the  galaxy with the lowest SFR,
the SFRs obtained with BC03, H$_\alpha$ and UV methods agree well,
with  differences of $\lesssim 0.2-0.4$ dex in logarithmic scale with respect to
the real SFR of the simulations. Note that  BC03 and 
H$_\alpha$  are based on the same method (conversion of the number of ionizing 
photons into SFR), but use two different models.

The other observational indicators ([OII] and FIR) predict in general
lower SFRs compared to SFR$^{\rm SIM}$, with
maximum differences of about 1 dex.
In particular, 
the [OII]-SFRs of metal-poor galaxies 
are significantly lower than SFR$^{\rm SIM}$, by $\sim 0.5$
dex. Note that 
the [OII] line intensity and [OII]/H$\alpha$ ratio strongly depend 
on the metallicity \citep{Kewley04} and has been calibrated by 
\citet{Kennicutt98} for around-solar metallicity; this
calibration is not accurate for the metal-poor galaxies of our sample. 
Also in the case of the SFR obtained with the FIR method differences
are expected, as the \citet{Kennicutt98} calibration is valid 
only for dusty metal-rich starburst galaxies (note that calibrations
that follow both dust-obscured and unobscured star formation can be found 
in the literature, see \citealt{Calzetti08}); for the metal-poor sample, 
since the amount of dust in {\sc sunrise} is proportional
to the metallicity, we have only 
small dust absorption (Sec.~\ref{subsec:gas_metallicity}) and 
reemission in the infrared (see also \citealt{Hayward14}). 
To further investigate this effect, we plot
in Fig.~\ref{fig:sfr_diff} the result obtained
using the FIR method of one of our metal-poor galaxies (C-CS, orange square) 
where the assumed 
dust-to-metal ratio in {\sc sunrise} was increased
10 times (as in Section~\ref{subsec:magnitudes}).
In this case, the SFR raises by $\sim 0.6$ dex, although it is
still moderately lower compared to the results of 
the other estimators.

Finally, note that SFRs derived observationally and in simulations 
assume different timescales, 
of about $10-100$ Myr for the former compared to
$0.2 - 0.5 $ Gyr for the latter.
This  can result in significant discrepancies among 
the two estimations if we are in the presence of recent starbursts. 
In our sample, however,  
we observe such a systematic difference only in the galaxy
with the lowest SFR (D-CS).

{\it The results of this section show that
the SFRs indicators
exhibit differences of $\lesssim 0.4$ dex in log(SFR),
which are caused both by the difference in methods and the different 
characteristic time-scales to which the methods are sensitive.
The $H_\alpha$ and UV methods
predict similar 
SFRs compared to the direct results of the simulations, 
except for the galaxy with the lowest SFR. 
For galaxies with low metallicities the [OII] method gives systematically lower
SFRs, with absolute differences lower than $0.5$ in 
logarithmic scale.
For the FIR method the differences are larger in metal-poor galaxies,
with SFRs systematically lower by $0.5-1$ dex in log(SFR), while the FIR estimation agrees
with the SFR of the simulations up to $0.1-0.2$ dex for metal-rich galaxies.
}


\section{Discussion and conclusions}

We have used a set of 15 simulated galaxies, of similar mass
to the Milky Way, to study biases and systematics in the
derivation of observables, and in the comparison
between them and observational data.
The aim of this study is twofold; first, help simulators
to be aware of systematics and be able to reliably
judge the agreement between simulated and observed galaxies
and, second, help observers to interpret observational
results by being able to better quantify
the differences between the observationally-obtained
galaxy properties and the real ones (that are known
in the simulations).

Our simulations comprise 15 galaxies with a
variety of merger, formation and accretion histories,
which results in a variety of final morphologies,
star formation rates, gas fractions and metallicities.
As these properties somewhat depend on the modelling
of feedback, the same 5 galaxies were simulated
using three different models for chemical/energy feedback;
 the strength of feedback and the amount of chemical yields varies from
moderate to strong. For moderate feedback we find
galaxies that 
are more metal-poor and form their stars
earlier, while for strong feedback we find younger galaxies
with a higher metal content. This diversity, typical of
real galaxies,  makes the
sample ideal to test biases and systematics in the
derivation of the synthetic spectra, and to study
the dependence of such effects on galaxy properties 
(with the caveat that with our sample we do not represent the more metal rich galaxies observed).

For our study we have computed the synthetic spectra
of the simulated galaxies at $z=0$, as at low-redshift many
databases of galaxy properties derived from large 
galaxy surveys (e.g. SDSS, 2dFGRS and 6dFGS) are available. 
For this purpose, we followed three different approaches:
($i$) Stellar Population Synthesis (SPS) models, which
give the spectra coming from stars; ($ii$) SPS models including
dust extinction with a simple recipe; and ($iii$) a full radiative
transfer calculation that gives the spectra including stellar and 
nebular emission, as well as the effects of dust which are
parametrized using the metallicity of the interstellar medium.
We have used the synthetic spectra to derive the observables
(magnitudes/colors, stellar masses and ages, stellar/gas metallicities
and star formation rates) in various ways, as to mimic
real observations, and we have compared the results with
the direct outputs of the simulations, i.e. the real
galaxy properties.

Biases and systematics appear at various stages in the
process of obtaining the observables from the simulations, due to:

\noindent - assumptions and parameters of SPS models, 

\noindent - dust/radiative transfer/projection effects, 

\noindent - weighting with the mass instead of luminosity for mean 
quantities,  

\noindent - observational biases, such as extrapolation to external 
regions of galaxies where no spectral/photometric data are 
available (e.g. fiber bias, Petrosian/Model magnitudes), 

\noindent - different parametrization of the star formation history  
and dust extinction when quantities are derived fitting a pre-constructed grid
of models,  

\noindent - in the particular case of gas metallicities and SFRs, the 
use of different calibrations.

We tested the effects of such biases on the magnitudes and colors of simulated
galaxies, their stellar masses, ages, stellar/gas metallicities and star formation 
rates.
Our results can be summarized as follows:

\begin{itemize}

\item {\it Magnitudes} 

\noindent - The galaxies'  magnitudes in the ({\it u,g,r,i,z}) SDSS bands 
derived 
from the five different SPS models used in our work show good
agreement, with differences lower than $0.1$ dex. Differences are larger 
for the most metal-rich galaxies, which also have a higher contribution of 
young and intermediate age stars, for which model uncertainties are the largest.

\noindent - Dust effects can be important if galaxies are 
seen edge-on. A simple
angle-averaged dust model predicts galaxies that are
$\sim 0.4-0.8$ dex  fainter compared to the no-dust case.
If full radiative transfer is considered, where dust is traced
by the metals, edge-on galaxies appear in general $\sim 0.3-1$
dex fainter, but only if these are relatively metal-rich.

\noindent - Estimating the magnitudes using a more observational approach
(as for instance the Petrosian and Model magnitudes of SDSS) exhibits some 
differences compared to the real magnitudes of the simulated galaxies,
with offsets $\sim 0.2-0.3$ dex for $60\%$ of
the systems and up to $0.6$ dex for the remaining $40\%$. 

\item {\it Stellar masses:} 

When observational techniques are applied to the simulated
galaxies to estimate their stellar masses, the results
vary depending on the treatment of mass loss in the simulations:

\noindent - If mass loss includes only SNe (which is a small
effect), all 
observational estimators
give systematically lower stellar masses compared to the
direct result of the simulation, as the fitted models include the 
full mass loss of Stellar Populations at each stage of evolution. The offset 
is similar for
all galaxies, of about a factor $0.3$ in log($M_*$) or, equivalently,
of 50$\%$ in $M_*$.

\noindent - If mass loss is properly treated in the simulations (e.g. adding
AGB stars),
the observational techniques recover the real stellar
masses with differences $0.1-0.2$ dex in log($M_*$).

\item {\it Stellar ages:}

\noindent - If stellar ages are estimated mimicking observational techniques
weighting with the luminosity (but ignoring the fiber bias),
galaxies appear younger, typically by $\sim 2$ Gyr, compared
to the direct result of simulations (mean age of stellar particles).
Some galaxies exhibit larger differences,
up to a  maximum of $5$ Gyr; among them, we find both very young
and very old galaxies. 

\noindent- If only stars in the nuclear part of galaxies are included (as 
to mimic single-fiber surveys such as SDSS), the
observationally-estimated galaxy ages are in general
higher compared to the direct result of the simulations
with differences of about $2-4$ Gyr for young
systems and of $\lesssim 1$ Gyr for old ones. However,
the results depend sensitively on the particular
properties of galaxies, namely the presence of significant/insignificant
age gradients and the resulting preferential sampling of old stars
within the fiber.

\item {\it Stellar metallicities:}

\noindent - Observationally-estimated stellar metallicities (ignoring the 
fiber bias)
are in general lower compared to the direct result 
of the simulations,
in particular if galaxies are metal-poor, with differences
up to $-0.3$ in logarithmic scale.
For more metal-rich systems, differences are similar
in absolute value, but can vary from positive to negative.
The differences in the behaviour of
metal-rich/metal-poor galaxies originates in
 the different weight of old/metal-poor and 
young/metal-rich stars in the different methods, in particular depending 
on the way the mean metallicity is calculated 
(mass/luminosity-weighted).  

\noindent - If only stars within the fiber are considered, stellar metallicities
tend to be higher compared to the direct result of the simulations 
with typical differences of $\sim 0.1-0.3$ dex and no strong dependence
with the real stellar metallicity.

\item {\it Gas metallicities:}

\noindent - Deriving gas metallicities using
different emission-line calibrations show large spread in the
resulting metallicity values, up to $0.5-0.8$ dex.
In general, metallicities lower than the real value
are predicted for very metal-poor systems, while for
more metal-rich galaxies they show a better agreement. 
Significant differences are found when the calibration
based on electron temperature is used; in this case
the derived metallicities are systematically lower
by differences of the order of $0.7 - 1$ dex.

\noindent - Gas metallicities are systematically higher than the direct
result of simulations when the fiber bias is included, 
due to a preferential sampling of metal-rich regions in the galaxy.
Differences can be up to $1$ dex for the most metal-poor galaxies,
while the for metal-rich sample 
differences are always smaller than $0.1$ dex.

\item {\it Star formation rates:}

\noindent - Observationally-derived SFRs using different
indicators present in general  offsets
of the order of  $\pm 0.4$ dex in log(SFR), 
mainly due to the
differences in methods and time-scales to which they are sensitive. 
The largest differences are found for the estimations
based on the FIR, which gives systematically lower
SFR values up to $1$ dex in logarithmic scale for the metal-poor
galaxies.

\end{itemize}

In summary, we have shown that it is important to properly
take into account different observational biases when
galaxy observations are interpreted and linked
to different underlying physical processes. Furthermore,
a meaningful comparison between observations and simulations
of galaxies requires a good understanding of the systematics;
if this is not considered it is not possible to properly
judge agreement/disagreement between simulations and observations,
and to decide which of the physical processes included
in the simulations are more relevant in the context
of the formation and evolution of galaxies in a cosmological
context.
 
In a companion paper (Guidi et al., in prep.), we compare the properties of our
simulated galaxies to observations of the SDSS survey,
in order to gain more insight in their agreement/disagreement
in relation to the treatment of feedback, star formation,
mass loss and chemical enrichment in the simulations.
We also quantity differences from the observationally-derived
quantities and the direct result of the simulations
to provide the relevant scalings and error bars for a meaningful
comparison, in the case of the different galaxy properties. 

Finally, we note that other possible sources of biases and systematics
have not been explored here, e.g. related to the inability to
resolve the typical height of gaseous discs (which could
affect the comparison between face-on and edge-on projections).
Moreover, how well
the simulated galaxies reproduce the observed sizes/concentrations/luminosity
profiles  can affect our findings related to the fiber bias and 
Petrosian/Model Magnitudes (although in previous papers we have
shown that our models produce galaxies
in broad agreement with observational results, e.g. \citealt{Scannapieco10}, 
\citealt{Scannapieco12}, \citealt{Aumer13}).
We hope that our work encourages other researches to explore all these
issues, as well as we will continue to test systematics/biases
in further analysis.

\section*{Acknowledgments}
We thank the reviewer for his/her useful comments and suggestions. 
We thank Anna Gallazzi for processing our mock observations with the SDSS 
pipeline, and Yago Ascasibar for useful discussions and comments. 
We also acknowledge Michael Aumer for providing his simulations, 
and P.-A. Poulhazan and P. Creasey for sharing the new chemical code.
GG and CS acknowledge support from the Leibniz Gemeinschaft,
through SAW-Project SAW-2012-AIP-5 129,
and from the High Performance Computer in Bavaria (SuperMUC) 
through Project pr94zo.
CJW acknowledges support through the Marie Curie Career Integration Grant 303912.

\bibliographystyle{mn2e}
\bibliography{biblio}

\appendix

\section[]{Gas metallicity calibrations}
\label{app:gas_metallicity}

The observational derivation of the gas metallicity of galaxies
requires a calibration of emission lines to recover the (O/H) ratio; different
ones are used in various observational campaigns
(e.g \citealt{Zaritsky94,Kobulnicky04,Kewley02,Pettini04}).
Previous work has already shown that large discrepancies, as
large as $\sim 0.7$ dex, arise when different
calibrations are used \citep{Pilyugin01}. 
The calibrations can be broadly classified as empirical 
(further subdivided into ``direct'' and  ``statistical'') 
and theoretical. The former derives metal 
abundances directly from electron temperature-sensitive lines or 
relations between temperature-estimated metallicities and strong emission 
lines, while the latter relies on photoionization models to calibrate the 
metallicity indicators (see the reviews by \citealt{Ferland03}; 
\citealt{Stasinska07} and \citealt{Kewley08}). 

In general, 
theoretical calibrations give, for a given galaxy, a higher metallicity compared
to electron temperature-based estimators \citep{Liang06,Kewley08}. 
The causes of these discrepancies are still unclear, 
although some authors conjecture either problems in the photoionization 
models \citep{Kennicutt03}, or temperature gradient fluctuations 
in the nebulae \citep{Stasinska05,Stasinska07,Bresolin08}. 
The effects of these disagreements can also 
bias the determination of the shape and the zero point in the 
mass-metallicity relation \citep{Andrews13}. 

\begin{figure}
  \centering
\includegraphics[width=8.5cm]{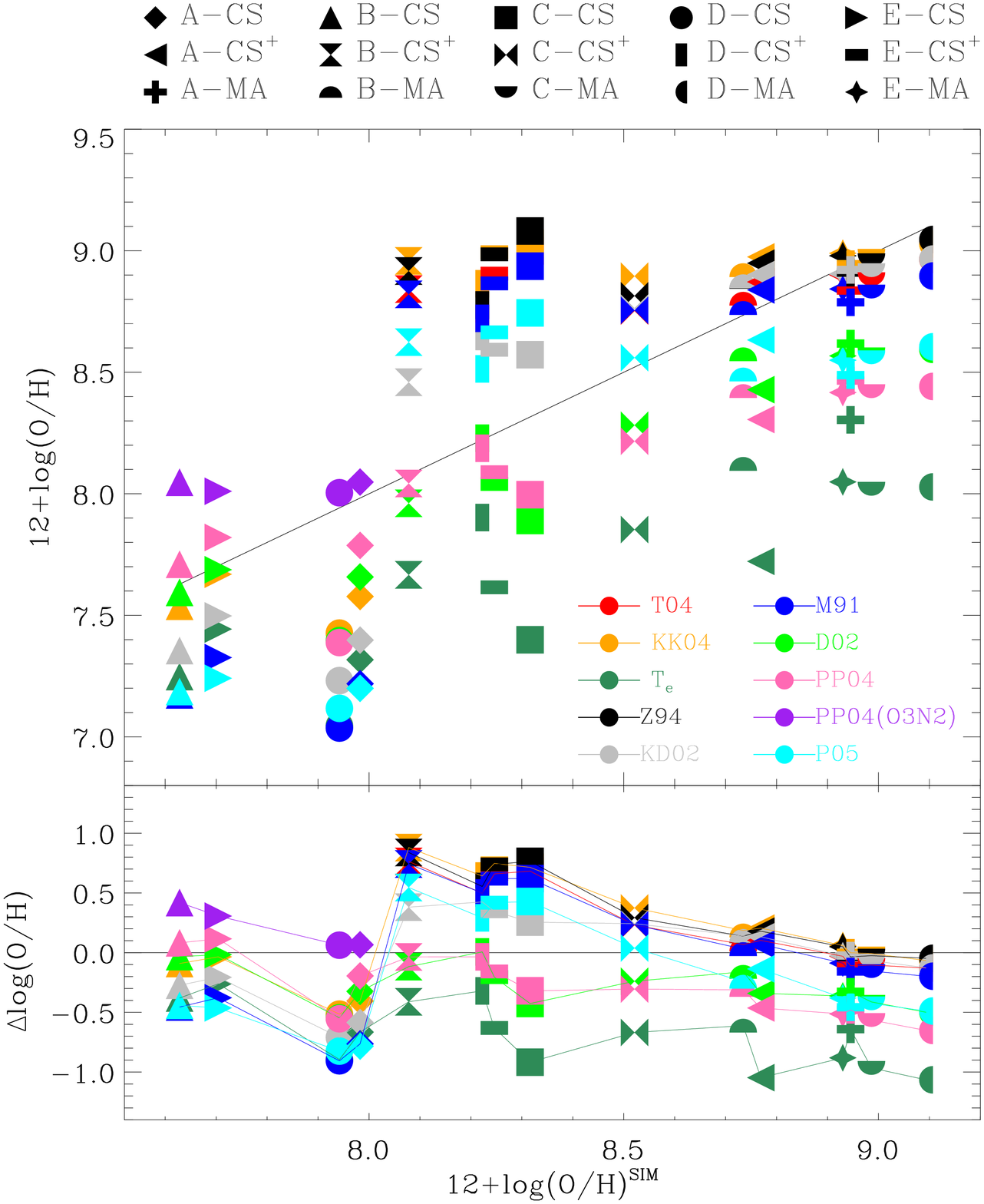}
\caption{Comparison of the gas metallicity of simulated galaxies
using different calibrations, as a function of the direct result
of the simulation.
}
\label{fig:gas_met_calibrations}
\end{figure}

Fig.~\ref{fig:gas_met_calibrations}
presents a comparison between the oxygen abundance
of our simulated galaxies assuming various calibrations,
that we summarize in Table~\ref{table:metallicity_calibrations}.
All results are derived after Balmer-correcting the emission
line ratios for dust extinction using the Calzetti law \citep{Calzetti94}.
Some calibrations have limited ranges of validity, and here
 we only show results for galaxies that satisfy these conditions
(see \citealt{Kewley08} for details).

\begin{table}
\centering
\caption{Main characteristics of the different metallicity calibrations (see \citealt{Kewley08} for more details).}
\begin{tabular}{ccc}
\hline
Method  &  Class &  Reference \\
\hline \hline
T04 & theoretical  & \citet{Tremonti04} \\
KK04 & theoretical & \citet{Kobulnicky04} \\
$T_e$ & direct & \citet{Izotov06} \\
Z94 & theoretical & \citet{Zaritsky94} \\
KD02 & theoretical & \citet{Kewley02} \\
M91  & theoretical & \citet{Mcgaugh91} \\
D02  & combined & \citet{Denicolo02} \\
PP04 & empirical & \citet{Pettini04} \\
PP04(O3N2) & empirical & \citet{Pettini04} \\
P05  & empirical & \citet{Pilyugin10} \\
\end{tabular}
\label{table:metallicity_calibrations}
\end{table}

It is clear that the metallicities for the various emission-line calibrations show 
larger differences, up to  $\sim \pm 1$ dex.
The use of different calibrations is certainly a source of concern,
although in general the discrepancies are similar for all galaxies
regardless their metallicities, although variations are smaller for
the more metal-rich galaxies. 
In particular, the calibration based on electron 
temperature is in general lower than the others (see 
Sec.~\ref{subsec:gas_metallicity}), manifesting itself as
a clear offset, particularly for galaxies with (direct) 
metallicities higher than $12 + \log(O/H) \sim 8.2$. 
On the other hand, T04 and Z94 give the highest metallicity values,
with the other calibrations lying in between. 
For the most metal-rich galaxies,
a number of indicators agree relatively well, particularly
 KK04, KD02,  M91, and  even  T04 and Z94, with
a small spread of $\sim 0.1 -0.2$ dex (note that all of these
are theoretical calibrations). 
For the most metal-poor galaxies, we also find that various
of the calibrations agree relatively well with each other, still
with variations of $\pm 0.5$ dex.

\bsp

\label{lastpage}

\end{document}